\documentclass[
 pre,twocolumn,
 aps,floatfix,
]{revtex4-2}
\usepackage{graphicx}
\usepackage{tikz}
\usepackage{dcolumn}
\usepackage{bm}
\usepackage{amsmath}
\usepackage{amssymb}
\usepackage{amsthm}
\usepackage{epsfig}
\usepackage{epsfig}
\usepackage{color}
\usepackage[colorlinks=true,link color=blue]{hyperref}
\usepackage{caption}
\usepackage{subcaption}
\usepackage[margin=10pt,font=small,labelfont=bf,
labelsep=endash]{caption}
\newcommand{\kk}{\mathrm{k}}

\begin{document}
\title{Kuramoto model subject to subsystem resetting: How resetting a part of the system may synchronize the whole of it}
\author{Rupak Majumder}
\email{Corresponding author; Email: rupak.majumder@tifr.res.in}
\author{Rohitashwa Chattopadhyay}
\author{Shamik Gupta} 
\affiliation{Department of Theoretical Physics, Tata Institute of Fundamental Research, Homi Bhabha Road, Mumbai 400005, India}
\date{\today}

\begin{abstract}
We introduce and investigate the effects of a new class of stochastic resetting protocol called subsystem resetting, whereby a subset of the system constituents in a many-body interacting system undergoes bare evolution interspersed with simultaneous resets at random times, while the remaining constituents evolve solely under the bare dynamics. Here, by reset is meant a re-initialization of the dynamics from a given state. We pursue our investigation within the ambit of the well-known Kuramoto model of coupled phase-only oscillators of distributed natural frequencies. Here, the reset protocol corresponds to a chosen set of oscillators being reset to a synchronized state at random times. We find that the mean $\omega_0$ of the natural frequencies plays a defining role in determining the long-time state of the system. For $\omega_0=0$, the system reaches a synchronized stationary state at long times, characterized by a time-independent non-zero value of the synchronization order parameter that quantifies macroscopic order in the system. Moreover, we find that resetting even an infinitesimal fraction of the total number of oscillators, in the extreme limit of infinite resetting rate, has the drastic effect of synchronizing the entire system, even in parameter regimes in which the bare evolution does not support global synchrony. By contrast, for $\omega_0 \ne 0$, the dynamics allows at long times either a  synchronized stationary state or an oscillatory synchronized state, with the latter characterized by an oscillatory behavior as a function of time of the order parameter, with a non-zero time-independent time average. Our results thus imply that the non-reset subsystem always gets synchronized at long times through the act of resetting of the reset subsystem. Our results, analytical using the Ott-Antonsen ansatz as well as those based on numerical simulations, are obtained for two representative oscillator frequency distributions, namely, a Lorentzian and a Gaussian. Given that it is easier to reset a fraction of the system constituents than the entire system, we discuss how subsystem resetting may be employed as an efficient mechanism to control attainment of global synchrony in the Kuramoto system.
\end{abstract}
\maketitle

\section{\label{sec:level1}Introduction}
The idea of stochastic resetting was first put forward in the context of classical diffusion in the seminal papers by Evans and Majumdar~\cite{evans2011diffusion}. Under stochastic resetting,  the bare dynamics of a system is repeatedly reset at random times to a given state. It was inspired by the simple yet innovative idea that an algorithm using diffusion and designed to search for a misplaced target when restarted over and over again would result in stray dynamical paths being effectively cut off. This in turn would reduce the time of detection of the target, as measured by the mean first-passage time through the target, resulting in an enhanced efficiency of the search algorithm. In the last few years, the basic paradigm of stochastic resetting has been studied in a wide variety of set-ups spanning across domains, e.g., in the context of diffusion processes~\cite{pal2016diffusion, PhysRevE.96.022130,schadschneider}, in unveiling first order transition for the optimal search time of L\'{e}vy flights~\cite{PhysRevLett.113.220602}, in study of telegraphic processes~\cite{masoliver2019telegraphic}, in analyzing Markov processes~\cite{PhysRevE.92.062148}, in random walk problems~\cite{boyer2019anderson, PhysRevE.101.062147,mendez,majumdar2021record, das2022discrete,PhysRevE.107.064141}, in considering optimization in first-passage resetting~\cite{redner}, in study of first-passage properties of a particle in a potential~\cite{dibyendu} and also in an interval~\cite{PhysRevE.99.032123}, in studying first-passage of a diffusing particle in bounded domains under stochastic resetting~\cite{PhysRevE.105.034109}, in discussing random walks on complex networks and subject to first-passage resetting~\cite{PhysRevE.103.062132}, in Brownian search process in a homogeneous topography interpersed with resetting~\cite{sar2023resetting}, to mitigate long transient time in deterministic systems~\cite{ray2021mitigating}, in discussing large deviations of ratio observables in reset processes~\cite{rosemary}, in experimental studies related to estimating the optimal mean time required by a free diffusing Brownian particle to reach a target~\cite{ciliberto}, in analysis of scaled Brownian motion~\cite{chechkin}, in investigations of dynamical systems~\cite{ray2021mitigating}, in studying predator-prey models~\cite{toledo2019predator,evans2022exactly,rksingh}, in studies focussed on complex chemical processes such as the Michaelis--Menten reaction scheme~\cite{PhysRevE.92.060101, PhysRevResearch.3.013273}, in investigating run and tumble models for active matter~\cite{evans2018run, santra2020run, PhysRevE.102.042135, PhysRevE.106.044127}, in analyzing efficacy of antiviral therapies~\cite{ramoso2020stochastic}, in finance models~\cite{stojkoski2022income,jolakoski2023first}, etc. 
 
\begin{figure*}
	\begin{tikzpicture}
		\definecolor{lightgrey}{RGB}{200,200,200}
		\definecolor{darkgrey}{RGB}{100,100,100}
		
		\begin{scope}[shift={(0,0)}]
			\draw[dashed] (0,0) circle (1.5cm);
			\fill[lightgrey] (1.5,0) circle (0.2cm);
			\fill[darkgrey] (1.5,0.06) circle (0.2cm);
			\fill[lightgrey] (1.5,-0.05) circle (0.2cm);
			
			\draw (0,-2.8) -- (0,-3.2) node[left,xshift=0.5cm,yshift=-0.3cm] {$t = 0$};
		\end{scope}
		
		\node at (2.4,0) {$\ldots$};
		
		\begin{scope}[shift={(4.6,0)}]
			\draw[dashed] (0,0) circle (1.5cm);
			\fill[lightgrey] (1,1.1) circle (0.2cm);
			\fill[darkgrey] (0.7,-1.3) circle (0.2cm);
			\fill[lightgrey] (0.3,-1.45) circle (0.2cm);
			\fill[lightgrey] (0.3, 1.45) circle (0.2cm);
			\fill[darkgrey] (-0.6,1.4) circle (0.2cm);
			\fill[darkgrey] (-1.0,1.1) circle (0.2cm);
			\fill[lightgrey] (-1.35, 0.6) circle (0.2cm);

		\end{scope}
		
		\begin{scope}[shift={(8.1,0)}]
			\draw[dashed] (0,0) circle (1.5cm);
			\fill[lightgrey] (1,1.1) circle (0.2cm);
			\fill[darkgrey] (1.5,0) circle (0.2cm);
			\fill[lightgrey] (0.3,-1.45) circle (0.2cm);
			\fill[lightgrey] (0.3, 1.45) circle (0.2cm);
			\fill[darkgrey] (1.5,0.0) circle (0.2cm);
			\fill[darkgrey] (1.5,-0.0) circle (0.2cm);
			\fill[lightgrey] (-1.35, 0.6) circle (0.2cm);
			
			\draw[->,>=stealth] (-2,-2) -- (-0.2,-2.8);
			\draw (-0.2,-2.8) -- (-0.2,-3.2) node[left, xshift=0.45cm, yshift=0.8cm] {$t_{1^-}$};
			
			\draw (0,-2.8) -- (0,-3.2) node[left,xshift=0.35cm,yshift=-0.3cm] {$t_{1}$};
		\end{scope}
		
		\node at (10.5,0) {$\ldots$};
		
		\begin{scope}[shift={(12.8,0 )}]
			\draw[dashed] (0,0) circle (1.5cm);
			\fill[lightgrey] (0.5,1.4) circle (0.2cm);
			\fill[darkgrey] (1,1.15) circle (0.2cm);
			\fill[lightgrey] (0.1,-1.45) circle (0.2cm);
			\fill[lightgrey] (-0.2, 1.5) circle (0.2cm);
			\fill[darkgrey] (1,-1.15) circle (0.2cm);
			\fill[darkgrey] (1.25,-0.8) circle (0.2cm);
			\fill[lightgrey] (1.45,0.4) circle (0.2cm);
			
			\draw (0,-2.8) -- (0,-3.2) node[left,xshift=0.5cm,yshift=-0.3cm] {$t_{2^-}$};
		\end{scope}
		
		\draw[->, very thick] (-2,-3) -- (15,-3);
		
		\node[below] at (6.3,-3.6) {Time};
	\end{tikzpicture}
	\caption{\textbf{Schematic diagram of subsystem resetting}: The dark and light grey circles represent individual oscillators together forming our system of Kuramoto oscillators, where the position of a given circle denotes the phase of the corresponding oscillator with respect to an arbitrary but fixed axis. The dark grey circles represent the oscillators of the reset subsystem, while the light grey circles represent the oscillators of the non-reset subsystem. The horizontal arrow denotes the arrow of time. As shown in the left most diagram, all the oscillators at $t=0$ are set at the same phase value (which we take to be zero) and begin to evolve following the bare Kuramoto dynamics given by Eq.~\eqref{eq:1}; Note that the circles, which should have fallen on top of each other, have been displaced for convenience of visualization. The time instant $t_1$ denotes the time at which the first reset event occurs, while the time instant $t_{1^-}$ denotes the instant just before the first reset event. We observe that till time $t_{1^-}$, the entire system is undergoing the bare Kuramoto dynamics. At time $t_1$, the reset subsystem (constituted by the dark grey circles) resets to the phase value zero, while the non-reset subsystem (constituted by the light grey circles) are kept untouched. After the reset event, the entire system again undergoes the evolution given by the bare Kuramoto dynamics till the time instant $t_{2^{-}}$, while at time $t_2$, the second reset event takes place in exactly the same manner as the first reset event. The time evolution for a fixed total time comprises such alternating sequences of bare evolution and reset events.}
	\label{fig:1}
\end{figure*}
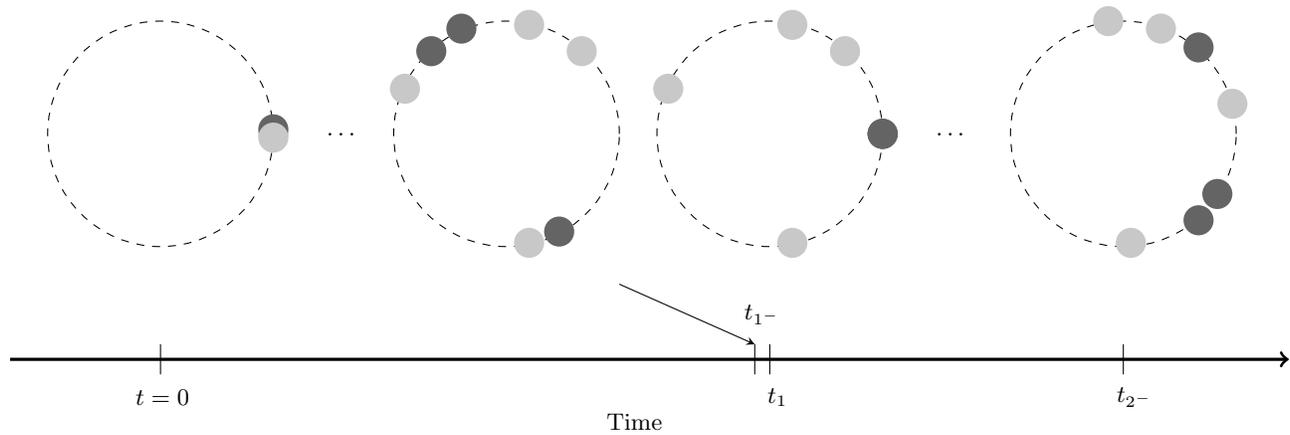 
 
 Stochastic resetting has also been studied in quantum systems, e.g., in the context of integrable and non-integrable closed quantum systems~\cite{mukherjee2018quantum}, and in studying the tight-binding model~\cite{das2022quantum,acharya2023tight} and the quantum random walk~\cite{yin2023restart}. Not only single-particle dynamics, stochastic resetting has also been implemented in many-body interacting systems, e.g., systems in the classical domain such as fluctuating interfaces~\cite{gupta2014fluctuating}, exclusion processes~\cite{basu2019symmetric,karthika2020totally}, classical spin systems~\cite{magoni2020ising}, coupled oscillators~\cite{sarkar2022,bressloff2024global}; on the quantum side, many-particle resetting  has been considered in the case of, e.g., quantum Ising chains~\cite{perfetto2021designing,fazio}.  A rather comprehensive discussion on stochastic resetting can be found in the recent reviews on the subject~\cite{evans2020stochastic, gupta2022stochastic, nagar2023stochastic}. 

In the case of many-body systems, resetting has been mostly studied in the particular context of global resetting, in which all the system constituents are \textit{simultaneously} reset to a given state at the time instances of reset~\cite{magoni2020ising,sarkar2022,perfetto2021designing}. A significant deviation from such a scenario comprises the case in which each constituent resets \textit{independently} of one other at random times, which goes under the name of local resetting~\cite{miron2021diffusion,bressloff2024global}. A dynamical scenario that lies between these two extremes is when only a given subset of the system constituents undergoes simultaneous resets at random times. Thus, we have a dynamical set-up, which we may refer to as subsystem resetting, wherein only a part of the system undergoes bare evolution interspersed with resets, while the rest of the system follows solely the bare evolution. Evidently then due to the interaction between the constituents in the reset and the non-reset part would one expect a strong interplay between bare evolution and reset moves. We may then already anticipate interesting emergent properties, both static and dynamic. Furthermore, in the case of global resetting, each reset event initiates afresh the dynamics of the entire system, resulting in all memory of dynamical evolution getting completely washed away each time a reset happens. Consequently, what matters in determining the state of the system at any given time instant is when did the last reset event take place~\cite{evans2011diffusion}. This may be contrasted with what happens under subsystem resetting, where evidently the non-reset part of the system retains memory of its entire dynamical evolution. The aforementioned scenario of subsystem resetting has not been explored in the literature, and which is the central theme of the current work. 

The set-up of subsystem resetting is particularly interesting to study in systems exhibiting an order-to-disorder transition. One may consider the paradigmatic classical Ising model comprising Ising spins, which with nearest-neighbor interactions exhibits a transition between a ferromagnetic (ordered) and a paramagnetic (disordered) phase in equilibrium and in dimensions greater than one. Another well-known model showing order-to-disorder transition, albeit from the domain of nonlinear dynamics and which will form the focus of our study in this work, is the so-called Kuramoto model~\cite{kuramoto1984chemical}. The model, which we discuss in detail below, involves globally-coupled phase-only oscillators of distributed natural frequencies that are interacting through the sine of the difference in phases between the oscillators. The Kuramoto model offers a novel framework to study spontaneous synchronization in many-body interacting systems~\cite{kuramoto1984chemical, strogatz2000kuramoto, acebron2005kuramoto, gupta2014kuramoto, rodrigues2016kuramoto, gupta2018statistical, pikovsky2015dynamics}. The utility of the model stems from the fact that it is simple enough to make analytical predictions while simultaneously capturing the bare essentials of the dynamics of synchronizing systems. Variations of this model have found applications in analyzing synchronization phenomena in widely disparate contexts, from yeast cell suspensions, flashing fireflies, firings of cardiac pacemaker cells, voltage oscillations in Josephson junction arrays to animal flocking behavior, pedestrians on footbridges, rhythmic applause in concert halls, and electrical power distribution networks~\cite{pikovsky2001universal,strogatz2004sync}. 

In the Kuramoto model, one observes a transition in the stationary state from an incoherent (disordered) to a synchronized (ordered) phase as one tunes the inter-oscillator interaction strength $K$ across a critical threshold $K_c$. Herein, order refers to a macroscopic fraction of oscillator phases evolving synchronously in time. The synchronization order parameter, which gives a measure of global phase coherence or synchrony in the system, has a time-independent value in the stationary state. The stationary order parameter has the value zero for $K$ values smaller than $K_c$, while it has a nonzero value for all $K\ge K_c$. In the setting of the Kuramoto model, one may ask the question: if we repeatedly reset a subset of the total number of oscillators to an ordered state, can this drive the rest of the oscillators towards order in the stationary state even when the bare dynamics fails to support such an ordering? Is there a critical size of the reset subsystem that ensures that one has ordering in the rest of the system? How does the rate of resetting play a role in inducing a given amount of synchronization in the non-reset subsystem? What is the fate of the synchronization transition exhibited by the bare model on including the effects of subsystem resetting? As possible applications of the introduced protocol, we may mention the following: (i) The case of the Kuramoto oscillators on complex networks~\cite{rodrigues2016kuramoto}, such as scale-free networks~\cite{Barabasi1999}, wherein one can reset only the hubs and investigate whether doing so allows to synchronize the entire system; (ii) The case of power grids~\cite{2012MSTWPRL} also provides an exciting avenue of application for subsystem resetting, wherein we can explore the conditions under which the failure of the complete grid can be avoided by resetting a small subset of the entire network.

It is pertinent to state at this point the specifics of the reset dynamics and the main results emerging from our study. Our resetting protocol involves all the oscillator phases of the reset subsystem being set to the value zero with a constant rate $\lambda > 0$. While the resetting protocol is obviously applied to the reset subsystem, our focus of study will be its effects on the non-reset subsystem. We have done our theoretical calculations in the continuum limit in which the total number of  oscillators in the system is infinite, $N \to \infty$, and have supported them with numerical simulations for $N = 10^4$ oscillators. In the limit $N\to \infty$, when the number of oscillators in both the reset  and the non-reset subsystem becomes infinite, a reasonable quantity denoting the size of the reset subsystem is the fraction $f$ of the total number of oscillators undergoing reset. Our main findings, pursued for  unimodal distributions for the natural frequencies of the oscillators, are the following: 
\begin{enumerate}
\item The mean $\omega_0$ of the oscillator frequencies plays a defining role in determining the long-time state of the system. This is unlike the bare Kuramoto dynamics, in which the effects of $\omega_0$ can be gotten rid off by viewing the dynamics in a co-rotating frame. 

\item Let us first discuss the limiting case $\lambda \to \infty$. In this limit, we unveil on the basis of exact analytical calculations a qualitative difference in the long-time behaviour between the case $\omega_0 = 0$ and the case $\omega_0 \neq 0$.  In the considered limit, we show that our system of interest, i.e., the non-reset subsystem, can be mapped to a version of the forced Kuramoto model studied in Ref.~\cite{Childs_2008}. The reset subsystem is on the other hand may be considered dynamically frozen in time in the reset configuration. This enables us to apply the analytically-tractable Ott--Antonsen (OA) ansatz~\cite{Ott_2008} to study the non-reset subsystem in terms of single-oscillator density functions; in this case, the dynamics of the non-reset subsystem remains confined to a low-dimensional manifold called the OA manifold. As a result, we are able to obtain an exact equation for the time evolution of the synchronization order parameter of the non-reset subsystem.

 (a) When one has $\omega_0=0$, our exact analytical results show that the non-reset subsystem is guaranteed to have a synchronized stationary state at long times; Quite remarkably, we find as a function of $K$ that the non-reset subsystem always exhibits synchrony, even for very small values of the parameter $f$ (see the plot for $f = 0.03$ in Fig.~\ref{fig:3}(b)). We see from our analysis  (specifically, Eq.~\eqref{eq:19} that the stationary value of the synchronization order parameter of the non-reset subsystem cannot be zero for any non-zero value of the parameters $K$ and $f$, thus suggesting existence of a synchronized stationary state for any $K$ and $f$. As a result, the synchronization transition as a function of $K$ exhibited by the bare dynamics gets converted into a crossover in presence of subsystem resetting in the limit $\lambda \to \infty$.
 
(b) In the case $\omega_0 \neq 0$, our exact analytical results again show that the non-reset subsystem always attains at long times a synchronized stationary state for $K \leq K_c$ irrespective of the value of $f$ (see the density plot in the $f-K$ plane in Fig.~\ref{fig:4}(a) and the plot for $f= 0.5$~and~$f = 0.8$ in Fig.~\ref{fig:4}(b)). However, for $K > K_c$, depending upon $f$, we may get for the non-reset subsystem at long times either a synchronized stationary state or an oscillatory synchronized state (specifically, an oscillatory behavior of the order parameter as a function of time, thus characterizing a non-stationary state of the dynamics, but which yields a non-zero time-independent time average). We have (i): $f$ is large, when one has a synchronized stationary state~(see the plot for $f = 0.5,K = 4.0$ in Fig.~\ref{fig:4}(c)), and (ii): $f$ is small, when one has an oscillatory synchronized state (see the plot for $f = 0.3,K = 4.0$ in Figs.~\ref{fig:4}(d) and ~\ref{fig:4}(e)). Thus, even with $\omega_0\ne 0$, the non-reset subsystem gets synchronized at long times through the act of resetting of the reset subsystem. 
 
 Interestingly, our analysis  done in the limit $\lambda\to\infty$ agrees well with our simulation results for $\lambda$ values as low as $25.0$ for the case of  $\omega_0 = 0$ (see Fig.~\ref{fig:3}(b), in which the results denoted by the black points and obtained from simulations for $f = 0.1$ and $\lambda = 25.0$ match reasonably well with the $\lambda \to \infty$ theoretical results) and as low as $100.0$ for the case of  $\omega_0 \neq 0$ (compare the agreement between the results denoted by the points and obtained from simulations with $\lambda = 100.0$ with the $\lambda \to \infty$ analytical results for $f = 0.8$ in Fig.~\ref{fig:4}(b) and for $f = 0.3$ in Fig.~\ref{fig:4}(d), all with $\omega_0 = 2.0$). 

On the basis of the above, we draw the following conclusion. As shown in Ref.~\cite{sarkar2022}, the Kuramoto model when subject to global resetting always reaches a stationary state at long times. By stark contrast, the dynamics in presence of subsystem resetting may or may not have a stationary state, depending on the values of the dynamical parameters and even in the limit in which resetting happens all the time (the limit $\lambda \to \infty$). 
 
\item For finite $\lambda$, one cannot map the non-reset subsystem to any version of the forced Kuramoto model. Here, the dynamical evolution of the two individual subsystems, namely, the reset and the non-reset subsystem, remains confined to respective OA manifolds. It is interesting though to note that the evolution of the system as a whole does not take place on an OA manifold. Thus, we have two time evolution  pertaining to the two subsystems taking place simultaneously on two different OA manifolds, which are dependent on one another and hence coupled.  Even in such a set-up, we are able to find approximate analytical evolution equations for the realization-averaged order parameters of the reset and the non-reset subsystem, for large $\lambda$. The problem of finding the value(s) of these order parameters in the stationary state, provided the latter exists, is reduced to finding the roots of two (for $\omega_0 = 0$) or four (for $\omega_0 \neq 0$) coupled nonlinear equations. 

(a) For $\omega_0 = 0$, our approximate analytical calculations predict the non-reset subsystem to have a synchronized stationary state at long times for any nonzero value of $f$ and $\lambda$, thereby converting the synchronization transition of the bare dynamics into a crossover, just as in the case of the $\lambda \to \infty$ limit discussed above. We have verified our approximate theory by a comparison with simulation results for $\lambda$ values as low as $1.0$ (see Fig.~\ref{fig:5}(a), in which the results denoted by the black points and obtained from simulations for $f = 0.1$ match reasonably well with our theoretical results). 

(b) For $\omega_0 \neq 0$,  our approximate analytical calculation predicts that the non-reset subsystem always attains at long times a synchronized stationary state for $K \leq K_c$. On the other hand, in the $K > K_c$ region, our approximate analytical calculation predicts the long-time behaviour of the non-reset subsystem to be either a synchronized stationary state at large $f$ or an oscillatory synchronized state at small $f$, much like the $\lambda \to \infty$ scenario. A comparison of theoretical with numerical results for $\lambda=5.0$ demonstrates for large $f$ values an agreement in the $K \leq K_c$ region, and not so good an agreement for $K>K_c$ (see Fig.~\ref{fig:6}(b), where the simulation results are denoted by squares and circles, and which correspond to $f = 0.5$ and $f=0.8$, respectively). For small $f$, see Fig.~\ref{fig:6}(a), our numerical simulation results display oscillations with decaying amplitude and eventual settling down to a synchronized stationary state, which is at variance with our theory that predicts an oscillatory synchronized state at long times.

\item All the above results are obtained by taking a Lorentzian distribution of natural frequencies for the oscillators.  Recently,  an approximation scheme has been developed that allows treating the bare Kuramoto model using the OA ansatz even for a Gaussian distribution of natural frequencies of the oscillators.~\cite{Campa_2022}. We apply the method of analysis in Ref.~\cite{Campa_2022} to our system that involves subsystem resetting. Here, in the $\lambda\to\infty$ limit and with a Gaussian frequency distribution, we are able to derive analytically an approximate equation for the time evolution of the order parameter of the non-reset subsystem. We verify the solution of this equation in the long-time limit by means of numerical simulations. For the small-$\lambda$ case and with a Gaussian distribution of frequencies, obtaining an analytical evolution equation for the order parameter of the non-reset subsystem becomes intractable, and thus, we are restricted to obtaining only simulation results. We find that our results for the Gaussian case are qualitatively similar to those for the Lorentzian distribution of the oscillator frequencies. 
\end{enumerate}

The paper is organized as follows. In Sec.~\ref{sec:level2}, we define in detail the Kuramoto model and the dynamics in the presence of stochastic resetting of a subsystem. In Sec.~\ref{sec:level3}, we analyze the dynamics for the case of a Lorentzian distribution of natural frequencies of the Kuramoto oscillators. In this section, we first discuss the $\lambda \to \infty$ case in Sec.~\ref{sec:level4} for both of the $\omega_0 = 0$ and $\omega_0 \neq 0$ cases, respectively. Then in Sec.~\ref{sec:level5}, we take up the finite-$\lambda$ case for both the choices $\omega_0 = 0$ and $\omega_0 \neq 0$, respectively. In Sec.~\ref{sec:level9}, we elaborate on the relevance of our analysis in controlling the extent of synchronization in our system. In Sec.~\ref{sec:Gaussian distribution}, we present our analysis of the dynamics by going beyond the Lorentzian distribution, namely, for a Gaussian distribution. Finally, the paper ends with conclusions in Sec.~\ref{sec:label10}. Some of the technical details are relegated to the three appendices.

\section{\label{sec:level2}The Kuramoto model in presence of subsystem resetting}
\subsection{\label{sec:level2.1}The bare Kuramoto model}
As mentioned in the Introduction, the Kuramoto model involves a system of $N$ globally-coupled phase-only oscillators. We denote by $\theta_j (t) \in [0, 2 \pi )$ the phase of the $j-$th oscillator at time $t$, with $j=1,2,\ldots,N$. The phase evolution in time is defined within the Kuramoto model as \cite{kuramoto1984chemical}
\begin{equation}
\frac{d \theta_j}{dt} = \omega_j + \frac{K}{N} \sum_{k = 1}^N \sin(\theta_k - \theta_j).
\label{eq:1}
\end{equation}
Here, the constant $K$ denotes the inter-oscillator coupling strength, which is scaled by $N$ to ensure effective competition between the first and the second term on the right-hand side of the above equation in the continuum limit $N \to \infty$. In this paper, we will consider the case $K \geq 0$ for all further calculations.
In Eq.~\eqref{eq:1}, the natural frequencies $\omega_j \in (-\infty, \infty)$ of the oscillators are quenched-disordered random variables distributed according to a given probability distribution $g(\omega)$, which has a finite mean $\omega_0 \geq 0$ and a finite width $\sigma \ge  0$. We assume the distribution to be unimodal and symmetric about the mean $\omega_0$.

The Kuramoto system is capable of exhibiting rich dynamics due to the interplay between randomness and coupling. The randomness originates from the variation in the natural frequency among the oscillators; in the absence of coupling, each oscillator phase tends to rotate uniformly and independently in time. This results in the individual phases being scattered uniformly and independently in $[0,2\pi)$ at large times, leading to an unsynchronized/incoherent state. On the other hand, the counter effect is provided by the coupling among the oscillators, which tends to make the oscillators acquire the same phase, thereby leading to a synchronized state. Depending upon the relative magnitude of these two competing effects, one observes within the Kuramoto dynamics in the limit $N \to \infty$ and in the stationary state a synchronization phase transition, or, more precisely, a supercritical bifurcation~ \cite{kuramoto1984chemical,strogatz2000kuramoto,gupta2018statistical}. The transition takes place between a low-$K$ unsynchronized/incoherent phase and a high-$K$ synchronized phase across a critical $K$, denoted by $K_c$.

The aforementioned phase transition is characterized by the synchronization order parameter $z(t)=r(t)e^{i\psi(t)}$, defined as \cite{kuramoto1984chemical,strogatz2000kuramoto}
\begin{equation}
    r(t)e^{i \psi (t)} \equiv \frac{1}{N} \sum_{j = 1}^N e^{i \theta_j(t)}.
    \label{eq:3}
\end{equation}
Thus, the complex quantity $z(t)$ is represented in the complex-$z$ plane by a vector of length $r(t)$ inclined at an orientation angle $\psi(t)$ with respect to an arbitrary reference. We refer to $z(t)$ as the synchronization vector.
The real quantity $r(t)$ measures the amount of global phase synchrony present in the system at time instant $t$, while the real quantity $\psi(t) \in [0,2\pi)$ quantifies the average phase. Clearly, we have $ 0 \leq r \leq 1$, with $r=0$ (respectively, $r=1$) implying incoherence (respectively, perfect synchrony). Any value of $r$ in the range $0<r<1$ implies partial synchrony. In terms of the quantities $r$ and $\psi$, the dynamics~\eqref{eq:1} takes the form
\begin{equation}
    \frac{d \theta_j}{d t} = \omega_j + K r(t) \sin(\psi(t) - \theta_j),
    \label{eq:2}
\end{equation}
which makes it evident the mean field nature of the Kuramoto dynamics: every oscillator phase evolves in the presence of a mean-field common to all, which is generated by the interaction among all the oscillators.

In this paper, we will consider the case $\sigma > 0$. This is because with $\sigma=0$, implying same $\omega_j$ for all the oscillators, the system will trivially synchronize. Specifically, we will be working with two representative frequency distributions, namely, a Lorentzian and a Gaussian, given respectively by
\begin{equation}
    g(\omega)=
\begin{cases}
\frac{\sigma}{\pi} \frac{1}{(\omega-\omega_0)^2 + \sigma^2}; &\hspace{0.3cm} (\mathrm{Lorentzian}),\\
\frac{1}{\sqrt{2 \pi \sigma^2}}\exp \left( - \frac{(\omega-\omega_0)^2}{2 \sigma^2} \right) ;&\hspace{0.3cm} (\mathrm{Gaussian}).
\end{cases}
\label{eq:11}
\end{equation}
For these distributions, the critical $K$ to observe the synchronization phase transition  can be calculated explicitly, as was first achieved by Kuramoto; One has \cite{kuramoto1984chemical,strogatz2000kuramoto,gupta2018statistical}
\begin{equation}
    K_c = \frac{2}{\pi g(\omega_0)}. \label{eq:39}
\end{equation}
For $K$ in the range $0 \leq K \leq K_c$, the system is in the unsynchronized phase with the stationary-state order parameter given by $r_\mathrm{st} \equiv r(t \to \infty ) = 0$. On the other hand, for $K > K_c$, the system is in the synchronized phase with $0 < r_\mathrm{st} \leq 1$. 

\subsection{The model in presence of subsystem resetting} \label{sec:Dynamics}
Following the analysis of Kuramoto, we know that even if we start from a fully-synchronized state, the system for $K \leq K_c$ will always desynchronize in time and the quantity $r$ will settle down at long times to the value zero in the continuum limit $N \to \infty $. Keeping this in mind, the situation we are interested in within the framework of the Kuramoto model is the following (see Fig.~\ref{fig:1}): We initialize the dynamics from a fully synchronized state, i.e., $\theta_j(0) = 0~\forall~j$. Then, at random times, the bare Kuramoto evolution following Eq.~\eqref{eq:2} is interrupted, whereby a given number of oscillators are reset back to their initial phase value, i.e., zero. The random times at which the system undergoes a reset vary from one realization of the dynamics to another. A priori we do not have any bias about which oscillators are to be reset. Specifically, we choose once and for all the subset of oscillators to be reset, and keep resetting the same subset across different realizations and different reset times. We construct this subset by choosing uniformly and independently $n$ distinct oscillators out of the total $N$ oscillators, with $n<N$. The details of the numerical implementation are summarized in Appendix~\ref{app:1}.

We define the reset subsystem as the one constituted by the oscillators that are being reset, while the rest of the oscillators form the non-reset subsystem. Then the question we are interested in is the following: By resetting the oscillators of the reset subsystem, are we able to synchronize the non-reset subsystem? Since the reset subsystem will get synchronized under the dynamics of resetting because of the choice of the resetting state, synchronization of the non-reset subsystem implies synchronization of the system as a whole. If the answer to the question just raised is yes, then this question is followed by a plethora of interesting questions: How does this inducing of synchronization in the non-reset subsystem through resetting of the reset subsystem depend on the resetting rate, the size $n$ of the reset subsystem, and the interaction strength $K$? For $K \leq K_c$, the bare dynamics does not support synchronization. Can we then make the system synchronized for $K \leq K_c$?  For $K>K_c$, even the bare dynamics supports synchronization, and then the question will be: Can resetting increase the amount of synchronization in the system than what is achieved in the bare dynamics? If it so turns out that the system in presence of resetting is synchronized for all values of $K$, the phase transition in the bare dynamics would turn into a crossover in stationary values of $r$  as a function of $K$. Thus, we may ask: Does the phase transition with respect to $K$ in the bare model persist even on including resetting in the bare dynamics? 

In this paper, we set out to answer the aforementioned questions. The random times at which resets happen are considered to follow a Poisson point process with rate $\lambda$. This implies that the random variable $\tau$, denoting the time interval between two successive resets, is distributed according to an exponential distribution
\begin{equation}
    p(\tau) = \lambda e^{-\lambda \tau}; \hspace{0.5cm} \lambda \geq 0, \hspace{0.5cm} \tau \in [0,\infty).
    \label{eq:ptau}
\end{equation}
The dynamics may then be defined as follows: During the infinitesimal interval $dt$ following any time instant $t$, the state $\{\theta_j(t)\}$ of the system evolves following the dynamics~\eqref{eq:2} with probability $(1-\lambda dt)$. On the other hand, with the complementary probability $\lambda dt$, the phases of the $n$ oscillators constituting the reset subsystem are all reset to the value zero, while the phases of the oscillators forming the non-reset subsystem evolve following the dynamics~\eqref{eq:2}. Here, the parameter $\lambda$ is the resetting rate, while the quantity $1/\lambda$ denotes the average time between two successive resets. Evidently then, on setting $\lambda$ to zero, the dynamics in presence of resetting reduces to the bare Kuramoto dynamics~\eqref{eq:2}. 

We will analyze the above-mentioned dynamics in the limit $N\to \infty$, where a more reasonable quantity than $n$ to characterize the size of the reset subsystem is the fraction $f\equiv n/N$ of the total number of oscillators undergoing the reset. From now on, we will use the quantities $f$ and $1-f$ to denote the size of the reset and the non-reset subsystem, respectively. Note that $f$ varies in the range $0\le f \le 1$. It is pertinent at this point to list down the various parameters that characterize our dynamics. These are the following: (i) the coupling $K$, (ii) the resetting rate $\lambda$, (iii) the fraction $f$ of the total number of oscillators undergoing the reset, and (iv) the parameters $\omega_0$ and $\sigma$ characterizing the frequency distribution $g(\omega)$.

As the reset and the non-reset subsystem are subjected to two different schemes of evolution, it is appropriate to track separately the order parameter $(r_\mathrm{r}(t),\psi_\mathrm{r}(t))$ of the reset subsystem as well as the order parameter $(r_\mathrm{nr}(t),\psi_\mathrm{nr}(t))$ of the non-reset subsystem. Given a set of $N$  Kuramoto oscillators, we will from now on label the oscillators in such a way that the ones with the label $j = 1, 2,\ldots , n$ form the reset subsystem, while the ones with the label $j=n+1,n+2,\ldots,N$  form the non-reset subsystem. Consequently, we have 
\begin{eqnarray}
    z_\mathrm{r}(t) \equiv r_\mathrm{r}(t)e^{i \psi_\mathrm{r} (t)} &\equiv & \frac{1}{n} \sum_{j = 1}^n e^{i \theta_j(t)},\\
    z_\mathrm{nr}(t) \equiv r_\mathrm{nr}(t)e^{i \psi_\mathrm{nr} (t)} &\equiv & \frac{1}{N-n} \sum_{j = n+1}^N e^{i \theta_j(t)}.
\end{eqnarray}
The quantities on the left, together with the global order parameter $(r(t),\psi(t))$ defined in Eq.~\eqref{eq:3} make up the observables of the problem at hand. For the later, it is straightforward to see that
\begin{eqnarray}
z(t)=fz_\mathrm{r}(t)+(1-f)z_\mathrm{nr}(t),
\end{eqnarray}
which immediately implies on rewriting the $z$'s in terms of $r$'s and $\psi$'s, as $r(t)e^{i\psi(t)}=fr_\mathrm{r}(t)e^{i\psi_\mathrm{r}(t)}+(1-f)r_\mathrm{nr}(t)e^{i\psi_\mathrm{nr}(t)}$ that
\begin{equation}
	r = \sqrt{f^2 r^2_\mathrm{r} + (1-f)^2 r^2_\mathrm{nr} + 2 f (1-f) r_\mathrm{r} r_\mathrm{nr} \cos(\psi_\mathrm{r}-\psi_\mathrm{nr})}.
\end{equation}

\begin{figure*}
	\begin{center}
		\begin{subfigure}[b]{0.24\textwidth}
			\includegraphics[width=\textwidth]{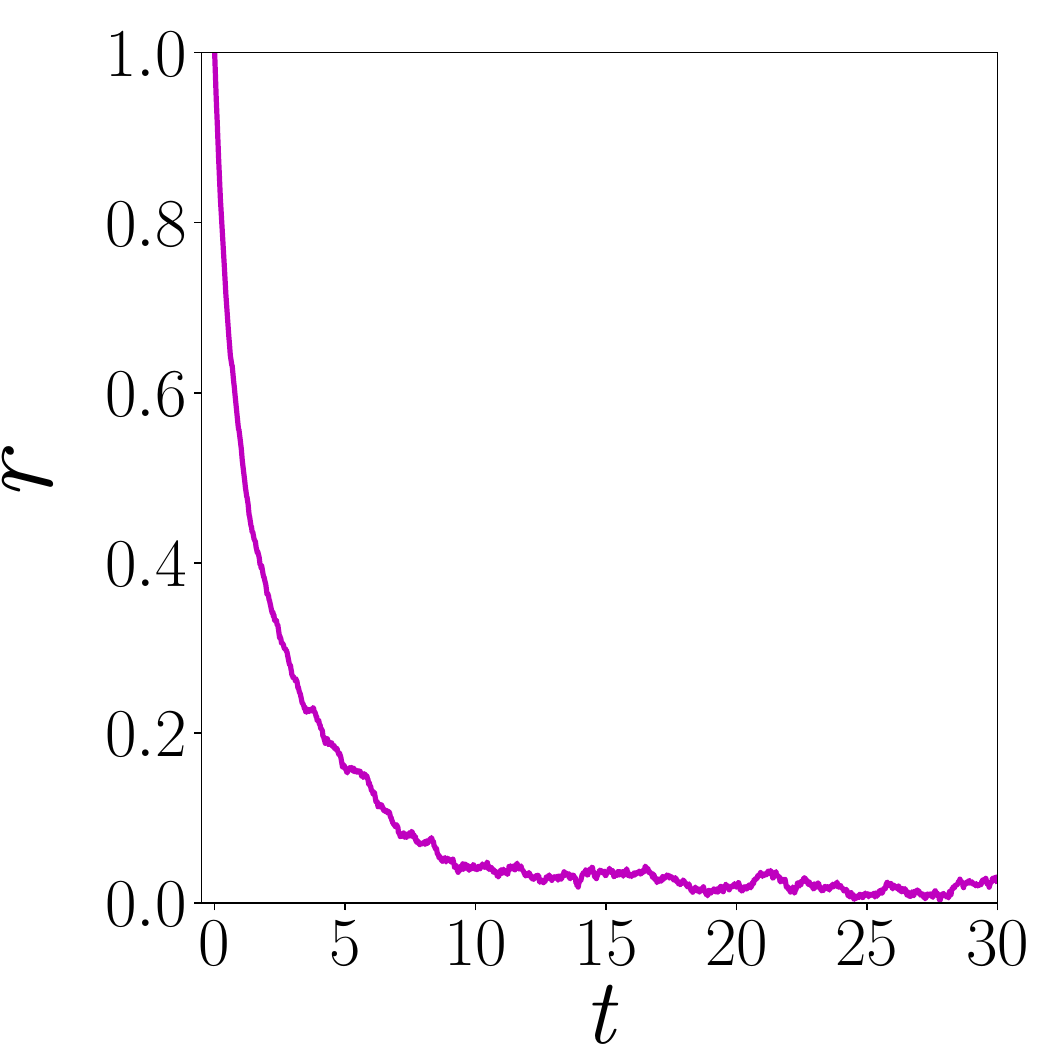}
			\caption{}
		\end{subfigure}
		\begin{subfigure}[b]{0.24\textwidth}
			\includegraphics[width=\textwidth]{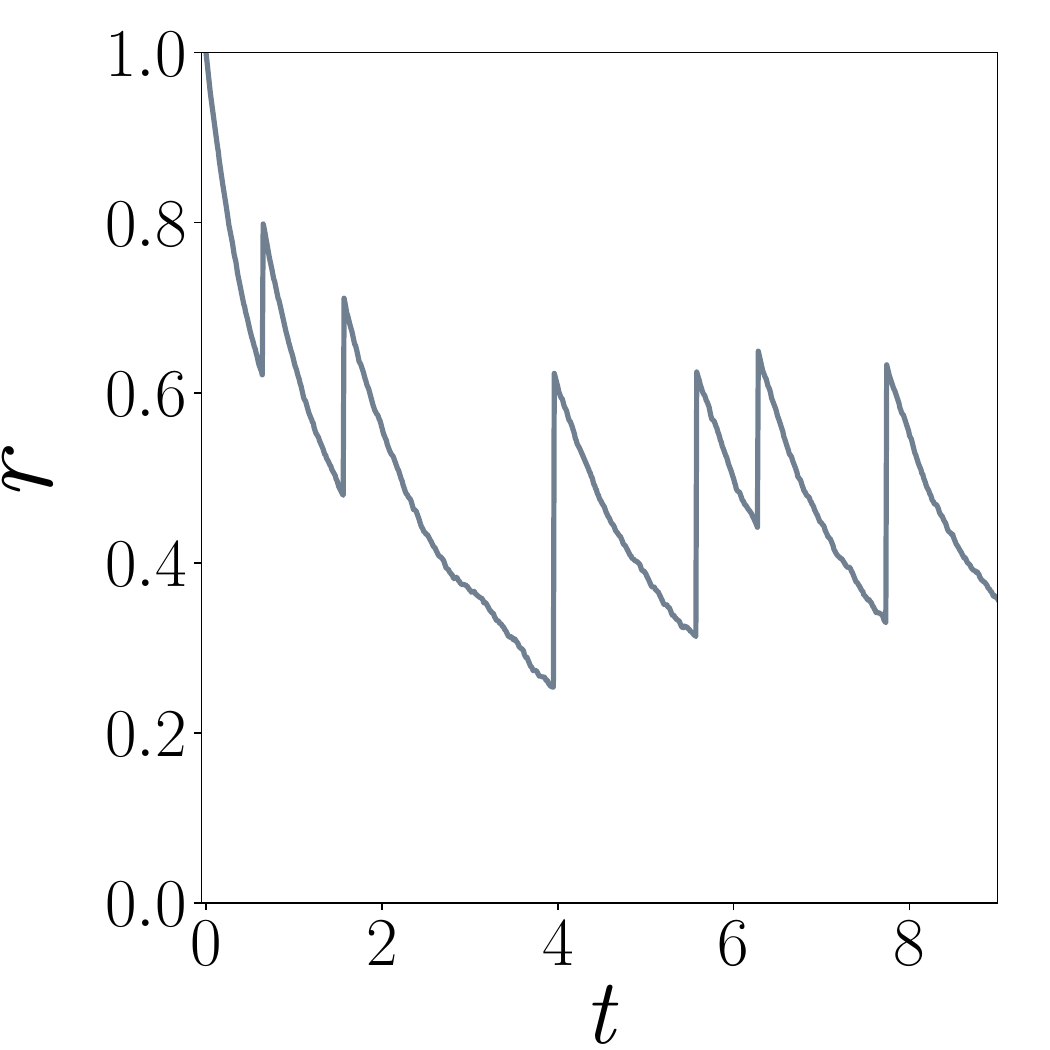}
			\caption{}
		\end{subfigure}
		\begin{subfigure}[b]{0.24\textwidth}
			\includegraphics[width=\textwidth]{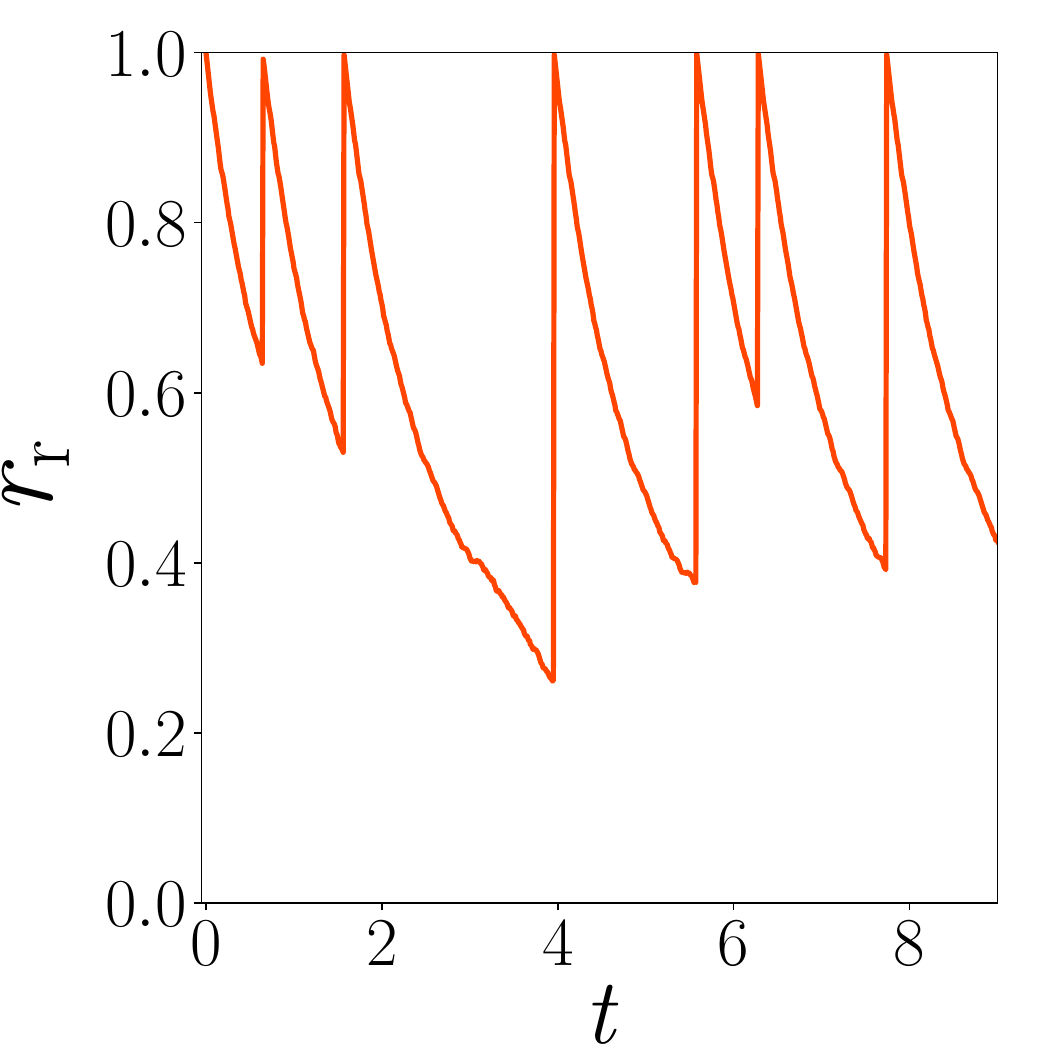}
			\caption{}
		\end{subfigure}
		\begin{subfigure}[b]{0.24\textwidth}
			\includegraphics[width=\textwidth]{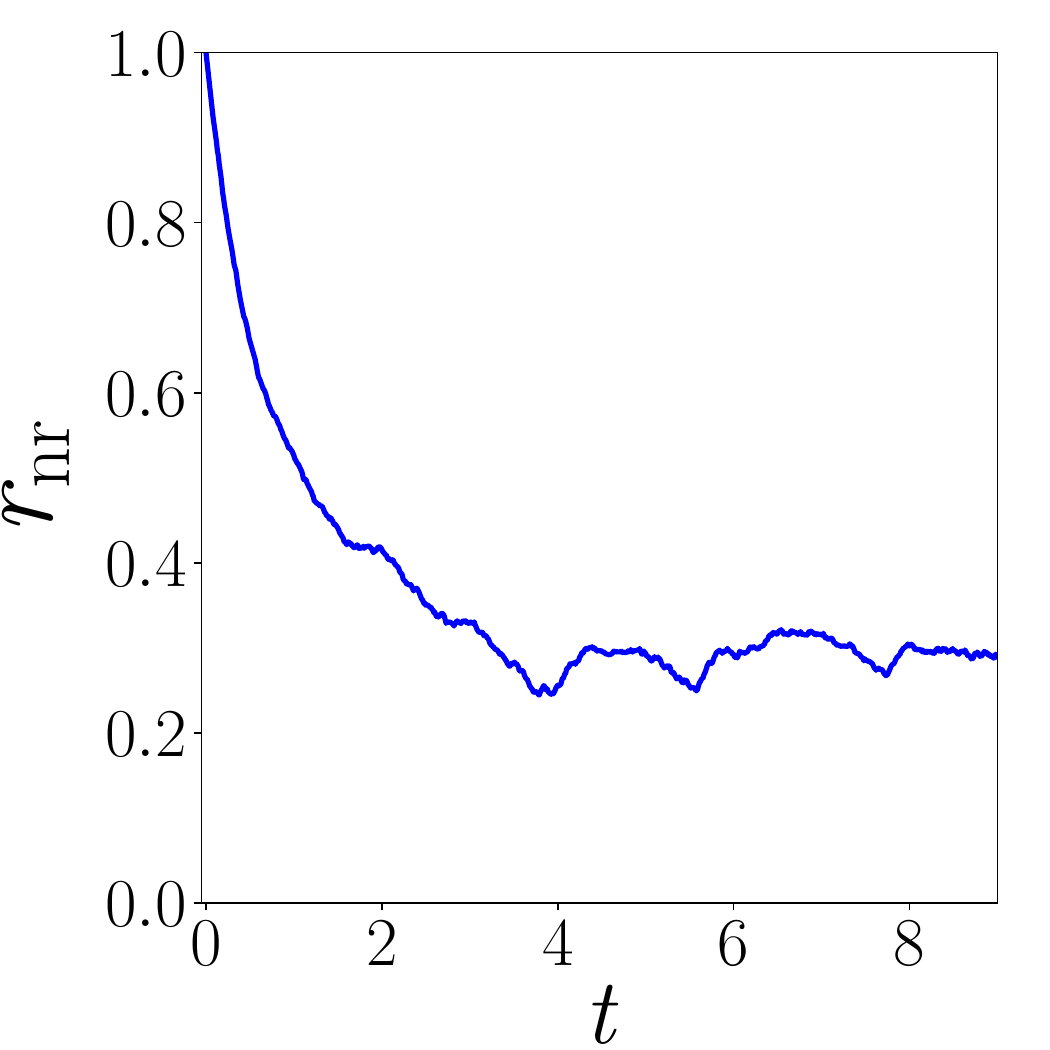}
			\caption{}
		\end{subfigure}          
		\caption{\textbf{Subsystem resetting with finite $\lambda$ and Lorentzian $g(\omega)$ with mean $\omega_0=0$ and width $\sigma=1.0$}:
			Time variation of the order parameter $r$ of the entire system in the case of bare dynamics (panel (a)) and in the case of bare dynamics interpersed with subsystem resetting (panel (b)), respectively. In the case of dynamics with resetting, time variation of the order parameters $r_\mathrm{r}$ (panel (c)) and $r_\mathrm{nr}$ (panel (d)) pertaining to the reset subsystem and the non-reset subsystem, respectively have also been shown. The values of the various parameters are: $K=1.5$, $\lambda=0.5$, $f=0.5$. The data correspond to a single realization of the dynamics, and are obtained from numerical simulation of the dynamics for a system of $N=10^4$ oscillators; the integration time step is taken to be $0.01$. The initial state has $\theta_j(0)=0~\forall~j$, and is the same for all figures in this paper that depict numerical simulation results.}
		\label{fig:2}
	\end{center}
\end{figure*}

From now on, we will drop the explicit time dependence in the quantities $z_\mathrm{r}(t),z_\mathrm{nr}(t),r_\mathrm{r}(t),\psi_\mathrm{r}(t),r_\mathrm{nr}(t),\psi_\mathrm{nr}(t)$, unless specified otherwise. Note that because of resetting of phases of all the reset oscillators to the same value, the quantity $r_\mathrm{r}$ hits the value of unity at the time instances of reset. A representative variation of the mentioned observables as a function of time in a single realization of the dynamics and for chosen values of the various parameters is shown in Fig.~\ref{fig:2}. In panel (c), the sudden changes in the value of $r_\mathrm{r}$ from its current value to unity denote the resetting instances. Note that all the plots in Fig.~\ref{fig:2} have been obtained for a value of $K$ (here $K = 1.5$) for which the bare dynamics fails to attain synchrony (see panel (a) of Fig.~\ref{fig:2}, where we may see the quantity $r$ settling to the zero value with time even when the oscillators are initiated from a perfectly synchronized state). On the other hand, when resetting is introduced in the dynamics, even the oscillators not getting reset become synchronized (see panel (d) of Fig.~\ref{fig:2}). Our goal is to explore synchronization of the oscillators even when the bare dynamics can not afford synchrony. The reset subsystem will always get synchronized because of the direct influence of resetting. If we can show that the non-reset subsystem is also getting synchronized because of resetting of the reset subsystem, it will in turn imply that the entire system is getting synchronized. That is why from now on we will only focus on the behaviour of the synchronization order parameter of the non-reset subsystem. Another reason for studying $r_\mathrm{r}$ and $r_\mathrm{nr}$ separately and not the quantity $r$ is that we are mainly interested in how resetting a part of the system influences the other part of the system.

\section{\label{sec:level3}Analysis for Lorentzian $g(\omega)$}
In this section, we proceed to analyze the Kuramoto model in presence of subsystem resetting, for the particular case of the Lorentzian frequency distribution given in Eq.~\eqref{eq:11}. We consider successively the case of infinite and finite resetting rate $\lambda$. In our analysis, we will use the fact that for the Lorentzian $g(\omega)$,  the critical $K$ to observe synchronization may be obtained from Eq.~\eqref{eq:39} as $K_c=2\sigma$.

\subsection{\label{sec:level4}$\lambda \to \infty$ Limit: Case for maximum possible induced-synchronization}
Before we embark on a detailed analysis of our system for arbitrary values of the resetting rate $\lambda$, let us ask the following question: For a given value of the parameter $f$, what is the maximum amount of synchronization that can be induced in the non-reset subsystem? One may expect that maximum synchronization in the non-reset subsystem will be achieved in the limit of infinite resetting rate, i.e., $\lambda \to \infty$. This is because, in this case, the reset subsystem remains perfectly synchronized almost at all times. Then, it is expected on the basis of the coupling between the reset and the non-reset subsystem that, if and when possible, the non-reset subsystem will acquire maximum possible synchrony. 

In the limit of infinite $\lambda$, the phases of all the oscillators in the reset subsystem can be considered to be fixed at the resetting angle, i.e., $\theta_j(t) = 0~\forall~t$, and $j = 1, \cdots, n$. Using this in Eq.~\eqref{eq:1}, we get the evolution equations for the oscillators in the non-reset subsystem, i.e., for $j = n+1, \ldots, N$, as
\begin{equation}
    \frac{d \theta_j}{dt} = \omega_j - Kf \sin(\theta_j) + \frac{K}{N}\sum_{k = n +1}^N \sin(\theta_k - \theta_j).
    \label{eq:12}
\end{equation}
Interestingly, this equation breaks the phase-shift symmetry that is present in the bare Kuramoto dynamics~\eqref{eq:1}. In other words, rotating every phase by an arbitrary amount (same for all), i.e., effecting the transformation $\theta_j \to \theta_j+\alpha~\forall~j$ and arbitrary $\alpha$, leaves the dynamics~\eqref{eq:1} invariant but not the dynamics~\eqref{eq:12}.  In terms of the quantities $r_\mathrm{nr}$ and $\psi_\mathrm{nr}$, Eq.~\eqref{eq:12} can be re-written as
\begin{equation}
    \frac{d \theta_j}{dt} = \omega_j - Kf \sin(\theta_j) + K(1-f) r_\mathrm{nr} \sin(\psi_\mathrm{nr} - \theta_j),
    \label{eq:4}
\end{equation}
with $j = n+1, \cdots , N$. From the above equation, it is evident that resetting at infinite rate effectively adds a constant forcing term of strength $Kf$ acting on the individual non-reset oscillators, represented by the second term on the right hand side of Eq.~\eqref{eq:4}. With the increase of the fraction $f$, the strength of this forcing term increases. On the other hand, the effective coupling  between the oscillators of the non-reset subsystem, given by $K(1-f)$, decreases with increasing $f$. Under the effect of the forcing term, the phases of the non-reset oscillators tend to become zero. This would imply fixing the orientation of the synchronization vector $z_\mathrm{nr}(t)$ of the non-reset subsystem at the value $\psi_\mathrm{nr}=0$. By contrast, the mutual interaction between the non-reset oscillators tends to rotate the same synchronization vector in time, provided one has $\omega_0\ne 0$. For $\omega_0 = 0$, the effects of forcing and mutual interaction thus go hand in hand in synchronizing the non-reset subsystem. For non-zero $\omega_0$, these two effects oppose each other, which we will show below to be resulting in a transition depending on the values of $\omega_0, f$ and $K$.

A variation of Eq.~\eqref{eq:12} has been studied by Childs and Strogatz \cite{Childs_2008} in the context of periodic forcing of the bare Kuramoto model and with a Lorentzian frequency distribution. We will follow their analysis, which makes use of the celebrated Ott-Antonsen (OA) ansatz. The latter offers a powerful exact method to study dynamics of coupled oscillator ensembles.  In the context of the bare Kuramoto model~\eqref{eq:1} with a Lorentzian distribution of the oscillator frequencies, the ansatz studies the evolution in phase space by considering a particular class of single-oscillator density functions that is defined on a low-dimensional manifold $\mathcal{M}\subset \mathcal{D}$ called the OA-manifold, the manifold being embedded in the space $\mathcal{D}$ of all possible density functions. A remarkable feature of the manifold is that when initialized on the manifold, the subsequent evolution of the system remains confined to it \cite{Ott_2008,Ott_2009}. This enables one to write down a single first-order ordinary differential equation for the evolution of the synchronization order parameter $r(t)$.  As we will discuss below, for our system~\eqref{eq:12}, the initial condition $\theta_j(0) = 0$ for $j = n+1, \cdots , N$ lies on the OA-manifold. Considering evolution under the dynamics~\eqref{eq:12} also to remain confined to this manifold~\cite{Ott_2008,Ott_2009,Childs_2008}, the order parameter $(r_\mathrm{nr},\psi_\mathrm{nr})$ may then be shown to satisfy a two-dimensional dynamical system. We now turn to details of the derivation of this dynamical system, following Ref.~\cite{Childs_2008}.

To proceed, we consider the dynamics~\eqref{eq:12} in the continuum limit $N \to \infty$. This implies that the size of both the reset subsystem, given by $Nf$, and the non-reset subsystem, given by $N(1-f)$, is also infinite. Then the state of the non-reset subsystem can be described using a single-oscillator density function $F(\theta, \omega, t)$. By definition, $F(\theta, \omega, t) d \theta d \omega$ denotes the fraction of non-reset oscillators at time $t$ that have their phase values between $\theta$ and $\theta 
+ d \theta$ and have their natural frequency between $\omega$ and $\omega + d \omega$. Normalization of the density function implies that one has
\begin{eqnarray}
    \int_0^{2 \pi} F(\theta, \omega, t) d \theta &=& g(\omega) \label{eq:5},\\
    \int_{- \infty}^{\infty} \int_0^{2 \pi} F(\theta, \omega, t) d \theta d \omega &=& 1.
\end{eqnarray}

It is evidently true that the number of oscillators with a given frequency is conserved under the dynamics~\eqref{eq:12}. As a consequence, $F(\theta, \omega, t)$ evolves in time according to the continuity equation
\begin{eqnarray}
    \frac{\partial F}{\partial t} + \frac{\partial}{\partial \theta} \left( F \frac{d \theta}{d t} \right) = 0,
    \label{eq:6}
\end{eqnarray}
wherein the quantity $d \theta/ d t$ is obtained from Eq.~\eqref{eq:4} in the continuum limit as
\begin{eqnarray}
    \nonumber
    \frac{d \theta}{d t} = \omega &+& \frac{1}{2 i} [ (K(1-f) z_\mathrm{nr} +  Kf) e^{-i \theta}\\
    &&- (K(1-f) z_\mathrm{nr}^{*} +  Kf) e^{i \theta} ] \label{eq:26},
\end{eqnarray}
where we have
\begin{equation}
    z_\mathrm{nr} = r_\mathrm{nr} e^{i \psi_\mathrm{nr}} = \int_{-\infty}^{\infty} \int_{0}^{2 \pi} e^{i \theta} F(\theta,\omega,t) d \theta d \omega, \label{eq:7}
\end{equation}
and the star denotes complex conjugation.  

Now, using the $2 \pi$ periodicity of $F(\theta,\omega, t )$ with respect to $\theta$, we obtain the Fourier expansion
\begin{eqnarray}
    F(\theta,\omega , t ) = \frac{g(\omega)}{2 \pi} \left[ 1 + \sum_{\substack{n = - \infty \\ n \ne 0 }}^{\infty} \tilde{F}_n(\omega,t)e^{i n\theta} \right],
    \label{eq:8}
\end{eqnarray}
where $\tilde{F}_n(\omega,t)$ is the $n$-th Fourier coefficient. Using $\int_0^{2 \pi} e^{i n \theta} d \theta = 2 \pi \delta_{n,0}$, it can be easily checked that the above expansion indeed satisfies Eq.~\eqref{eq:5}. As $F(\theta,\omega , t )$ is a real quantity, we have a further condition on the Fourier coefficients:
\begin{equation}
    \tilde{F}_{-n}(\omega,t) = \left[ \tilde{F}_n(\omega,t) \right]^{*}.
\end{equation}
Substitution of Eqs.~\eqref{eq:8}~and~\eqref{eq:26} in Eq.~\eqref{eq:6} yields an infinite number of coupled non-linear differential equations for the coefficients $\tilde{F}_{n}(\omega,t)$, which are analytically intractable. At this point, Ott and Antonsen suggested that if we only consider the special class of density functions $F(\theta,\omega,t)$ for which
\begin{equation}
    \tilde{F}_n(\omega,t) = \left[ \alpha(\omega,t) \right]^n
    \label{eq:22}
\end{equation}
for all $n \geq 1$, then the aforementioned infinite number of equations are automatically satisfied so long as $\alpha(\omega,t)$ evolves according to the following equation obtained by substituting Eqs.~\eqref{eq:26},~\eqref{eq:8}, and~\eqref{eq:22} in Eq.~\eqref{eq:6}:
\begin{eqnarray}
	\nonumber
	\frac{\partial \alpha}{\partial t} &=& \frac{K}{2 } [(1-f)z_\mathrm{nr}^{*} + f] - i \omega \alpha \\
	&&- \frac{K}{2} [(1-f) z_\mathrm{nr} + f ] \alpha^{2},
	\label{eq:9}
\end{eqnarray}
where we have dropped the functional dependence in $\alpha$ for ease of notation. Here $z_\mathrm{nr}$ is obtained from Eq.~\eqref{eq:7} as
\begin{equation}
    z_\mathrm{nr}= \int_{-\infty}^{\infty} \alpha^{*}(\omega , t) g(\omega) d \omega.
    \label{eq:10}
\end{equation}
Thus, instead of solving the mentioned infinite set of coupled non-linear differential equations for the $\tilde{F}_n(\omega,t)$'s, we only have to solve the equation for $\alpha(\omega, t)$ and thereby obtain the entire density function $F(\theta,\omega,t)$.

Since we start from the initial condition $\theta_j(0) = 0$ for $j = n+1, \cdots , N$, the initial  density function for the oscillators in the non-reset subsystem can be written as $F(\theta, \omega, t = 0) = g(\omega) \delta(\theta)=g(\omega)(1/(2\pi))\sum_{n=-\infty}^\infty e^{in\theta}$,  which is consistent with Eq.~\eqref{eq:22} and which yields 
\begin{equation}
 \tilde{F}_n(\omega,0)=1~\forall~n\ne 0.   
 \label{eq:tildeFn-t0}
\end{equation}
Correspondingly, we take
\begin{equation}
    \alpha(\omega,0)=1.
    \label{eq:alpha-t0}
\end{equation}
Therefore, our dynamics is initialized on the OA manifold, and Eqs.~\eqref{eq:9} and~\eqref{eq:10} together would describe the time evolution of the order parameter on this manifold.

In the case of the Lorentzian frequency distribution given in Eq.~\eqref{eq:11},  Eqs.~\eqref{eq:9} and~\eqref{eq:10} can be further simplified if $\alpha(\omega,t)$ satisfies certain conditions in the complex $\omega$-plane. Namely, $\alpha(\omega,t)$ can be analytically continued from real $\omega$  into the complex $\omega$-plane for all $t \geq 0$, and that $|\alpha(\omega,t)| \to 0$ as Im$(\omega) \to - \infty$ and $|\alpha(\omega,t)| \leq 1$ for real $\omega$~\cite{Ott_2008}. With these conditions, 
one may compute Eq.~\eqref{eq:10} by the use of contour integration to obtain $z_\mathrm{nr}(t)=\alpha^*(\omega_0-i\sigma,t)$, and thereby derive the exact evolution equation of the order parameter $z_\mathrm{nr}$ from Eq.~\eqref{eq:9} as
\begin{eqnarray}
	\nonumber
	\frac{d z_\mathrm{nr}}{d t} &=& \frac{K}{2 }\{ [(1-f) z_\mathrm{nr} + f ] \\
		&&- [(1-f) z_\mathrm{nr}^{*} + f] z_\mathrm{nr}^{2} \}
	- \left( \sigma - i\omega_0 \right)z_\mathrm{nr}.
	\label{eq:13}
\end{eqnarray}

At this point, we can reduce the number of dynamical parameters by non-dimensionalizing the parameters through the transformation $t \to \sigma t$, $K \to K/\sigma$, and $\omega_0 \to \omega_0 / \sigma$. Note that for all analysis and results related to the infinite resetting rate case, we will assume from now on that such a non-dimensionalization of parameters to have been made. Next, using $z_\mathrm{nr} = r_\mathrm{nr}e^{i \psi_\mathrm{nr}}$ in Eq.~\eqref{eq:13} and comparing real and imaginary parts on both sides of the equation, we obtain the desired two-dimensional dynamical system:
\begin{eqnarray}
    \nonumber
    r_\mathrm{nr}^{'} &=& \frac{K(1-f)}{2} r_\mathrm{nr} \left( 1 - r_\mathrm{nr}^2 \right ) - r_\mathrm{nr}\\
    &&+ \frac{Kf}{2} \left( 1 - r_\mathrm{nr}^2 \right ) \cos {\psi_\mathrm{nr}}
    \label{eq:14}
\end{eqnarray}
and
\begin{eqnarray}
    r_\mathrm{nr} \psi_\mathrm{nr}^{'} = - \left[ - \omega_0 r_\mathrm{nr} + \frac{Kf}{2} \left( 1 + r_\mathrm{nr}^2 \right) \sin {\psi_\mathrm{nr}} \right], 
    \label{eq:15}
\end{eqnarray}
where the prime denotes differentiation with respect to dimensionless time.

It is pertinent to ask: Does the synchronization phase transition of the bare model persist when considering resetting events in the infinite resetting rate limit? In other words, is the incoherent state with $r_\mathrm{nr} = 0$ still a possible stationary-state solution of Eqs.~\eqref{eq:14} and~\eqref{eq:15}? In order to obtain an answer, we put $r_\mathrm{nr}^{'} = r_\mathrm{nr} = 0$ in Eq.~\eqref{eq:14}, which gives $\cos {\psi_\mathrm{nr}} = 0$, implying $\sin {\psi_\mathrm{nr}} = 1$. Using this along with $r_\mathrm{nr} = 0$ in Eq.~\eqref{eq:15}, we get
\begin{eqnarray}
    \frac{Kf}{2} = 0,
\end{eqnarray}
which is satisfied only with $f = 0$. We thus conclude that the incoherent state is a stationary state of the dynamics only when the system undergoes bare Kuramoto evolution (the case $f=0$). On the other hand,  inclusion of resetting in the dynamics at an infinite rate always synchronizes the non-reset subsystem, even when one resets a
vanishing fraction of the total number of oscillators. Thus, the synchronization phase transition in the bare Kuramoto dynamics gets converted into a crossover with infinite resetting rate because of the non-existence of the incoherent stationary state. 

As concluded above, $r_\mathrm{nr}=0$ cannot be a stationary-state solution of the dynamics, Eqs.~\eqref{eq:14} and~\eqref{eq:15}, and hence, we can rewrite Eq.~~\eqref{eq:15} as~\cite{Childs_2008}
\begin{eqnarray}
    \psi_\mathrm{nr}^{'} = - \left[ - \omega_0  + \frac{Kf}{2} \left(  r_\mathrm{nr} + \frac{1}{r_\mathrm{nr}} \right) \sin {\psi_\mathrm{nr}} \right]. 
    \label{eq:16}
\end{eqnarray}

Next, we are interested in the stationary-state value of $r_\mathrm{nr}$ from Eqs.~~\eqref{eq:14} and~\eqref{eq:16}, and hence, we need to put $r_\mathrm{nr}^{'} = 0$ in these equations. Now,  $\psi_\mathrm{nr}^{'} = \mathrm{constant} \ne 0$ cannot be a stationary-state solution, as $r_\mathrm{nr}^{'} = 0 $ in Eq.~\eqref{eq:14} will require $\cos{\psi_\mathrm{nr}}$ to have a constant value in the stationary state, which in turn implies $\psi_\mathrm{nr}$ to attain a time independent value at long times. This latter fact implies that one must have $\psi_\mathrm{nr}^{'} = 0$ in the stationary state. Using these facts in Eqs.~\eqref{eq:14} and~~\eqref{eq:16}, we get the stationary-state equations as 
\begin{eqnarray}
	\frac{K(1-f)}{2} r_\mathrm{nr}^\mathrm{st}\left[1 - (r_\mathrm{nr}^\mathrm{st})^2 \right ] - r_\mathrm{nr}^\mathrm{st}\nonumber \\
	+\frac{Kf}{2} \left[ 1 - (r_\mathrm{nr}^\mathrm{st})^2 \right ] \cos {\psi_\mathrm{nr}^\mathrm{st}} &=& 0,
	\label{eq:17}\\
	\omega_0 - \frac{Kf}{2} \left( r_\mathrm{nr}^\mathrm{st} + \frac{1}{r_\mathrm{nr}^\mathrm{st}} \right) \sin {\psi_\mathrm{nr}^\mathrm{st}} &=& 0.
	\label{eq:18}
\end{eqnarray}
Here, ``st" denotes the stationary-state value. It may be worthwhile to compare the above situation with the one for the bare Kuramoto dynamics, for which setting $f=0$ in Eq.~\eqref{eq:16} implies that one has $\psi_\mathrm{nr}^{'} =\omega_0=\mathrm{constant} \ne 0$ in the stationary state. This means that in the stationary state of the bare Kuramoto model, the synchronization order parameter $z_\mathrm{nr}$ can have a time-independent value of its length $r_\mathrm{nr}$, while its orientation angle may change uniformly in time with angular velocity $\omega_0$, for $\omega_0 \neq 0$. Unlike that, in our case, in order to have a time-independent $r_\mathrm{nr}$ in the stationary state, the orientation angle of $z_\mathrm{nr}$ must not also change as a function of time, even when $\omega_0 \neq 0$.

\begin{figure*}
    \begin{center}
        \begin{subfigure}[b]{0.37\textwidth}
                \includegraphics[width=\textwidth]{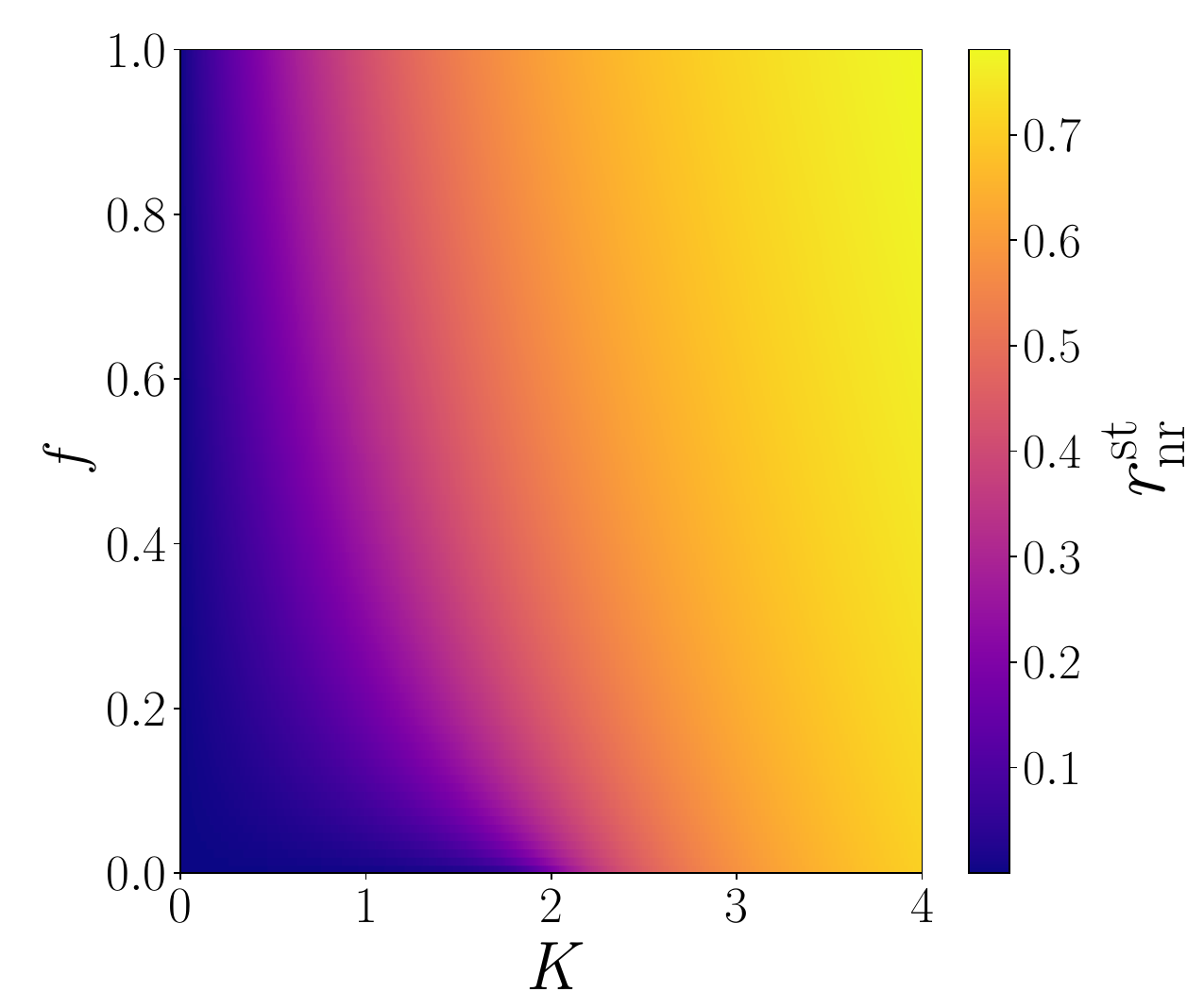}
                \caption{}
                \label{fig:omega_0_phase_diag}
        \end{subfigure}
        \begin{subfigure}[b]{0.305\textwidth}
                \includegraphics[width=\textwidth]{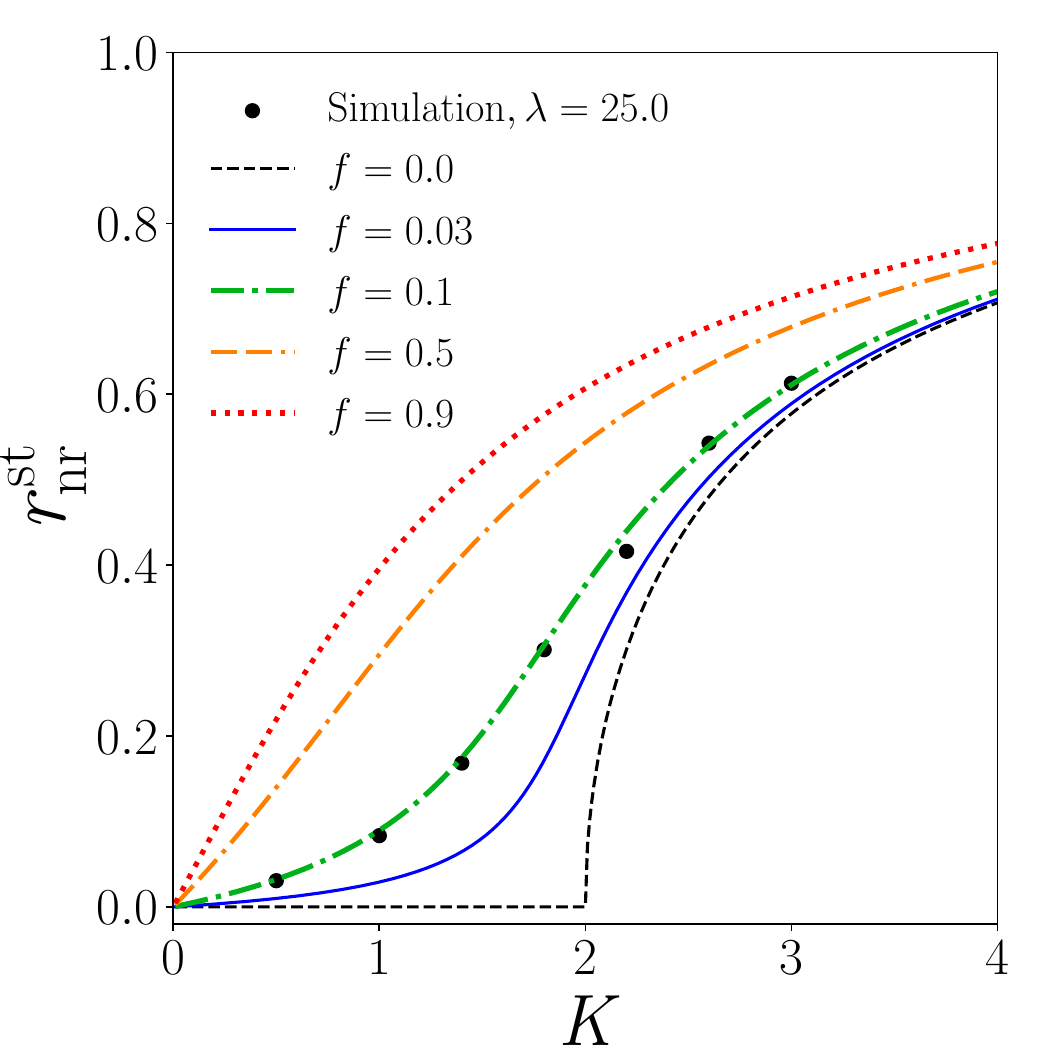}
                \caption{}
                \label{fig:omega_0_order_para}
        \end{subfigure}
        \begin{subfigure}[b]{0.305\textwidth}
                \includegraphics[width=\textwidth]{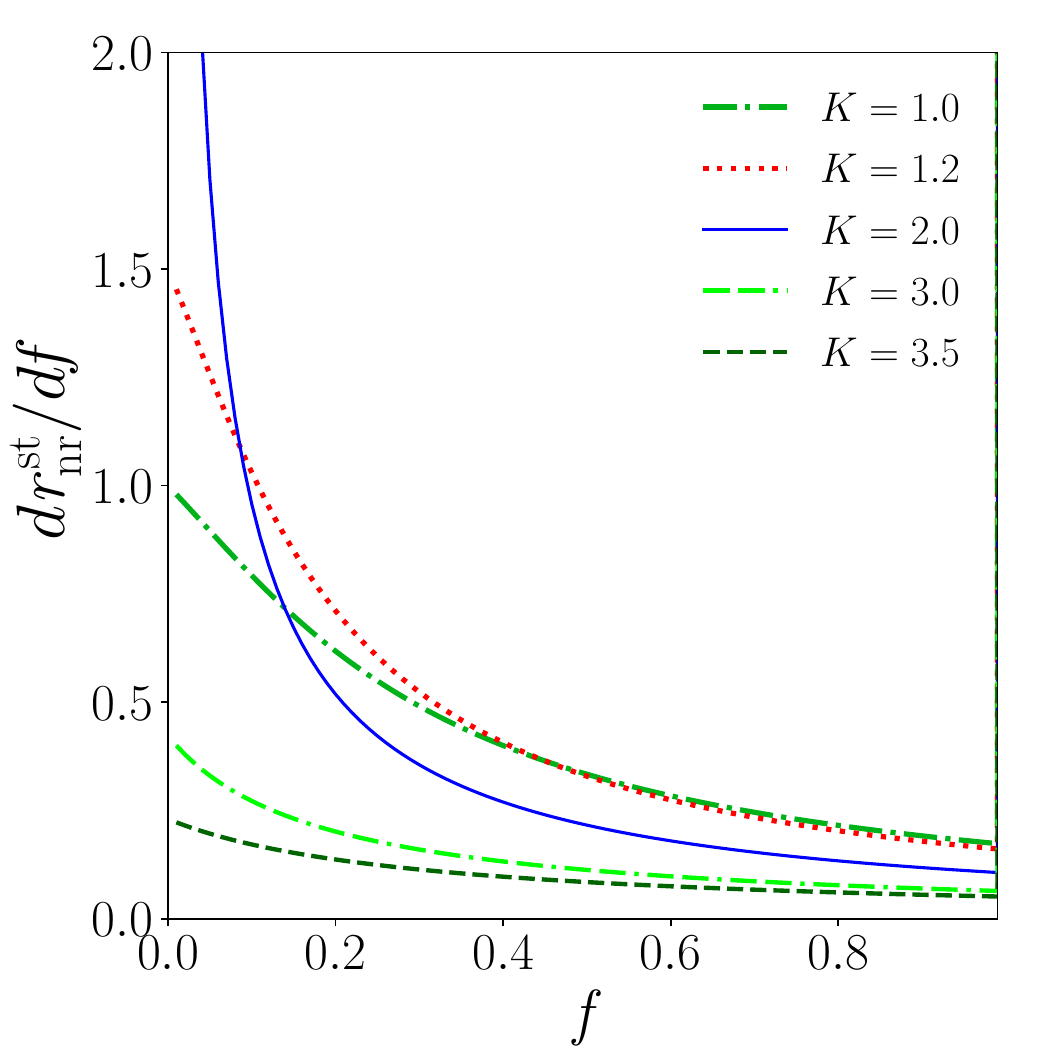}
                \caption{}
                \label{fig:omega_0_increase}
        \end{subfigure}          
    \caption{\textbf{Subsystem resetting with $\lambda \to \infty$ and Lorentzian $g(\omega)$ with mean $\omega_0=0$ and width $\sigma=1.0$}: Panel (a) shows analytical results for the stationary-state synchronization order-parameter $r^\mathrm{st}_\mathrm{nr}$ of the non-reset subsystem, depicted in terms of a density plot in the $f-K$ plane. The analytical results are computed by solving the roots of Eq.~\eqref{eq:19}. As one tunes $K$ at a fixed $f\ne 0$, the quantity $r^\mathrm{st}_\mathrm{nr}$ undergoes a crossover from low to high values. From the density plot in panel (a), variation of $r^\mathrm{st}_\mathrm{nr}$ with $K$ for representative values of $f$ is presented in panel (b). Here, the black dashed line corresponds to the bare Kuramoto dynamics, for which one has $r^\mathrm{st}_\mathrm{nr}=0$ for $K \le K_c$, and one has a synchronization transition as a function of $K$. In this panel, we also show by points numerical simulation results for $f=0.1$ and $\lambda=25.0$, demonstrating agreement with theory; the data correspond to the stationary state of a single realization of the dynamics for a system of $N=10^4$ oscillators with the integration time step equal to $0.005$. Panel (c) shows as a function of $f$ the quantity $d r^\mathrm{st}_\mathrm{nr}/df$, for different $K$ values given in the panel.} \label{fig:3}
    \end{center}
\end{figure*}

In the rest of this section, we will put our efforts into solving Eqs.~\eqref{eq:17} and~\eqref{eq:18} and deduce the ensuing physical picture. We will consider successively the cases $\omega_0=0$ and $\omega_0 \ne 0$. 

\subsubsection{Resetting with $\omega_0=0$}
In the particular case of infinite resetting rate under consideration, let us obtain the average phase $\psi_\mathrm{nr}$ of the non-reset subsystem. Putting $\omega_0 = 0$ in Eq.~\eqref{eq:18}, and since we have $r^\mathrm{st}_\mathrm{nr} \neq 0$ for any nonzero $f$, the only way Eq.~\eqref{eq:18} is satisfied will be by having
\begin{equation}
    \sin{\psi^\mathrm{st}_\mathrm{nr}} = 0. \label{eq:sin-in-omega-0}
\end{equation}
The above conclusion could have also been arrived at by analyzing Eq.~\eqref{eq:16} for $\omega_0=0$. In this case, noting that $r_\mathrm{nr}$ is an intrinsically positive quantity, Eq.~\eqref{eq:16} has the form $d\theta/dt=-\alpha \sin \theta$, with $\alpha>0$ and $\theta\in[0,2\pi)$, which evidently has one stable fixed point at $\theta=0$. We thus conclude that the long-time solution of Eq.~\eqref{eq:16} is given by $\sin{\psi^\mathrm{st}_\mathrm{nr}} = 0$ and $\cos{\psi^\mathrm{st
}_\mathrm{nr}} = 1$. As a result, Eq.~\eqref{eq:17} reduces to a cubic equation for $r^\mathrm{st}_\mathrm{nr}$:
\begin{eqnarray}
    \nonumber
    &&\left( r^\mathrm{st}_\mathrm{nr} \right)^3 + \left( \frac{f}{1-f} \right) \left( r^\mathrm{st}_\mathrm{nr} \right)^2 \nonumber \\
    &&+ \left[ \frac{2}{K (1 - f)} 
    -1 \right]r^\mathrm{st}_\mathrm{nr}- \left( \frac{f}{1-f} \right) = 0.
    \label{eq:19}
\end{eqnarray}

The above equation can be solved exactly for $r^\mathrm{st}_\mathrm{nr}$, see Appendix~\ref{app:2}. Interestingly, it is clear from Eq~\eqref{eq:19} itself that, for any nonzero $f$, one cannot have $r^\mathrm{st}_\mathrm{nr} = 0$ as a solution of the equation. For now, we are going to discuss the main features of the solution of Eq.~\eqref{eq:19}, depicted in the plots in Fig.~\ref{fig:3}.

Figure~\ref{fig:3}(a) depicts the density plot for $r^\mathrm{st}_\mathrm{nr}$ in the $f-K$ plane, wherein the synchronization transition of the bare Kuramoto model appears along the $K$ axis for $f =  0$. As the figure suggests and as has been explained earlier, the mentioned transition becomes a crossover as soon as $f$ becomes non-zero. We thus conclude that the non-reset subsystem is synchronized, i.e., $r^\mathrm{st}_\mathrm{nr} \ne 0$, for any nonzero value of $f$, however small. It can be seen from the figure that as we increase $f$ at a fixed $K$, the quantity  $r^\mathrm{st}_\mathrm{nr}$ takes up higher and higher values for $K$ values in the region $K \leq K_c = 2$, where the bare dynamics does not support any synchronization. Note that as mentioned in the paragraph following Eq.~\eqref{eq:13}, we are here working with non-dimensionalized parameters, and hence have the critical value of the coupling as equal to $2$ instead of $2\sigma$. To bring out better the salient features of this induced synchronization in the non-reset subsystem, we plot in panel (b) the quantity $r^\mathrm{st}_\mathrm{nr}$ as a function of $K$ for several representative values of $f$. The black dashed line represents the case of the bare Kuramoto dynamics, showing no synchronization, i.e., $r^\mathrm{st}_\mathrm{nr} = 0$ for $K \leq K_c$. We see from panel (b) that we can induce a significant amount of synchronisation in the non-reset subsystem even by resetting as low as $3 \%$ of the total number of oscillators ($f=0.03$). The effect of resetting becomes more prominent near the transition point of the bare Kuramoto model, i.e., near $K=K_c$. Interestingly, we see from panel (b) that the change in the value of $r_\mathrm{nr}^\mathrm{st}$ does not take place proportionately to the change in the value of $f$. Namely, for the same amount of change in $f$, once from $10\%$ to $50\%$ and again from $50\%$ to $90\%$, the corresponding change in $r_\mathrm{nr}^\mathrm{st}$ is different in the two cases. In fact, one has a greater change in $r_\mathrm{nr}^\mathrm{st}$ when $f$ is changed at a lower than at a higher value. This fact is more clearly established on plotting the quantity $d r^\mathrm{st}_\mathrm{nr}/d f$ as a function of $f$ for several values of $K$, see panel (c). The panel shows that the derivative $d r^\mathrm{st}_\mathrm{nr}/d f$ takes up its maximum value when $f$ is close to zero, and which monotonically decreases as we increase $f$. On the face of it, this result may appear counter-intuitive, as one would expect the effect of resetting $90 \%$ of the total number of oscillators to be more pronounced in inducing synchronization in the non-reset subsystem than resetting say $50 \%$ of the total number of oscillators.

Remarkably, for a given value of $K$ and as a function of $f$, there is a bound on the amount of synchronization that can be achieved in the non-reset subsystem through resetting as $f$ hits its maximum allowed value, i.e., in the limit $f\to1^-$. We now derive this bound. Multiplying both sides of Eq.~\eqref{eq:19} with $(1-f)$ and taking  the limit $f \to 1^-$, we get the desired upper bound as
\begin{eqnarray}
    (r^\mathrm{st}_\mathrm{nr})_\mathrm{max} = \frac{\sqrt{1+K^2}-1}{K}.
    \label{eq:rstnr-max}
\end{eqnarray}
An important point to note from Fig.~\ref{fig:3}(c) is that the quantity $\lim\limits_{f \to 0} d r^\mathrm{st}_\mathrm{nr}/d f$  increases with $K$ in the region $K < K_c$, diverges at $K =K_c$, and then decreases as we increase $K$ further, beyond $K=K_c$. This is a signature of the synchronization phase transition in the bare Kuramoto model, which is retrieved in the $f= 0$ case. 

From Eq.~\eqref{eq:rstnr-max}, we see that for a particular $K$ value, $r_\mathrm{nr}^\mathrm{st}$ cannot exceed $(r^\mathrm{st}_\mathrm{nr})_\mathrm{max}$. This upper bound on $r_\mathrm{nr}^\mathrm{st}$ implies that $r_\mathrm{nr}^\mathrm{st}$ versus $f$ at a fixed $K$ has to behave in such a way as to saturate to the value $(r^\mathrm{st}_\mathrm{nr})_\mathrm{max}$ as $f$ hits the value of unity from below. This would in turn imply that the slope $d r^\mathrm{st}_\mathrm{nr}/d f$ changes faster at smaller values of $f$ than at higher values, explaining the observation mentioned above regarding the behavior of the quantity $d r^\mathrm{st}_\mathrm{nr}/d f$ as a function of $f$.

Now, one may question the physical relevance of studying the case $\lambda \to \infty$, which  is evidently experimentally unattainable. One may argue in favor of such a study by mentioning that it gives a bound on how much synchronization can be induced via resetting for fixed $K$~and~$f$, in the extreme case in which the reset subsystem remains completely synchronized almost at all times. For any finite $\lambda$, the reset subsystem cannot remain synchronized at all times, and one may expect only a smaller amount of synchrony to get induced in the non-reset subsystem through its interaction with the reset subsystem. Secondly, as shown in Fig.~\ref{fig:3}(b), our theoretical predictions obtained in the $\lambda \to \infty$ limit correctly reproduce numerical simulation results for finite $\lambda$, for $\lambda$ values as low as $25.0$. 

\subsubsection{\label{sec:infinite nonzero omega}Resetting with $\omega_0 \ne 0$}
In the non-zero $\omega_0$ case, Eq.~\eqref{eq:18} gives the average phase $\psi_\mathrm{nr}$ of the non-reset subsystem in the stationary state as
\begin{equation}
    \sin {\psi^\mathrm{st}_\mathrm{nr}} = \frac{\omega_0}{\frac{Kf}{2} \left( r^\mathrm{st}_\mathrm{nr} + \frac{1}{r^\mathrm{st}_\mathrm{nr}} \right) }, \label{eq:sin-omega-nonzero}
\end{equation}
which implies that $\sin {\psi^\mathrm{st}_\mathrm{nr}}$ is always positive in the stationary state. Correspondingly, $\cos {\psi^\mathrm{st}_\mathrm{nr}}$ may have either sign. To decide on the sign, let us consider Eq.~\eqref{eq:16}. As $r_\mathrm{nr}$ is a positive quantity, Eq.~\eqref{eq:16} has the form $d\theta/dt= \omega_0-\alpha \sin \theta$, with $\omega_0 >0,\alpha>0$ and $\theta\in[0,2\pi)$, which evidently has a stable fixed point in the first quadrant and an unstable fixed point in the second quadrant, provided we have $\alpha > \omega_0$. We thus conclude that for the dynamics $d\theta/dt= \omega_0-\alpha \sin \theta$, when the stable fixed point exists, it lies necessarily in the first quadrant. In our system, Eq.~\eqref{eq:16}, when the fixed point exists, the stable fixed point (which by definition coincides with the stationary solution, Eq.~\eqref{eq:sin-omega-nonzero}) lies necessarily in the first quadrant. Consequently, we have in the stationary state that $\cos {\psi^\mathrm{st}_\mathrm{nr}}$ is positive and equals
\begin{equation}
    \cos {\psi^\mathrm{st}_\mathrm{nr}} = \sqrt{1 - \frac{\omega_0^2}{\frac{K^2f^2}{4} \left( r^\mathrm{st}_\mathrm{nr} + \frac{1}{r^\mathrm{st}_\mathrm{nr}} \right)^2 }},
\end{equation}
which together with Eq.~\eqref{eq:17} gives the quantity $r^\mathrm{st}_\mathrm{nr}$ as solving the equation
\begin{eqnarray}
    \nonumber
    \frac{1}{1-\left(r^\mathrm{st}_\mathrm{nr}\right)^2} = \frac{K(1-f)}{2} + \sqrt{\frac{K^2f^2}{4} \frac{1}{\left(r^\mathrm{st}_\mathrm{nr}\right)^2} - \frac{\omega_0^2}{\left( 1 + \left(r^\mathrm{st}_\mathrm{nr}\right)^2 \right )^2}}.\\\label{eq:20}
\end{eqnarray}

\begin{figure*}
	\centering
	\begin{subfigure}{0.38\textwidth}
		\includegraphics[width=\linewidth]{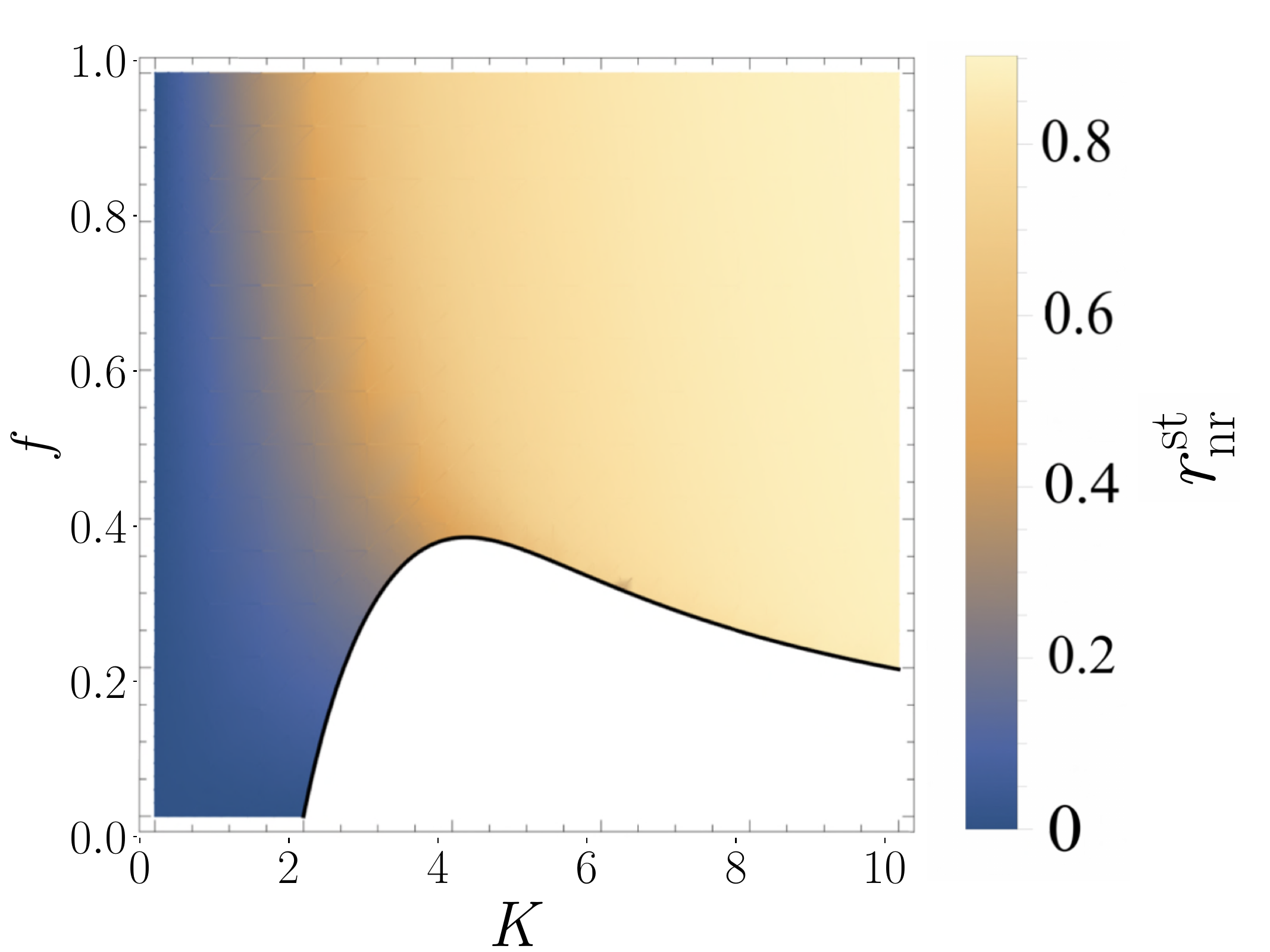}
		\caption{}
		\label{fig:4a}
	\end{subfigure}
	\begin{subfigure}{0.31\textwidth}
		\includegraphics[width=\linewidth]{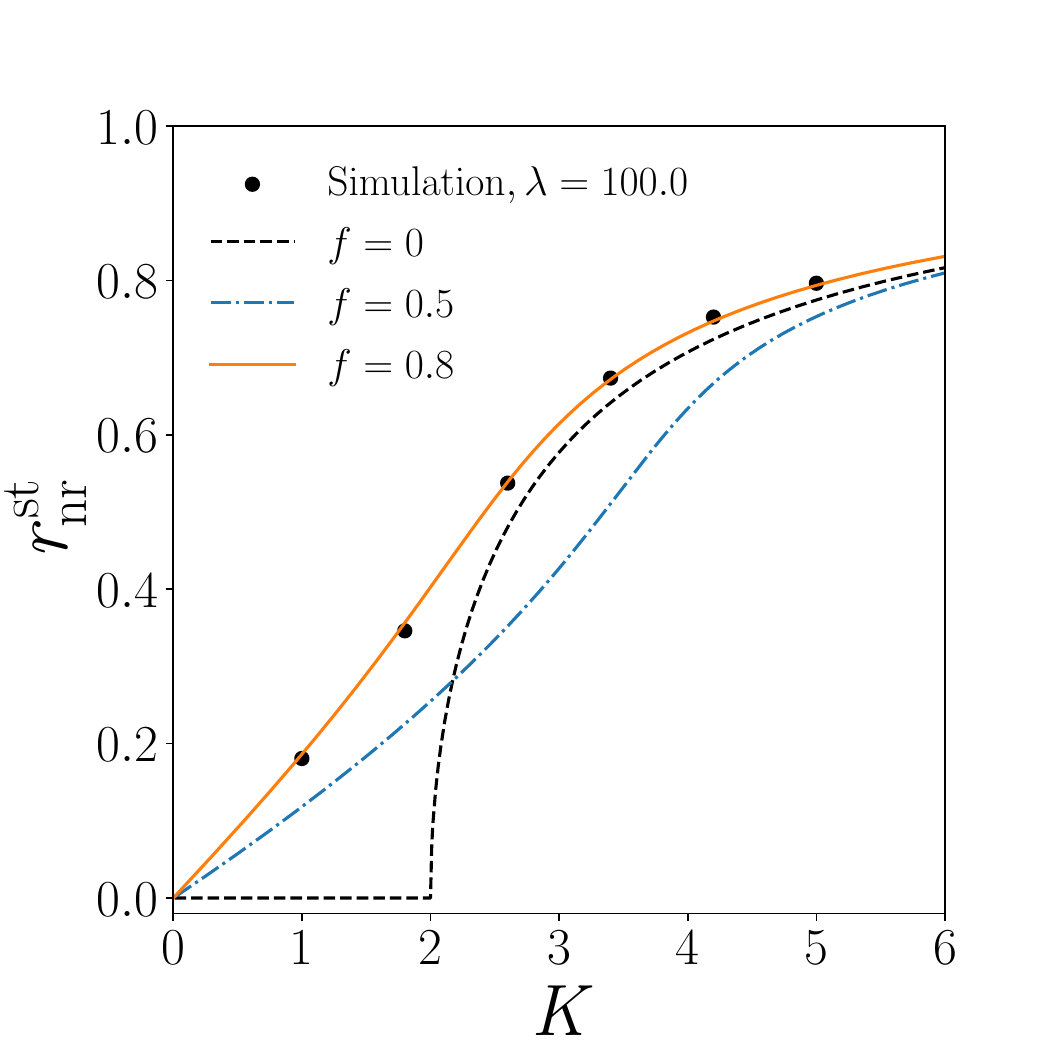}
		\caption{}
		\label{fig:4b}
	\end{subfigure}
	\\
	\begin{subfigure}{0.31\textwidth}
		\includegraphics[width=\linewidth]{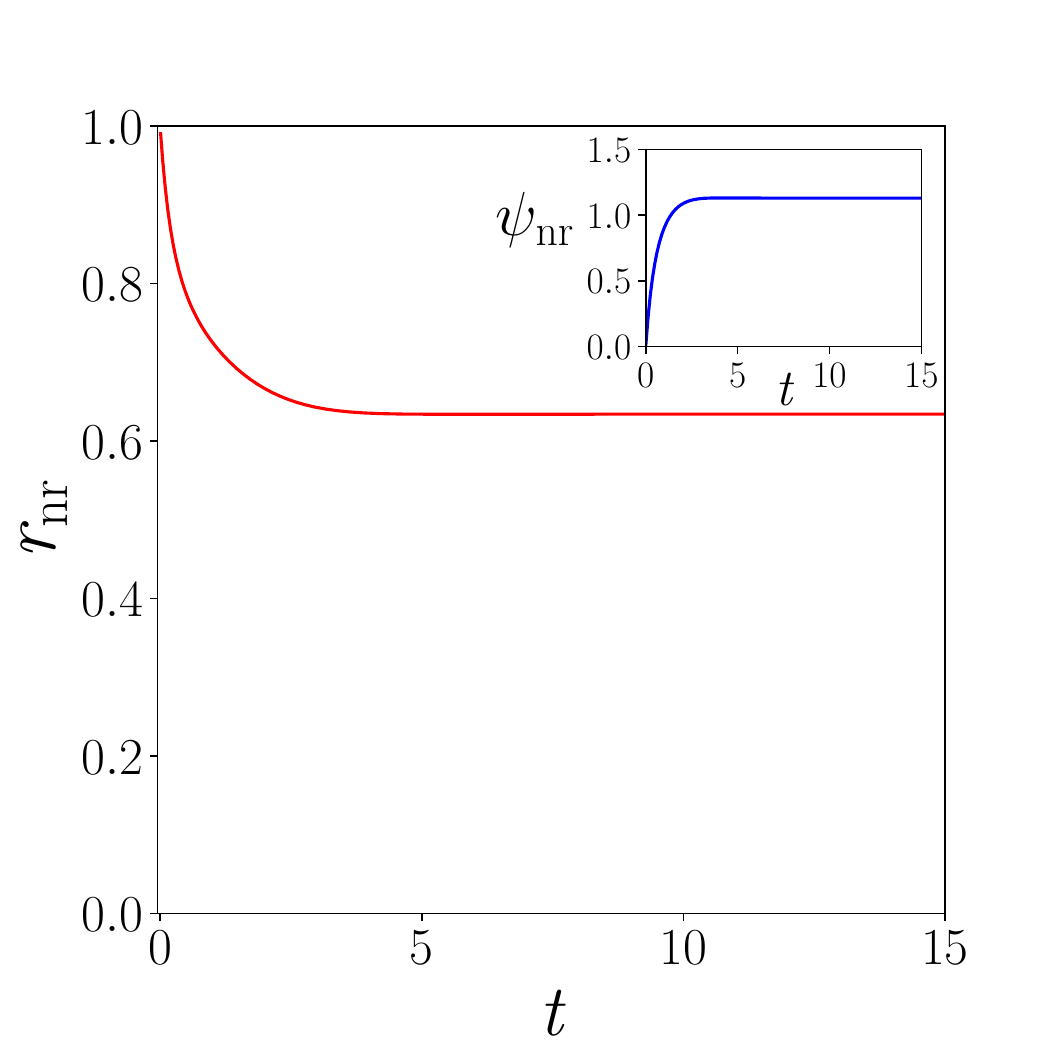}
		\caption{}
		\label{fig:4c}
	\end{subfigure}
	\begin{subfigure}{0.31\textwidth}
		\includegraphics[width=\linewidth]{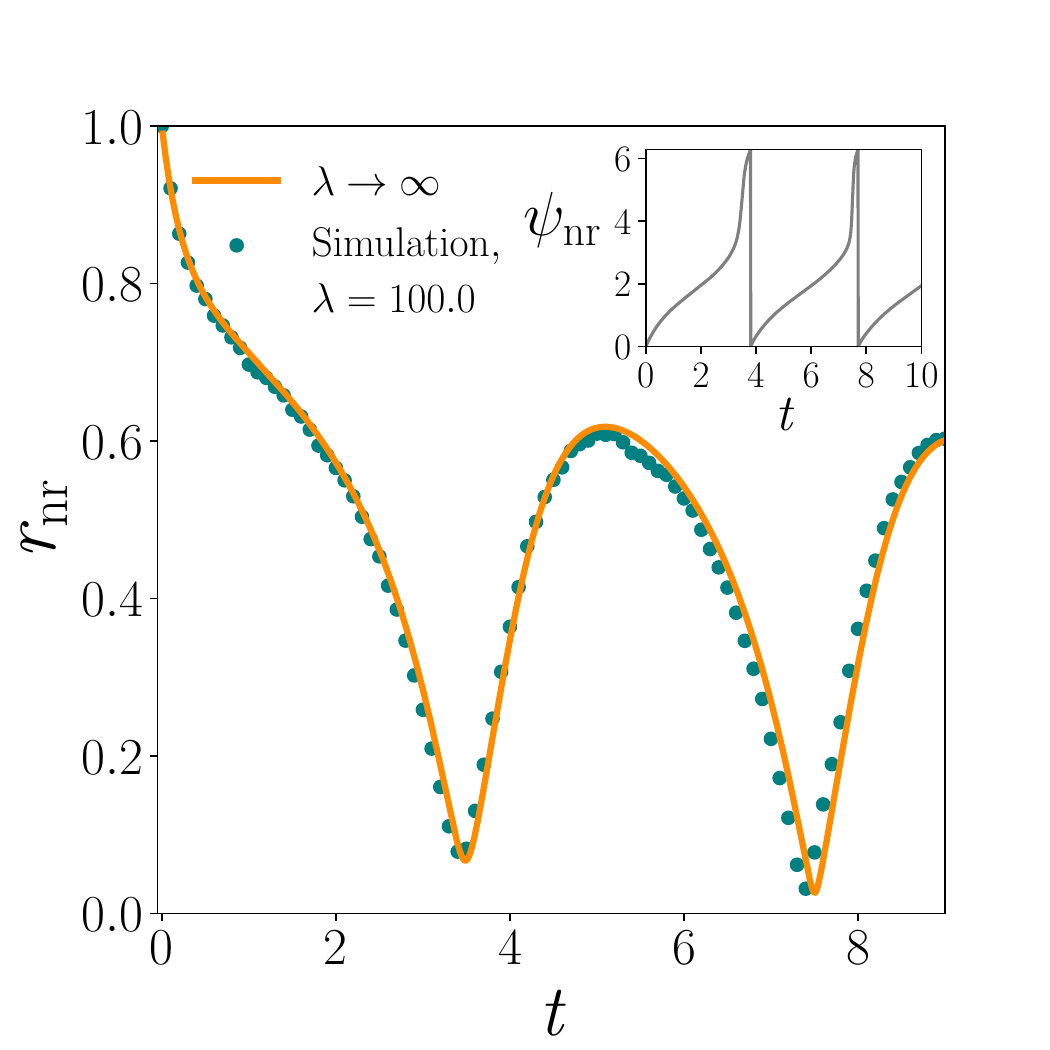}
		\caption{}
		\label{fig:4d}
	\end{subfigure}
	\begin{subfigure}{0.31\textwidth}
		\includegraphics[width=\linewidth]{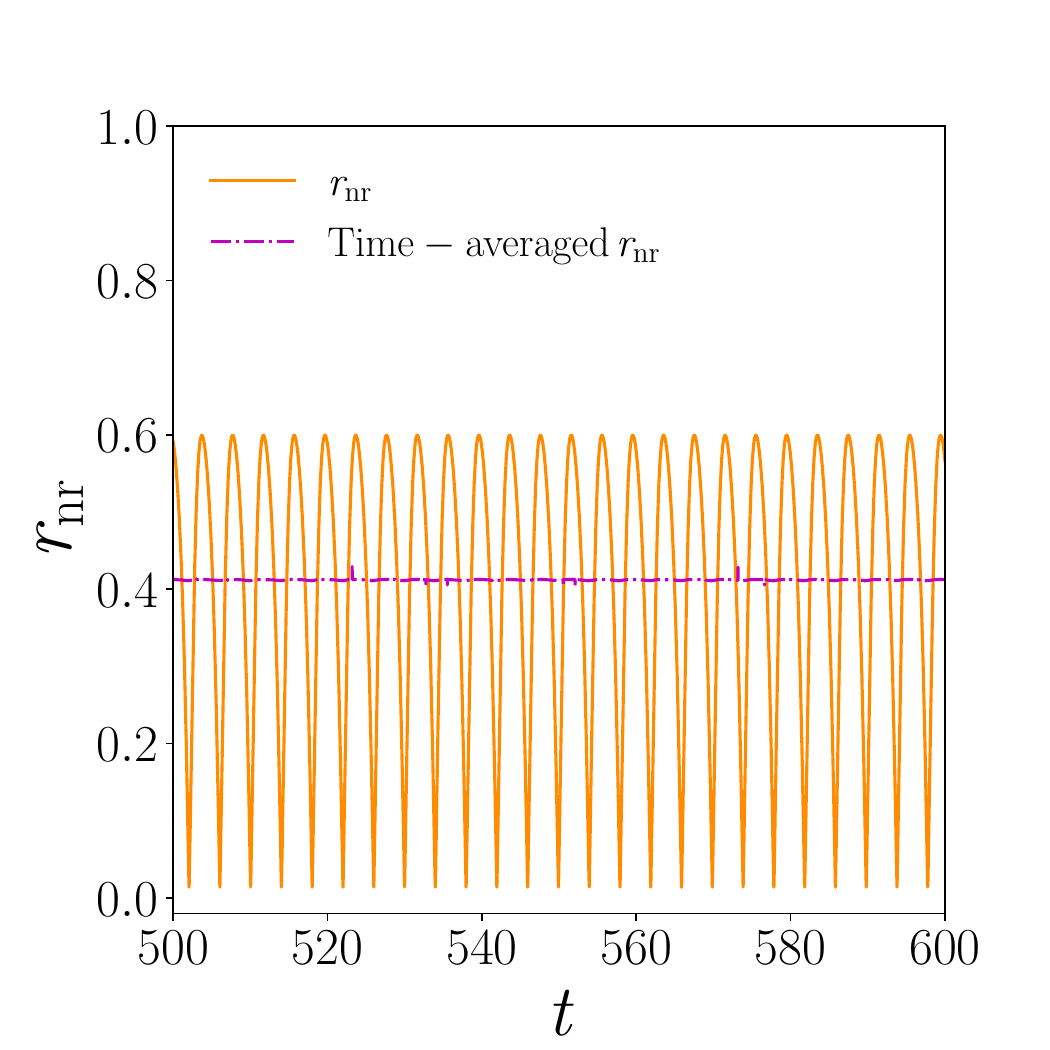}
		\caption{}
		\label{fig:4e}
	\end{subfigure}
	
	\caption{\textbf{Subsystem resetting with $\lambda \to \infty$ and Lorentzian $g(\omega)$ with mean $\omega_0\ne 0$ (specifically, $\omega_0=2.0$) and width $\sigma=1.0$}: Panel (a) shows analytical results for the stationary-state synchronization order-parameter $r_\mathrm{st}^\mathrm{nr}$ of the non-reset
		subsystem, depicted in terms of a density plot in the $f-K$ plane. The analytical results are obtained by solving numerically the roots of Eq.~\eqref{eq:20}. In the white region of the plot, when roots do not exist, $r_\mathrm{nr}$ instead of reaching a stationary value at long times, oscillates as a function of time, and $\psi_\mathrm{nr}$ remains time dependent even at long times. Nevertheless, $r_\mathrm{nr}$ at long times yields a non-zero time-independent time average, thereby defining an oscillatory synchronized state (representative plots for this case are shown in panels (d) and (e) for $K=4.0$ and $f=0.3$; here, $r_\mathrm{nr}$ and $\psi_\mathrm{nr}$ are obtained by numerically solving Eqs.~\eqref{eq:14} and~\eqref{eq:16} simultaneously; for our choice of the initial state as given in Sec.~\ref{sec:Dynamics}, we have $r_\mathrm{nr}=1$ and $\psi_\mathrm{nr}=0$ at the initial time instant $t=0$). In contrast, in the colored region, when roots do exist, the system reaches a stationary state, in which both the quantities $r_\mathrm{nr}$ and $\psi_\mathrm{nr}$ attain time-independent values at long times (a representative plot is shown in panel (c) for $K=4.0$ and $f=0.5$; for our choice of the initial state as given in Sec.~\ref{sec:Dynamics}, we have $r_\mathrm{nr}=1$ and $\psi_\mathrm{nr}=0$ at the initial time instant $t=0$). In the coloured region, as one tunes $K$ at a fixed $f\ne 0$, the quantity $r^\mathrm{st}_\mathrm{nr}$ undergoes a crossover from low to high values. The black line in panel (a) is the curve given by Eq.~\eqref{eq:21}, denoting the boundary between regions with stationary and non-stationary behavior of $r_\mathrm{nr}$ at long times.
		From the density plot in panel (a), variation of $r^\mathrm{st}_\mathrm{nr}$ with $K$ for representative values of $f$ is presented in panel (b). Here, the black dashed line corresponds to the bare Kuramoto dynamics, for which one has $r^\mathrm{st}_\mathrm{nr}=0$ for $K \le K_c$, and one has a synchronization transition as a function of $K$. In this panel, we also show by points numerical simulation results for $f=0.8$ and $\lambda=100.0$, demonstrating agreement with theory. A similar agreement between theory and simulation is also demonstrated in panel (d). The simulation data correspond to a single realization of the dynamics for a system of $N=10^4$ oscillators with the integration time step equal to $0.005$.}
	\label{fig:4}
\end{figure*}

In Fig.~\ref{fig:4}(a), we have presented the solution of Eq.~\eqref{eq:20}. Interestingly, for $K \geq K_c=2$, i.e., for $K$ values larger than the critical threshold of the bare Kuramoto model for the case of Lorentzian $g(\omega)$, there exist values of $f$ for which the dynamics does not support a stationary state.  The latter corresponds to the white region in the density plot of Fig.~\ref{fig:4}(a). In this region, instead of reaching a time-independent value at long times, $r_\mathrm{nr}$ as well as  $\psi_\mathrm{nr}$ is time dependent; the quantity $r_\mathrm{nr}$ oscillates as a function of time, with a non-zero time-independent time average (defining what we refer to as an oscillatory synchronized state). Thus, the white region corresponds to non-stationary states. In contrast, in the colored region of Fig.~\ref{fig:4}(a), the system reaches a stationary state at long times. In this region, both the quantities $r_\mathrm{nr}$ and $\psi_\mathrm{nr}$ attain time-independent values at long times. Referring to Fig.~~\ref{fig:4}(a), as one tunes $f$ at a fixed $K$, one observes a transition between a region supporting stationary states and another  supporting non-stationary states, for $K$ values in the range $K >2$. Moreover, we may conclude based on the results discussed above that similar to the case with $\omega_0= 0$, the non-reset subsystem gets synchronized at long times through the act of resetting of the reset subsystem. 

It has been established in Ref.~\cite{sarkar2022} that when subjected to global resetting, with other details of the resetting protocol identical to the ones considered in the current work, the dynamics always reaches a stationary state at long times.
On the basis of our analysis presented above, we conclude that unlike the global resetting case, on performing resetting of a subsystem at an infinite rate, one may or may not have a stationary state depending on the values of the dynamical parameters. Another interesting conclusion to draw from Fig.~\ref{fig:4}(a) is that, in the region where a stationary state exists (coloured region of Fig.~\ref{fig:4}(a)), the synchronization phase transition of the bare Kuramoto model gets replaced by a crossover in presence of resetting. On the other hand, a new transition between stationary and non-stationary states appears, depending on the values of the parameters $f$ and $K$.

The origin of the aforementioned transition between stationary and non-stationary states lies in the existence of the broken phase-shift symmetry in the dynamics, Eq.~\eqref{eq:12}, compared to the bare Kuramoto dynamics (see the discussion following Eq.~\eqref{eq:12}). In the case of the bare Kuramoto dynamics, Eq.~\eqref{eq:1} possesses phase-shift symmetry. As a result, for the bare Kuramoto dynamics with $\omega_0 \ne 0$, we can go to a reference frame rotating with angular velocity $\omega_0$ with respect to an inertial frame; in such a frame, the dynamics transforms into that which has $\omega_0 = 0$. This is the reason why the order parameter shows identical behaviour in the case of the bare Kuramoto dynamics with $\omega_0 = 0$ and $\omega_0 \ne 0$. Unlike that, the dynamics under subsystem resetting at infinite rate as given by Eq.~\eqref{eq:12} does not have phase-shift symmetry, the reason for which is the following. As in this case, reset happens at an infinite rate, the phases of the oscillators in the reset subsystem remain frozen at the value zero. As a result, the attractive interaction between the oscillators in the reset and the non-reset subsystem tries to make the phases of the oscillators of the non-reset subsystem take up the value zero. In other words, the interaction between the oscillators in the reset and non-reset subsystem tries to freeze the orientation of the synchronization order parameter $z_\mathrm{nr}$ of the non-reset subsystem at the angle $\psi_\mathrm{nr}=0$. This effect is captured by the second term on the right hand side of Eq.~\eqref{eq:4}, whose impact increases with increasing of the reset fraction $f$. On the other hand, because of $\omega_0$ being non-zero, mutual interaction between the oscillators of the non-reset subsystem tries to rotate the oscillator phases with frequency $\omega_0$. In other words, interaction among the oscillators of the non-reset subsystem tries to rotate the synchronization order parameter $z_\mathrm{nr}$ of the non-reset subsystem with frequency $\omega_0$. This effect is captured by the third term on the right hand side of Eq.~\eqref{eq:4}, whose impact decreases with increasing of the reset fraction $f$ and increases with increasing of the magnitude of $z_\mathrm{nr}$, i.e., of $r_\mathrm{nr}$. The interplay of these two opposing effects generates the non-trivial density plot in Fig.~\ref{fig:4}(a). For $K \leq K_c = 2$, the value of $r_\mathrm{nr}$ is small. As a result, the effect of mutual interaction is very weak and unable to counter the effects of resetting, leading the system into a stationary state. On the other hand, for $ K > 2$, one has a more involved situation. For smaller $f$, the third term on the right hand side of Eq.~\eqref{eq:4} dominates over the second term. Thus, the effect of resetting being small, mutual interaction between the non-reset oscillators is able to overcome the effect of resetting. As a result, the quantity $z_\mathrm{nr}$ keeps rotating in time, without reaching a stationary state. We see from Eqs.~\eqref{eq:14}~and~\eqref{eq:16} that the time evolution of $r_\mathrm{nr}$ and $\psi_\mathrm{nr}$ are coupled to each other. As a rotating $z_\mathrm{nr}$ results in $\psi_\mathrm{nr}$ changing with time, it would in turn mean $r_\mathrm{nr}$ to also change with time, and one has a non-stationary state. This explains the white region of Fig.~\ref{fig:4}(a). On the other hand, as we increase $f$, the effect of resetting increases, whereas the effect of the mutual interaction, being a function of $(1-f)$, starts decreasing (see Eq.~\eqref{eq:4}). As a result, if we keep increasing $f$ keeping $K$ fixed, effect of resetting becomes dominant after crossing a particular critical value $f = f_c$, beyond which the system has a stationary state. 

Let us now obtain an equation relating $f_c$ and $K$. The solution of this equation will provide us with the boundary that separates the parameter regions in Fig.~\ref{fig:4}(a)  corresponding to stationary and non-stationary states. A stationary state exists if Eq.~\eqref{eq:20} has a real root. Clearly, the equation will not have a real root if the term inside the square root is negative. Hence, in order to obtain the boundary, we set the quantity inside the square root equal to zero, yielding
\begin{eqnarray}
    \frac{1}{1-r_\mathrm{nr}^2} = \frac{K(1-f_c)}{2}, \hspace{0.5cm}
    \frac{K^2f_c^2}{4} \frac{1}{r_\mathrm{nr}^2} = \frac{\omega_0^2}{\left( 1 + r_\mathrm{nr}^2 \right )^2} ,
\end{eqnarray}
solving which we get the equation for the boundary as
\begin{equation}
    K \left( \frac{f_c^2}{1-f_c} \right) = \omega_0^2 \frac{K (1-f_c) - 2}{\left( K (1-f_c) - 1 \right)^2},
    \label{eq:21}
\end{equation}
while the amount of stationary-state synchronization at the boundary is given by 
\begin{equation}
    r^\mathrm{st}_\mathrm{nr} = \sqrt{\frac{K (1-f) - 2}{K (1-f)}}.
\end{equation}

Lastly, as in the $\omega_0=0$ case, our theoretical predictions obtained in the limit $\lambda \to \infty$ match reasonably well with numerical simulation results for finite $\lambda$ as low as $100.0$, see Fig.~\ref{fig:4}, panels (b) and (d).

\subsection{\label{sec:level5}The case of finite-$\lambda$}
Until now, we have considered the case of infinite resetting rate, i.e., the limit $\lambda \to \infty$. We now move on to consider the case of finite $\lambda$. At first glance, a straightforward analysis for our case of subsystem resetting seems significantly non-trivial than that for global resetting. In the case of bare evolution, as mentioned earlier, in so far as the behavior of the synchronization order parameter is concerned, a great simplification is offered in the continuum limit $N \to \infty$ by the OA ansatz \cite{Ott_2008}. Analysis using the OA ansatz can be extended to apply in the case of global resetting of the Kuramoto model, as long as the resetting protocol resets the system globally to a state that remains on the OA manifold \cite{sarkar2022}. This situation may be contrasted with subsystem resetting, wherein even if we initialize the dynamics on the OA manifold, it no longer remains on the manifold following a reset event: As we will show, following a reset, the single-oscillator density function for the entire system does not any longer satisfy the condition given in Eq.~\eqref{eq:22}, which defines the OA ansatz. 

We will see below that in the case of bare evolution, we may consider the system to be made up of two subsystems. Then, if the initial density function of the two subsystems separately follows the OA condition~\eqref{eq:22}, we can apply the OA ansatz separately to find the time evolution of the order parameters of the two subsystems, which obviously will be dependent on one another. We will demonstrate that such an analysis will pave the way for treating the case of the Kuramoto model in presence of subsystem resetting. 

\subsubsection{\label{sec:level6}Groundwork for subsystem resetting: Analysis in absence of resetting}
Here, we will consider the bare Kuramoto evolution, and understand the conditions for the applicability of the following two methods of analysis: (i) applying the OA ansatz to the entire system, and (ii) applying the OA ansatz separately to two subsystems making up the entire system. 

Let us consider the dynamics of our system of Kuramoto oscillators to be initialized with the condition $\theta_j(0) = 0~\forall~j$. The time evolution follows the dynamics defined in Eq.~\eqref{eq:1}. In the continuum limit, we can write the single-oscillator density function of the system at time $t = 0$ as
\begin{eqnarray}
    F(\theta, \omega, 0) = g(\omega) \delta (\theta) = \frac{g(\omega)}{2 \pi} \sum_{n = -\infty}^{\infty} e^{i n\theta}.
\end{eqnarray}
All the Fourier coefficients of this initial density function are evidently equal, $\tilde{F}_n (\omega,0) = 1~\forall~n$, with $\tilde{F}_n (\omega,0)$ being the $n$-th Fourier coefficient of $F(\theta, \omega, 0)$. It is evident, that this initial condition lies on the OA manifold, as it satisfies the condition given in Eq.~\eqref{eq:22}:
\begin{eqnarray}
    \tilde{F}_n (\omega,0) = \left[ \tilde{F}_1 (\omega,0) \right]^n = 1~\forall~ n.
\end{eqnarray}
Provided the density function continues to remain on the OA manifold under the dynamical evolution, we can apply the OA analysis to study the density function of the entire system and obtain a first-order differential equation for the order parameter $r(t)$ of the entire system~\cite{ Ott_2008}.

On the other hand, if our system dynamics is initialized with the initial condition in which $f$ fraction of the total number of oscillators have the phase value zero, whereas the rest of the oscillators have the nonzero phase value equal to $\alpha $, then the initial density function writes as
\begin{eqnarray}
    F(\theta, \omega,0) &=& g(\omega) \left[ f \delta(\theta) + (1-f) \delta (\theta - \alpha) \right],\nonumber\\
    &=& \frac{g(\omega)}{2 \pi} \sum_{n = -\infty}^{\infty} \left[ f + (1-f) e^{-i n\alpha}\right]e^{i n\theta}.
\end{eqnarray}
In this case, the $n$-th Fourier coefficient, $\tilde{F}_n (\omega,0)$ does not satisfy in general the condition given in Eq.~\eqref{eq:22}:
\begin{eqnarray}
    \tilde{F}_n (\omega,0) = \left[ f + (1-f) e^{-i n\alpha}\right] \ne \left[ \tilde{F}_1 (\omega,0) \right]^n.
\end{eqnarray}
Thus, in this scenario, the initial condition of the dynamics does not lie on the OA manifold. As a result, we cannot apply the OA ansatz to write a single differential equation for the order parameter $r(t)$ of the entire system. On the contrary, if we confine our attention to the oscillators in the individual subsystems separately, their initial density functions, given by $F_1(\theta,\omega,0)$ and  $F_2(\theta,\omega,0)$, respectively, satisfy independently the condition given in Eq.~\eqref{eq:22} because of their mathematical forms as given below:
\begin{eqnarray}
    F_1(\theta, \omega, 0) &=& g(\omega) \delta (\theta) = \frac{g(\omega)}{2 \pi} \sum_{n = -\infty}^{\infty} e^{i n\theta},\\
    \nonumber
    F_2(\theta, \omega, 0) &=& g(\omega) \delta (\theta - \alpha) = \frac{g(\omega)}{2 \pi} \sum_{n = -\infty}^{\infty} \left(e^{-i \alpha} \right)^n e^{i n\theta}.\\
\end{eqnarray}
Consequently, in this case,  we can apply the OA ansatz separately to the two subsystems, to obtain three coupled differential equations for the order parameters of the two subsystems. This is discussed in detail below.

The aforementioned analysis will prove useful in  situations in which the initial density function of the entire system does not satisfy the OA condition given in Eq.~\eqref{eq:22}, but nevertheless the initial density functions of the two subsystems satisfy the OA condition.
As resetting a part of the system initialized on its OA manifold lands the system exactly onto such a state, the mentioned analysis is going to be useful to analyze the case of subsystem resetting. Let us now delve into the details of the mentioned analysis.

We consider the system of $N$ oscillators evolving according to Eq.~\eqref{eq:1}, as discussed in Sec.~\ref{sec:level2.1}. Among these $N$ oscillators, the ones with the label $j=1, \cdots , n$ form our subsystem marked $1$ constituted by the fraction $f = n/N$ of the total number of oscillators. The rest of the oscillators form our subsystem $2$ containing a fraction $1-f = (N-n)/N$ of the total number of oscillators.  
In the continuum limit $N \to \infty$, the oscillator frequency distribution of the two subsystems will be the one referring to the entire system, i.e.,
\begin{eqnarray}
    g_1(\omega) = g_2(\omega) = g(\omega).
\end{eqnarray}
In this limit, we characterize the state of the subsystem $1$ by the density function $F_1(\theta^{(1)}, \omega^{(1)}, t)$ and that of subsystem $2$ by the density function $F_2(\theta^{(2)}, \omega^{(2)}, t)$. The condition of normalization reads as 
\begin{eqnarray}
    \int_0^{2 \pi} F_1(\theta^{(1)}, \omega^{(1)}, t) d \theta^{(1)} &=& g_1(\omega^{(1)}) = g(\omega^{(1)}) \label{eq:23},\\
    \int_0^{2 \pi} F_2(\theta^{(2)}, \omega^{(2)}, t) d \theta^{(2)} &=& g_2(\omega^{(2)}) = g(\omega^{(2)}) \label{eq:24}.
\end{eqnarray}

Next,  we define the order parameters of the two subsystems  as
\begin{eqnarray}
    z_1 &\equiv& r_1 e^{i \psi_1 } \equiv \int_{-\infty}^{\infty} \int_0^{2 \pi} e^{i \theta^{(1)} }  F_1(\theta^{(1)}, \omega^{(1)}, t) d \theta^{(1)} d \omega^{(1)}, \hspace{0.5cm} \label{eq:29}\\
    z_2 &\equiv& r_2 e^{i \psi_2 } \equiv \int_{-\infty}^{\infty} \int_0^{2 \pi} e^{i \theta^{(2)} }  F_2(\theta^{(2)}, \omega^{(2)}, t) d \theta^{(2)} d \omega^{(2)}. \label{eq:30}
\end{eqnarray}
The continuum limit of Eq.~\eqref{eq:1} yields for the two subsystems the equations 
\begin{eqnarray}
	\nonumber
	\frac{d \theta^{(\kk)}}{d t} = \omega^{(\kk)} &+& \frac{K}{2 i} e^{- i \theta^{(\kk)} } \left[ f z_1 + (1-f) z_2 \right]\\
	&-&\frac{K}{2 i} e^{ i \theta^{(\kk)} } \left[ f z^{*}_1 + (1-f) z^{*}_2 \right] \label{eq:25},
\end{eqnarray}
with $\kk = 1,2$. 

 Similar to what was done in Sec.~\ref{sec:level4},  here also we can write a continuity equation for each of the density functions $F_{\kk}(\theta^{(\kk)}, \omega^{(\kk)}, t)$. Using the $2 \pi$-periodicity in $\theta^{(\rm k)}$ satisfied by the two functions, we can expand each of the $F_{\kk}(\theta^{(\kk)}, \omega^{(\kk)}, t)$'s in Fourier series. Let $\tilde{F}^{(\kk)}_n(\omega^{(\kk)},t)$ be the $n$-th Fourier coefficient of the function $F_{\kk}(\theta^{(\kk)}, \omega^{(\kk)}, t)$. Putting the Fourier expansions into the two continuity equations, we get an infinite number of coupled nonlinear differential equations for the $\tilde{F}^{(\kk)}_n(\omega^{(\kk)},t)$'s. For the special choice defining the OA ansatz~\cite{Laing_2009}, 
\begin{eqnarray}
    \tilde{F}^{(\kk)}_n(\omega^{(\kk)},t) = \left[ \alpha_{\kk} (\omega^{(\kk)},t)\right]^n,
\end{eqnarray}
the mentioned set of differential equations for the two subsystems reduces to  two partial differential equations governing the dynamics of $\alpha_{\kk} (\omega^{(\kk)},t)$ with $\kk=1,2$:
\begin{eqnarray}
    \nonumber
    \frac{\partial \alpha_{\kk}}{\partial t} = \frac{K}{2 }  [f z^{*}_{1} &+ &(1-f) z^{*}_{2}] - i \omega \alpha_{\kk} \\
    &-& \frac{K}{2}  [f z_{1} + (1-f) z_{2} ] \label{eq:31}   \alpha_{\kk}^{2}.
\end{eqnarray}

In the case of the Lorentzian frequency distribution given in Eq.~\eqref{eq:11}, the integrals in Eqs.~\eqref{eq:29} and~\eqref{eq:30} can be evaluated analytically, as discussed in Sec.~\ref{sec:level4}; one finally gets $z_{\kk} = \alpha^{*}_{\kk} \left( \omega_0 - i \sigma , t \right)$. Consequently,  Eq.~\eqref{eq:31} yields 
\begin{eqnarray}
    \nonumber
    \frac{d z_{\kk}}{d t}  &=& \frac{K}{2 } [ f z_1 + (1-f)z_2 ] \\
    \nonumber
    &&- \frac{K}{2 } [ f z^{*}_1+ (1-f) z^{*}_2 ] z^2_{\kk} 
     -  \left( \sigma - i\omega_0 \right)z_{\kk}\\
    \label{eq:32}
\end{eqnarray}
for each $\kk=1,2$. Comparing the real and the imaginary part on both sides of these two equations, we get the evolution equations of the order parameters $(r_1, \psi_1)$ and $(r_2,\psi_2)$ as
\begin{eqnarray}
    \frac{ d r_1}{d t} &=& - \sigma r_1 + K \left( \frac{1 - r_1^2}{2} \right) \left[ f r_1 + (1-f) r_2 \cos{\psi} \right], \hspace{0.5cm} \label{eq:33}\\
    \frac{ d r_2}{d t} &=& - \sigma r_2 + K \left( \frac{1 - r_2^2}{2} \right) \left[ f r_1 \cos{\psi} + (1-f) r_2  \right], \hspace{0.5cm}\label{eq:34}\\
     \frac{d\psi_1}{dt} &=&\omega_0 - K (1-f) \sin{\psi} \left(\frac{1 + r_1^2}{2 r_1}\right)  r_2, \label{eq:54}\\
    \frac{d\psi_2}{dt} &=&\omega_0 + K f \sin{\psi} \left(\frac{1 + r_2^2}{2 r_2}\right)  r_1 ,\label{eq:55}
\end{eqnarray}
where we have $\psi \equiv \psi_1 - \psi_2$. As discussed in Appendix~\ref{app:3} in the context of Eqs.~\eqref{eq:54} and~\eqref{eq:55}, and for our choice of the initial condition (namely, $\theta_j(0)=0~\forall~j$) together with $\omega_0 = 0$, one has $\psi_1(t) = 0 $ and $\psi_2(t) = 0~\forall~t$. We then obtain the stationary-state values of $r_1$ and $r_2$ to be satisfying
\begin{eqnarray}
	\nonumber
	(r_1^{\rm st})^3 &+& \left[\frac{1-f}{f} \right] \left[(r_1^{\rm st})^2 r_2^{\rm st}-r_2^{\rm st}\right]\\
	&+& \left[\frac{2 \sigma}{Kf} -1\right]r_1^{\rm st} = 0, \label{eq:bare-steady-twO-sustem1}\\
	\nonumber (r_2^{\rm st})^3 &+& \left[\frac{f}{1-f} \right] \left[ r_1^{\rm st}(r_2^{\rm st})^2 -r_1^{\rm st}\right]\\
	&+& \left[\frac{2 \sigma}{K(1-f)} -1\right]r_2^{\rm st} = 0. \label{eq:bare-steady-twO-sustem2}
\end{eqnarray}
It may be noted that $r_1^{\rm st} = r_2^{\rm st}$ is a solution of the above two coupled equations, thereby reducing them into a single equation for $r_1^{\rm st} = r_2^{\rm st} = r^{\rm st}$.

If the dynamics of both the subsystems is initialized with identical initial conditions, we can intuitively write $r_1(t) = r_2(t)$ and $\psi_1(t)=\psi_2(t)$ (yielding $\psi(t)=0$) at all times $t \ge 0$, valid in the limit $N \to \infty$. In such a case, the dynamics of the entire system gets initialized also from the same initial condition, and hence, the order parameter $r(t)$ for the entire system will be identical to the quantities $r_1(t)$ and $r_2(t)$ at all times. 
Using these facts in either of the two equations~\eqref{eq:33} and~\eqref{eq:34}, we get
\begin{eqnarray}
    \frac{d r}{d t} + \left[ \sigma - \frac{K}{2} \right] r + \frac{K}{2} r^3 = 0, \label{eq:36}
\end{eqnarray}
which is exactly the equation derived in Ref.~\cite{Ott_2008}. Thus, in the case of identical initial conditions for the two subsystems, we can reduce the four equations, i.e., Eqs.~\eqref{eq:33},~\eqref{eq:34},~\eqref{eq:54}, and~\eqref{eq:55}, into one single equation describing the evolution of the order parameter for the entire system.

\subsubsection{\label{sec:level7}Resetting with $\omega_0=0$ }
We now incorporate resetting into our system dynamics, as detailed in Sec.~\ref{sec:Dynamics}. We will consider subsystem 1 to be the reset subsystem and subsystem 2 to be the non-reset subsystem. Hence, from now on, while using Eqs.~\eqref{eq:33},~\eqref{eq:34},~\eqref{eq:54}, and~\eqref{eq:55}, we are going to use $r_\mathrm{r}(t)$ in place of $r_1(t)$, $\psi_\mathrm{r}(t)$ in place of $\psi_1(t)$, $r_\mathrm{nr}(t)$ in place of $r_2(t)$, and $\psi_\mathrm{nr}(t)$ in place of $\psi_2(t)$.

Evidently, resetting at finite $\lambda$ promotes the status of the order parameters, $(r_\mathrm{r}(t),\psi_\mathrm{r}(t))$ and $(r_\mathrm{nr}(t),\psi_\mathrm{nr}(t))$, to that of a random variable (in the case of infinite $\lambda$, by contrast, one has $r_\mathrm{r}(t)=1, \psi_\mathrm{r}(t)=0$, and also deterministic evolution given by Eqs.~\eqref{eq:14} and~\eqref{eq:16} for the quantities $r_\mathrm{nr}$ and $\psi_\mathrm{nr}$, so that   the latter are not random variables). Therefore, it is appropriate to consider these quantities when averaged over dynamical realizations, i.e., the quantities $\bar{r}_\mathrm{r}(t)$, $\bar{r}_\mathrm{nr}(t)$, $\bar{\psi}_\mathrm{r}(t)$ and $\bar{\psi}_\mathrm{nr}(t)$. Let us first consider the case $\omega_0 = 0$.

As discussed above for the case of bare evolution with the initial condition $\psi_\mathrm{r}(0)=\psi_\mathrm{nr}(0) = 0$ and with $\omega_0=0$, we have  $\psi_\mathrm{r}(t) = 0 $ and $\psi_\mathrm{nr}(t) = 0~\forall~t$. When considering effects of resetting, the following picture is true. Suppose we start with a state in which $\psi_\mathrm{r}(0)=\psi_\mathrm{nr}(0)=0$. Until the next reset instant, when the system follows bare evolution, one will continue to have both $\psi_\mathrm{r}=\psi_\mathrm{nr}=0$. At the instant of reset, the reset oscillators are all reset to the phase value zero, while leaving untouched the non-reset oscillators. Hence, at the instant of reset, one will continue to have $\psi_\mathrm{r}=\psi_\mathrm{nr}=0$. In summary, we conclude that $\psi_\mathrm{r}(t) = 0 $ and $\psi_\mathrm{nr}(t) = 0$ for all times $t$. As a result, Eqs.~\eqref{eq:33},~\eqref{eq:34},~\eqref{eq:54}, and~\eqref{eq:55} yield the following equations for the bare evolution between two successive resets:
\begin{eqnarray}
    \frac{ d r_\mathrm{r}}{d t} &=& - \sigma r_\mathrm{r} + K \left( \frac{1 - r_\mathrm{r}^2}{2} \right) \left[ f r_\mathrm{r} + (1-f) r_\mathrm{nr}  \right], \hspace{0.5cm} \label{eq:40}\\
    \frac{ d r_\mathrm{nr}}{d t} &=& - \sigma r_\mathrm{nr} + K \left( \frac{1 - r_\mathrm{nr}^2}{2} \right) \left[ f r_\mathrm{r} + (1-f) r_\mathrm{nr}  \right]. \hspace{0.5cm}\label{eq:41}
\end{eqnarray}\\

Let us say that at time $t$, the order parameters of the two subsystems equal $r_\mathrm{r}(t)$ and $r_\mathrm{nr}(t)$ in a typical dynamical realization. In the time interval $[t,t+dt]$, the quantity $r_\mathrm{r}$ evolves following Eq.~\eqref{eq:40} with probability $(1-\lambda d t)$, while with probability $\lambda d t$, it gets reset to the value unity. This yields the realization average of the change in the order parameters within the time interval $[t,t+d t]$, given that their values were $r_\mathrm{r}(t)$ and $r_\mathrm{nr}(t)$ at time $t$, as
\begin{eqnarray}
    d \tilde{r}_\mathrm{r} &=& \left(1-\lambda d t\right)d r_\mathrm{r} + \lambda d t \left(1- r_\mathrm{r}\right), \label{eq:42}\\
    d \tilde{r}_\mathrm{nr} &=& d r_\mathrm{nr}, \label{eq:43}
\end{eqnarray}
where the tilde denotes average over all those realizations for which the order parameters have values $r_\mathrm{r}$ and $r_\mathrm{nr}$ at time $t$. Using Eqs.~\eqref{eq:40} and~\eqref{eq:41} in Eqs.~\eqref{eq:42} and~\eqref{eq:43}, we get to leading order in $dt$ that
\begin{eqnarray}
    \nonumber
    \frac{d \tilde{r}_\mathrm{r}}{d t} &=& - \sigma r_\mathrm{r} + K \left( \frac{1 - r_\mathrm{r}^2}{2} \right) \left[ f r_\mathrm{r} + (1-f) r_\mathrm{nr}  \right] + \lambda \left(1- r_\mathrm{r}\right), \\
    \label{eq:44}\\
    \frac{d \tilde{r}_\mathrm{nr}}{d t} &=&  - \sigma r_\mathrm{nr} + K \left( \frac{1 - r_\mathrm{nr}^2}{2} \right) \left[ f r_\mathrm{r} + (1-f) r_\mathrm{nr}  \right]. \label{eq:45}
\end{eqnarray}

Now, the values $r_\mathrm{r}, r_\mathrm{nr}$ at time $t$ being also random variables, in order to get the realization-averaged order parameters, we have to consider the average of these random variables. This gives
\begin{widetext}
\begin{eqnarray}
    \frac{d \bar{r}_\mathrm{r}}{d t} &=& - \sigma \bar{r}_\mathrm{r} + \frac{K}{2} \left[f \bar{r}_\mathrm{r} + (1-f)\bar{r}_\mathrm{nr} - f \overline{r_\mathrm{r}^3}-(1-f) \overline{r_\mathrm{r}^2 r_\mathrm{nr}} \right] 
    + \hspace{0.1cm}\lambda \hspace{0.1cm}(1 - \bar{r}_\mathrm{r}), \label{eq:46}\\
    \frac{d \bar{r}_\mathrm{nr}}{d t} &=& - \sigma \bar{r}_\mathrm{nr} + \frac{K}{2} \left[f \bar{r}_\mathrm{r} + (1-f)\bar{r}_\mathrm{nr}-f \overline{r_\mathrm{r} r_\mathrm{nr}^2} - (1-f) \overline{r_\mathrm{nr}^3} \right]. \label{eq:47}
\end{eqnarray}
\end{widetext}
Equations~\eqref{eq:46} and~\eqref{eq:47} give the exact evolution equation for the realization-averaged order parameters of the reset and non-reset subsystems for the case at hand. Note that Eqs.~\eqref{eq:46} and~\eqref{eq:47} involve correlations between the quantities $r_\mathrm{r}$ and $r_\mathrm{nr}$ as well as their higher-order moments. In order to solve Eqs.~\eqref{eq:46} and~\eqref{eq:47}, one would require time evolution equations for the quantities $\overline{r_\mathrm{r}^3}$, $\overline{r_\mathrm{nr}^3}$, $\overline{r_\mathrm{r}^2 r_\mathrm{nr}}$ and $\overline{r_\mathrm{r} r_\mathrm{nr}^2}$, which would involve even higher-order moments.  To make these equations analytically tractable requires approximations as detailed below. 

\begin{figure*}
	\centering
	\begin{subfigure}[b]{0.45\textwidth}
		\centering
		\includegraphics[width=\textwidth]{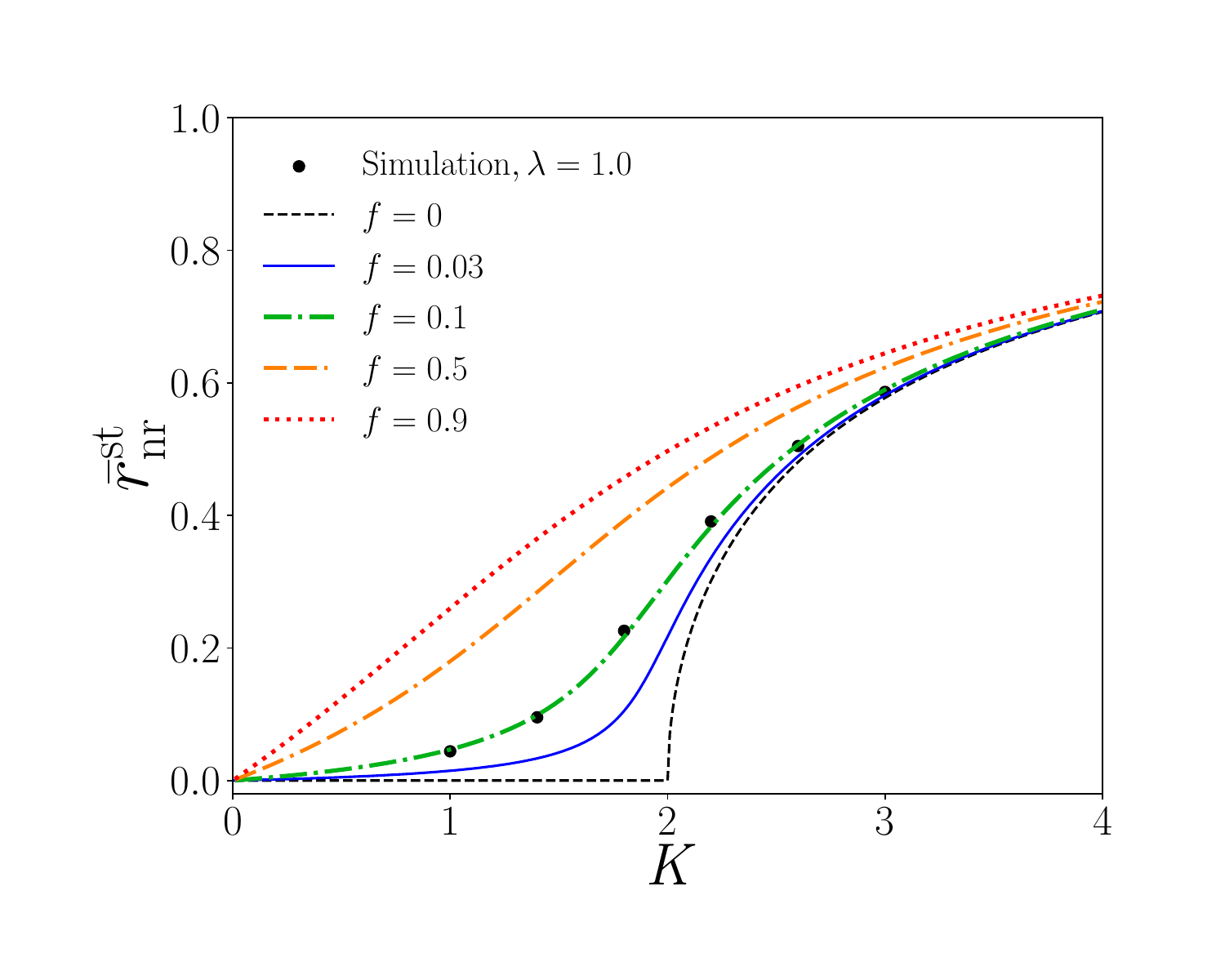}
		\caption{$\lambda = 1.0$}
		\label{fig:5a} 
	\end{subfigure}
	\begin{subfigure}[b]{0.45\textwidth}
		\centering
		\includegraphics[width=\textwidth]{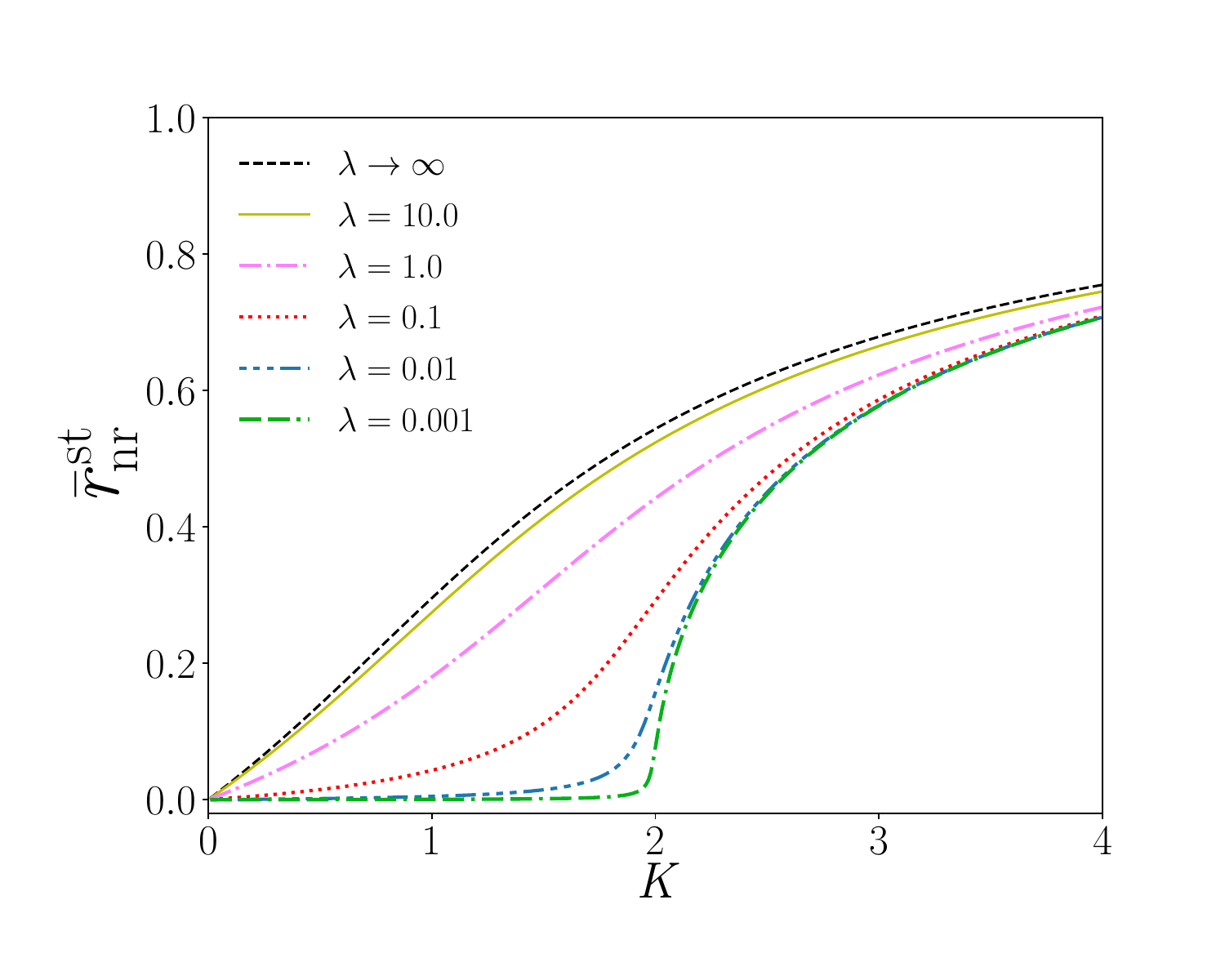}
		\caption{$f = 0.5$}
		\label{fig:5b}
	\end{subfigure}
	\caption{\textbf{Subsystem resetting with finite $\lambda$ and Lorentzian $g(\omega)$ with mean $\omega_0=0$ and width $\sigma=1.0$}: Analytical results showing variation of $r^\mathrm{st}_\mathrm{nr}$ with $K$ for representative values of $f$ are presented in panel (a) for $\lambda=1.0$.  Here, the black dashed line corresponds to the bare Kuramoto dynamics. In this panel, we also show by points numerical simulation results for $f=0.1$, demonstrating agreement with theory; the data correspond to the stationary-state value averaged over $100$ realizations of the dynamics for a system of $N=10^4$ oscillators with the integration time step equal to $0.05$. Panel (b) shows analytical results for the variation of $r^\mathrm{st}_\mathrm{nr}$ with $K$ for representative values of $\lambda$ and with $f=0.5$. Here, the black dashed line corresponds to the limit $\lambda \to \infty$ result reported in Fig.~\ref{fig:3}(b).  The analytical results in both the panels are obtained by simultaneously solving Eqs.~\eqref{eq:49} and~\eqref{eq:50}.}
	\label{fig:5}
\end{figure*}

When $\lambda$ is large, the dynamics of the reset subsystem will be dominated by resetting, which happens over a very small time scale of order $1/ \lambda$. On the other hand, the effect of this resetting will propagate into the non-reset subsystem over the time scale set by the strength of the interaction between the reset and the non-reset subsystem, i.e., over the time scale of order $1/K$. For large $\lambda$, when we have a separation of time scales between these two processes, assuming the random variables $r_\mathrm{r}$ and $r_\mathrm{nr}$ to have sharply-peaked distributions, let us invoke the following approximations:
\begin{eqnarray}
    \overline{r_\mathrm{r} r_\mathrm{nr}^2} \approx \bar{r}_\mathrm{r} \bar{r}_\mathrm{nr}^2; \hspace{0.3cm} \overline{r_\mathrm{r}^2 r_\mathrm{nr}} \approx \bar{r}_\mathrm{r}^2 \bar{r}_\mathrm{nr}; \hspace{0.3cm} \overline{r_\mathrm{nr}^3} \approx \bar{r}_\mathrm{nr}^3; \hspace{0.3cm} \overline{r_\mathrm{r}^3} \approx \bar{r}_\mathrm{r}^3. \hspace{0.8cm} \label{eq:48}
\end{eqnarray}
Under the above approximations, we obtain from Eqs.~\eqref{eq:46} and~\eqref{eq:47} the stationary-state values $\bar{r}^\mathrm{st}_\mathrm{r}$ and $\bar{r}^\mathrm{st}_\mathrm{nr}$ to be satisfying 
\begin{eqnarray}
	\nonumber
	\left(\bar{r}^\mathrm{st}_\mathrm{r}\right)^3 &+& \left[\frac{1-f}{f} \right] \left[ \left(\bar{r}^\mathrm{st}_\mathrm{r}\right)^2 \bar{r}^\mathrm{st}_\mathrm{nr} - \bar{r}^\mathrm{st}_\mathrm{nr}\right]\\
	&+& \left[\frac{2 (\lambda+\sigma)}{K f} - 1 \right] \bar{r}^\mathrm{st}_\mathrm{r} = \frac{2 \lambda}{Kf}, \label{eq:49}\\
	\nonumber \left(\bar{r}^\mathrm{st}_\mathrm{nr}\right)^3 &+& \left[\frac{f}{1-f} \right] \left[ \bar{r}^\mathrm{st}_\mathrm{r} \left( \bar{r}^\mathrm{st}_\mathrm{nr}\right)^2 - \bar{r}^\mathrm{st}_\mathrm{r}\right] \\
	&+& \left[\frac{2 \sigma}{K (1-f)} - 1 \right] \bar{r}^\mathrm{st}_\mathrm{nr} = 0. \label{eq:50}
\end{eqnarray}

As we move on, let us emphasize that for the case at hand, $\bar{r}^\mathrm{st}_\mathrm{nr} = 0$ cannot be a solution of Eqs.~\eqref{eq:49} and~\eqref{eq:50}, for any nonzero $f$ and $\lambda$. As a result, under the approximation scheme~\eqref{eq:48}\ invoked by us, the synchronization phase transition of the bare model converts into a crossover for any nonzero $f$ and $\lambda$. 

Interestingly, in the $\lambda \to \infty$ limit, the only solution of Eq.~\eqref{eq:49} turns out to be $\bar{r}_\mathrm{r}^\mathrm{st} = 1$. Using it in Eq.~\eqref{eq:50}, we retrieve Eq.~\eqref{eq:19}, which we had derived as an exact equation in the infinite resetting rate limit. Thus, at least in the limit $\lambda \to \infty$, our approximation scheme~\eqref{eq:48} is correct; this is consistent with the reasoning behind the approximation that we give in the paragraph preceding Eq.~\eqref{eq:48}. On the other hand, setting $\lambda = 0$ in Eqs.~\eqref{eq:49} and~\eqref{eq:50} obtains the stationary state of the bare evolution given by Eqs.~\eqref{eq:bare-steady-twO-sustem1}~and~\eqref{eq:bare-steady-twO-sustem2}. 

The results of our analysis are presented in Fig.~\ref{fig:5}.
We see from Figure~\ref{fig:5}(b) that, even for $\lambda = 10.0$, the quantity $\bar{r}^\mathrm{st}_\mathrm{nr}$ is close to the infinite resetting rate result, denoted by the black dashed line. We have checked that for $\lambda$ values around $25.0$, the results almost superpose on the $\lambda \to \infty$ result.  Figure~\ref{fig:5}(b) also shows that as we decrease $\lambda$, the amount synchronization induced in the non-reset subsystem for fixed $K$ and $f$ also decreases. Moreover, as $\lambda \to 0$, the results approach those for the bare evolution. 

In Fig.~\ref{fig:5}(a), we plot the finite-$\lambda$ (here we plotted for representative $\lambda$ value of $1.0$) counterpart of the corresponding infinite resetting rate plot, Fig.~\ref{fig:3}(b). In this panel, we also demonstrate agreement between our approximate theory and results from direct numerical simulations for $f = 0.1$. 

Interestingly, Eqs.~\eqref{eq:49}~and~\eqref{eq:50} show that different combinations of $\lambda, f$~and~$K$ can give identical solutions of $\bar{r}^\mathrm{st}_\mathrm{nr}$. For example, both the combinations $\lambda = 1.0, f= 0.4, K = 1.5$~and~$\lambda = 10.0, f= 0.2, K = 1.5$ produce $\bar{r}^\mathrm{st}_\mathrm{nr} \approx 0.28$ as the solution of Eqs.~\eqref{eq:49}~and~\eqref{eq:50}. This implies that for a given target value of $\bar{r}^\mathrm{st}_\mathrm{nr}$ to be achieved, there is a trade-off between how fast the resetting events take place, i.e., the value of the resetting rate $\lambda$, and how big a part of the system does one choose to reset, i.e., the value of the reset fraction $f$.

\subsubsection{\label{sec:level8}Resetting with $\omega_0\neq 0$ }

In the case of $\omega_0 \neq 0$, the relevant equations are Eqs.~\eqref{eq:42} and~\eqref{eq:43} together with the following equations for $\tilde{\psi}_\mathrm{r}$ and $\tilde{\psi}_\mathrm{nr}$:
\begin{eqnarray}
    d \tilde{\psi}_\mathrm{r} &=& \left(1-\lambda d t\right)d \psi_\mathrm{r} - \lambda d t \psi_\mathrm{r}, \label{eq:tilde-psir}\\
    d \tilde{\psi}_\mathrm{nr} &=& d \psi_\mathrm{nr}. \label{eq:tilde-psinr}
\end{eqnarray}
Using Eqs.~\eqref{eq:33},~\eqref{eq:34},~\eqref{eq:54},~\eqref{eq:55} with the obvious substitution $1 \leftrightarrow \mathrm{r}$ and $2 \leftrightarrow \mathrm{nr}$, and proceeding as in Sec.~\ref{sec:level7}, one finally obtains 
\begin{eqnarray}
    \nonumber \frac{d \bar{r}_\mathrm{r}}{d t}  &=& - \sigma \bar{r}_\mathrm{r} + \frac{K}{2} \left[f \bar{r}_\mathrm{r} + (1-f) \overline{r_\mathrm{nr} \cos{(\psi_\mathrm{r} - \psi_\mathrm{nr})}} - f \overline{r^3_\mathrm{r}}\right.\\
    &&  \left. - (1-f) \overline{r^2_\mathrm{r} r_\mathrm{nr} \cos{(\psi_\mathrm{r} - \psi_\mathrm{nr})}} \right] + \lambda( 1 - \bar{r}_\mathrm{r}), \label{eq:56}\\
   \nonumber \frac{d \bar{r}_\mathrm{nr}}{d t} &=& - \sigma \bar{r}_\mathrm{nr} + \frac{K}{2} \left[f \overline{r_\mathrm{r} \cos{(\psi_\mathrm{r} - \psi_\mathrm{nr})}} + (1-f)\bar{r}_\mathrm{nr} \right.\\
    && \hspace{1cm} \left. -f \overline{r_\mathrm{r} r_\mathrm{nr}^2 \cos{(\psi_\mathrm{r} - \psi_\mathrm{nr})}} - (1-f) \overline{r_\mathrm{nr}^3} \right], \label{eq:57}\\
\nonumber \frac{d \bar{\psi}_\mathrm{r}}{d t} &=& \omega_0 - K(1-f) \overline{\sin{(\psi_\mathrm{r} - \psi_\mathrm{nr})} \left(\frac{1 + r_\mathrm{r}^2}{2 r_\mathrm{r} }\right)r_\mathrm{nr}} - \lambda \bar{\psi_\mathrm{r}},\\
\label{eq:58}\\
 \frac{d \bar{\psi}_\mathrm{nr}}{d t} &=& \omega_0 + Kf \overline{\sin{(\psi_\mathrm{r} - \psi_\mathrm{nr})} \left(\frac{1 + r_\mathrm{nr}^2}{2 r_\mathrm{nr} }\right)r_\mathrm{r}}. \label{eq:59}
\end{eqnarray}

Equations~\eqref{eq:56},~\eqref{eq:57},~\eqref{eq:58}, and~\eqref{eq:59} give the exact evolution equation for the realization-averaged order parameters of the reset and non-reset subsystems for finite resetting rate $\lambda$ and with $\omega_0 \neq 0$. Solving the above equations exactly is a difficult task owing to the presence of terms involving higher-order moments and correlations, e.g., the term $\overline{r_\mathrm{nr} \cos{(\psi_\mathrm{r} - \psi_\mathrm{nr})}}$. In the hope of extracting  useful analytical results, we now invoke an approximation scheme, similar to the scheme~\eqref{eq:48} invoked for $\omega_0=0$, which ignores correlations and converts the above set of equations into one that is closed. As for the $\omega_0=0$ case, this approximation will be valid at high $\lambda$, where a timescale separation is possible, as discussed in Sec.~\ref{sec:level7}. Using the mentioned approximation, the stationary-state values of the averaged order parameters may be seen to satisfy the equations
\begin{eqnarray}
    \nonumber \frac{d \bar{r}_\mathrm{r}}{d t}  &=& - \sigma \bar{r}_\mathrm{r} + \frac{K}{2} \left[f \bar{r}_\mathrm{r} + (1-f) \bar{r}_\mathrm{nr} \cos{(\bar{\psi}_\mathrm{r} - \bar{\psi}_\mathrm{nr})} - f \bar{r}^3_\mathrm{r}\right.\\
    &&  \left. - (1-f) \bar{r}^2_\mathrm{r} \bar{r}_\mathrm{nr} \cos{(\bar{\psi}_\mathrm{r} - \bar{\psi}_\mathrm{nr})}\right] + \lambda( 1 - \bar{r}_\mathrm{r}), \label{eq:60}\\
   \nonumber \frac{d \bar{r}_\mathrm{nr}}{d t} &=& - \sigma \bar{r}_\mathrm{nr} + \frac{K}{2} \left[f \bar{r}_\mathrm{r} \cos{(\bar{\psi}_\mathrm{r} - \bar{\psi}_\mathrm{nr})} + (1-f)\bar{r}_\mathrm{nr} \right.\\
    && \hspace{1cm} \left. -f \bar{r}_\mathrm{r} \bar{r}_\mathrm{nr}^2 \cos{(\bar{\psi}_\mathrm{r} - \bar{\psi}_\mathrm{nr})} - (1-f) \bar{r}_\mathrm{nr}^3 \right], \label{eq:61}\\
\nonumber \frac{d \bar{\psi}_\mathrm{r}}{d t} &=& \omega_0 - K(1-f) \sin{(\bar{\psi}_\mathrm{r} - \bar{\psi}_\mathrm{nr})} \left(\frac{1 + \bar{r}_\mathrm{r}^2}{2 \bar{r}_\mathrm{r} }\right)\bar{r}_\mathrm{nr} - \lambda \bar{\psi_\mathrm{r}},\\
\label{eq:62}\\
 \frac{d \bar{\psi}_\mathrm{nr}}{d t} &=& \omega_0 + Kf \sin{(\bar{\psi}_\mathrm{r} - \bar{\psi}_\mathrm{nr})} \left(\frac{1 + \bar{r}_\mathrm{nr}^2}{2 \bar{r}_\mathrm{nr} }\right)\bar{r}_\mathrm{r}. \label{eq:63}
\end{eqnarray} 
Before we move on to discuss the solution of these equations, let us note that within our approximation scheme, we have  $\overline{\cos{(\psi_\mathrm{r} - \psi_\mathrm{nr})}} \approx \cos{(\bar{\psi}_\mathrm{r}-\bar{\psi}_\mathrm{nr})}$. This makes the approximation at hand more restrictive to be true than the approximation~\eqref{eq:48} invoked for the $\omega_0 = 0$ case.

\begin{figure*}
	\centering
	\begin{subfigure}{0.584\textwidth}
		\includegraphics[width=\linewidth]{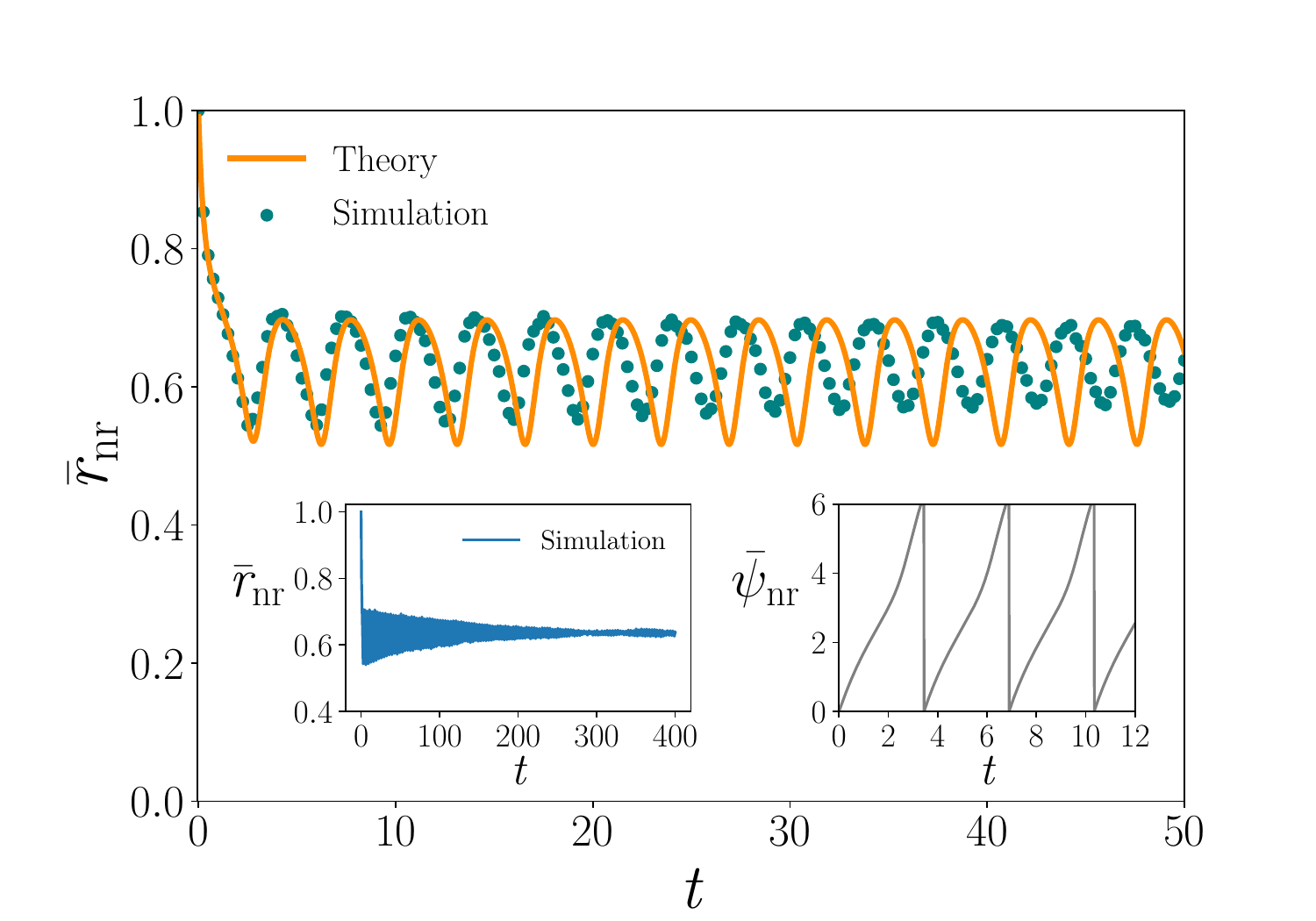}
		\caption{$f = 0.2,\lambda = 5.0$}
		\label{fig:6a}
	\end{subfigure}
	\hfill
	\begin{subfigure}{0.41\textwidth}
		\includegraphics[width=\linewidth]{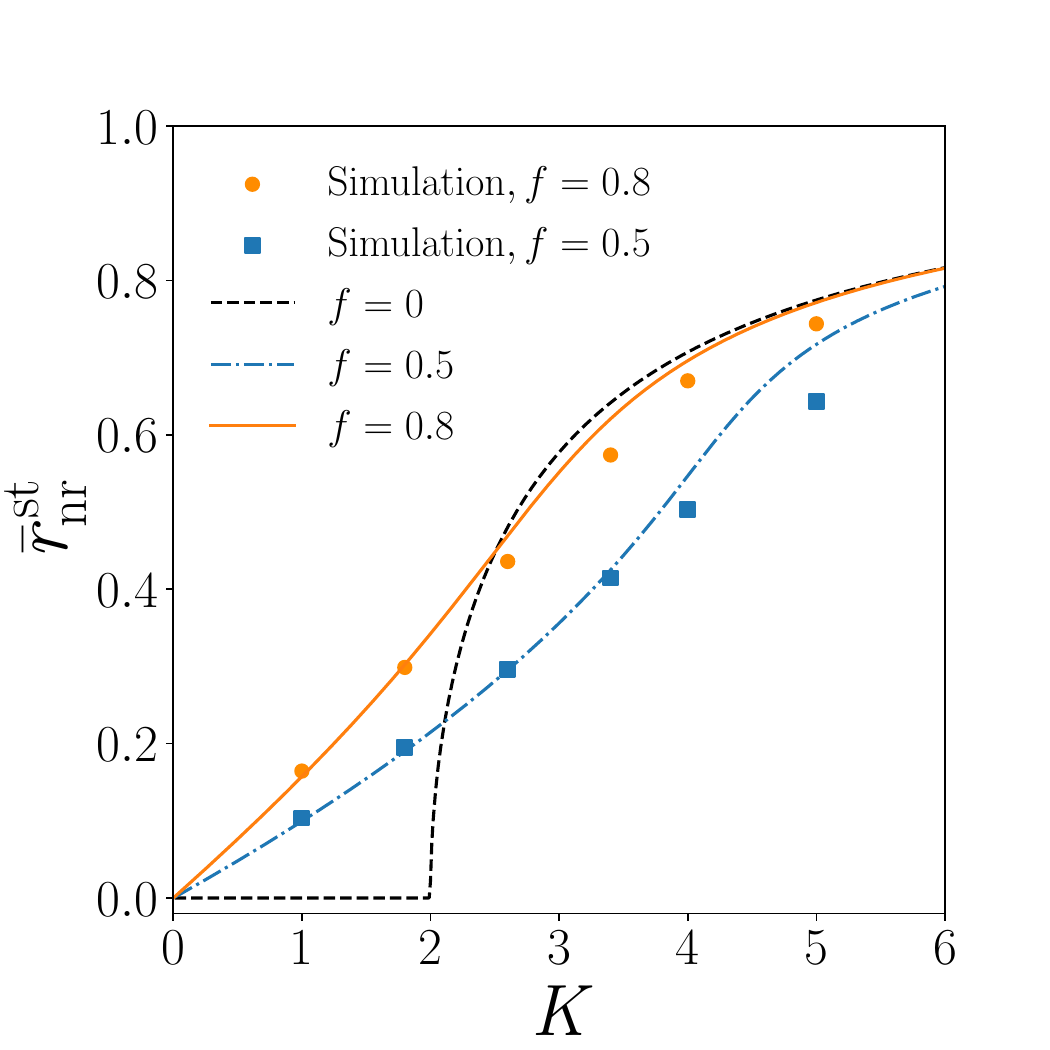}
		\caption{$\lambda = 5.0$}
		\label{fig:6b}
	\end{subfigure}
	\caption{\textbf{Subsystem resetting with finite $\lambda$ (specifically, $\lambda=5.0$) and Lorentzian $g(\omega)$ with mean $\omega_0 \ne 0$ (specifically, $\omega_0=2.0$) and width $\sigma=1.0$}: For $K=4.0$, panel (a) compares in the main plot theoretical and simulations results for the realization-averaged order parameter $\bar{r}_\mathrm{nr}$ as a function of time $t$ for $f=0.2$. The theoretical results are obtained by solving Eqs.~\eqref{eq:60},~\eqref{eq:61},~\eqref{eq:62},~\eqref{eq:63} simultaneously, while simulations correspond to $N=10^4$ number of oscillators with integration time step equal to $0.05$ and averaging over $100$ realizations. Theoretical results in panel (a) suggest that for small $f$, the quantity $\bar{r}_\mathrm{nr}$ instead of reaching a stationary value at long times oscillates as a function of time, while $\psi_\mathrm{nr}$ remains time dependent even at long times (see the inset of panel (a)). On the other hand, our simulation results suggest  $\bar{r}_\mathrm{nr}$ to display oscillations with decaying amplitude and attaining a stationary value at long times (see the inset of panel (a)). In contrast, for higher $f$ values, our theoretical calculations predict that the system reaches a stationary state, in which both the quantities $r_\mathrm{nr}$ and $\psi_\mathrm{nr}$ attain time-independent values at long times. Variation of $r^\mathrm{st}_\mathrm{nr}$ with $K$ for representative values of $f$ is presented in panel (b). Here, the black dashed line corresponds to the bare Kuramoto dynamics. In this panel, we also show by points numerical simulation results for $f=0.5$ and $f = 0.8$, demonstrating good agreement with theory in the $K \leq K_c$ region; the agreement gets worsened for higher $K$ values in the $K >K_c$ region.}
	\label{fig:6}
\end{figure*}

The results of the above analysis are presented in Fig.~\ref{fig:6}. The main conclusions are the following: (i) for $K\le K_c$, both theory and simulations agree on the fact that the system reaches a synchronized stationary state at long times, for all $f$. (ii) for $K>K_c$, our approximate theory predicts a synchronized stationary state at large $f$, which is in qualitative agreement with numerical simulation results. For small $f$, simulations display $\bar{r}_\mathrm{nr}$ oscillating with decaying amplitude and eventually settling to a synchronized stationary state. This is at variance with our approximate theory that predicts an oscillatory synchronized state at long times. The reason for this mismatch between analytical and simulation in certain parameter regimes, as seen in Fig.~\ref{fig:6}, could be attributed to the approximation invoked in obtaining Eqs.~\eqref{eq:60},~\eqref{eq:61},~\eqref{eq:62}~and~\eqref{eq:63} from Eqs.~\eqref{eq:56},~\eqref{eq:57},~\eqref{eq:58}~and~\eqref{eq:59}, namely, we have ignored the correlations between the quantities $r_\mathrm{r}$ and $r_\mathrm{nr}$. A better match is expected on developing an analytical framework that accounts for the mentioned correlations. This is an issue not easy to address given the many-body character of the Kuramoto dynamics, and which can be pursued as part of future studies.

\subsection{\label{sec:level9}Discussion of our results: Controlling the amount of synchronization}
Here, we discuss the implications of the $\lambda \to \infty$ results in answering the following question of practical relevance: What should be the fraction $f$ of reset oscillators that ensures enhanced synchronization in the non-reset subsystem compared to its bare evolution? We consider separately the cases $\omega_0=0$ and $\omega_0\neq0$.

For $\omega_0=0$, the quantity $r_{\mathrm{nr}}^\mathrm{st}$ as a function of $K$ has a value greater than that for bare evolution for any reset fraction $f$, however small, see  Fig.~\ref{fig:3}(b). The implication is that at any $K$, one can induce enhanced synchronization compared to the bare evolution by choosing $f$ as small as one wishes.
Thus, subsystem resetting offers an efficient mechanism aimed at controlling phase coherence in a synchronizing system through manipulation of as small a number of the system constituents as the available resources allow for.    
 
\begin{figure}
	\begin{subfigure}[b]{0.25\textwidth}
		\includegraphics[width=\linewidth]{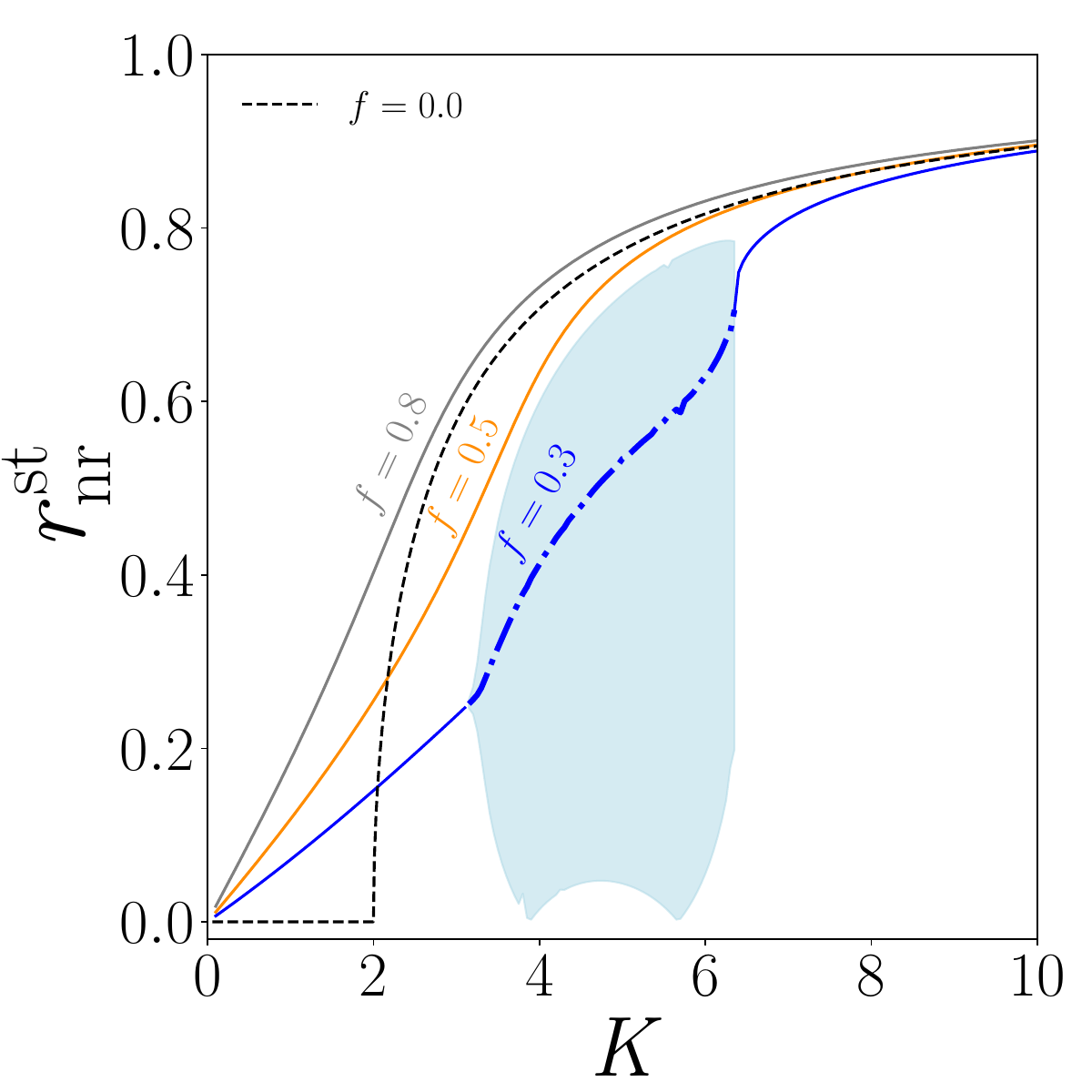}
		\caption{$\omega_0 = 2.0$}
		\label{fig:omegaeq2}
	\end{subfigure}%
	\begin{subfigure}[b]{0.25\textwidth}
		\includegraphics[width=\linewidth]{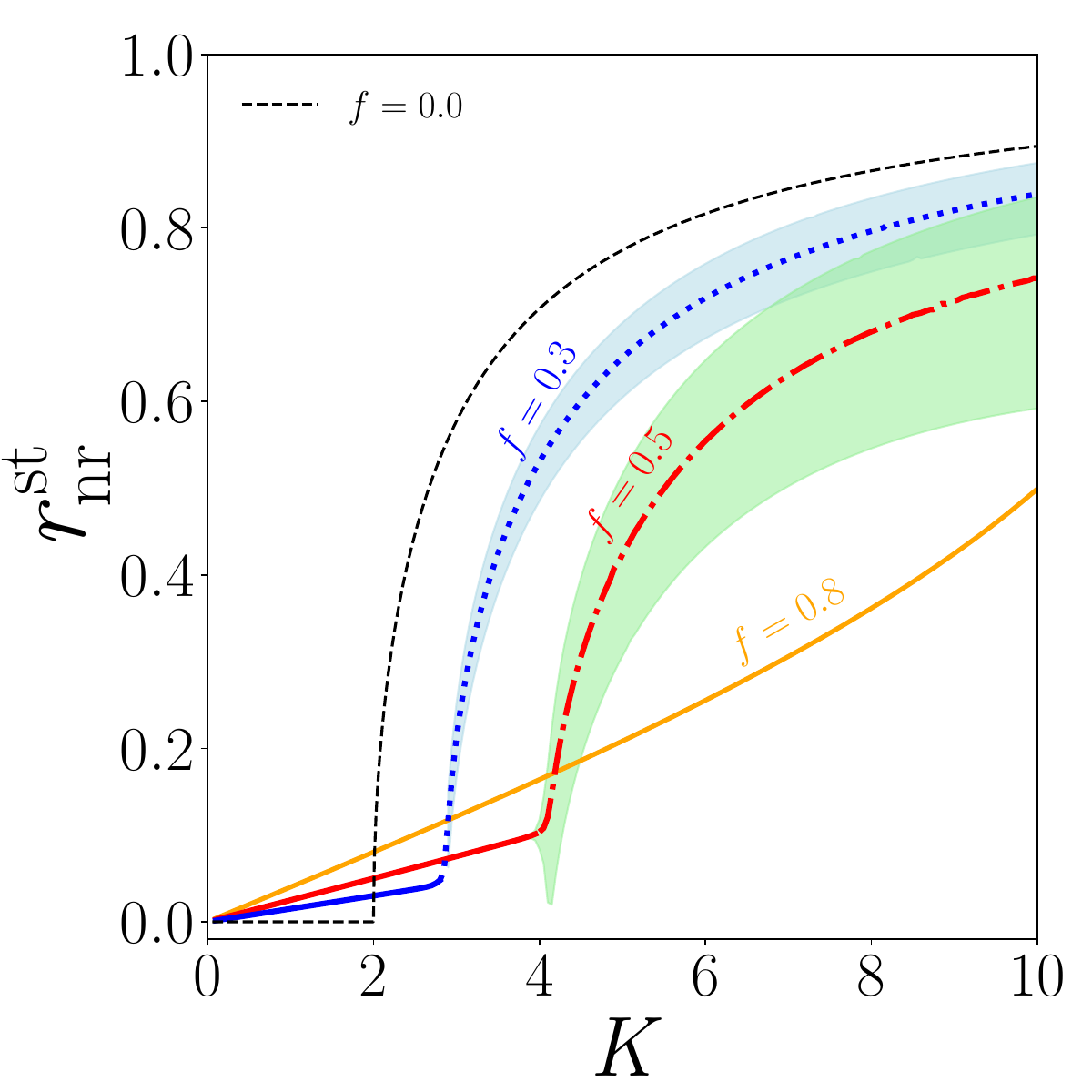}
		\caption{$\omega_0= 10.0$}
		\label{fig:omeganeq10}
	\end{subfigure}
	\caption{\textbf{Subsystem resetting with $\lambda \to \infty$ and Lorentzian $g(\omega)$ with mean $\omega_0 \ne 0$ and width $\sigma=1.0$}: The plots show for two values of $\omega_0$ (namely $\omega_0 = 2.0$ in panel (a) and $\omega_0 = 10.0$ in panel (b)) and in terms of solid lines the stationary value $r_\mathrm{nr}^\mathrm{st}$ of the synchronization order parameter in the non-reset subsystem; the data are obtained by solving numerically the roots of Eq.~\eqref{eq:20} 
		On the other hand, the shaded regions correspond to oscillatory behavior of $r_\mathrm{nr}$ as a function of time at long times, with the extent of the shaded region at a fixed $K$ denoting the amplitude of oscillations, and broken lines inside the shaded region denoting the non-zero time-independent value of the time average of $r_\mathrm{nr}$ at long times; the data are obtained by numerically solving Eqs.~\eqref{eq:14} and~\eqref{eq:16} simultaneously. The values of the parameters $\omega_0$ and $f$ are marked in the figure.}
	\label{fig:8}
\end{figure}

The case of $\omega_0\neq0$ is more intricate. Here, the optimal choice of $f$ depends on the specific values of $\omega_0$ and $K$. Indeed, referring to Fig.~\ref{fig:8}(a) for $\omega_0=2.0$, one finds that choosing a large $f$ results in enhanced synchrony, while a similar choice proves detrimental for larger $\omega_0$, see Fig.~\ref{fig:8}(b) for $\omega_0=10.0$. From the foregoing discussions, we conclude that enhancement of $r_\mathrm{nr}^\mathrm{st}$ compared to the bare evolution is more easily achieved with $\omega_0=0$. For $\omega_0\neq0$, enhanced synchronization demands resetting the subsystem oscillators to the phase value $\omega_0 t$, as this would result in a situation analogous to the $\omega_0=0$ case.

\section{\label{sec:Gaussian distribution}Results for Gaussian $g(\omega)$}
In the preceding section, we had taken the distribution function $g(\omega)$ for our oscillators to be Lorentzian as given in Eq.~\eqref{eq:11}. Here, we analyze our system with a $g(\omega)$ that is Gaussian instead, see Eq.~\eqref{eq:11}. We will take the width $\sigma$ to be unity, since results for any other value of $\sigma$ may be obtained through rescaling of variables, see the discussion following Eq.~\eqref{eq:13}. We start with the case of infinite resetting rate, i.e., $\lambda \to \infty$. Our analysis using the OA ansatz will remain unchanged up to Eq.~\eqref{eq:9}, which we re-write here:
\begin{eqnarray}
	\nonumber
	\frac{\partial \alpha}{\partial t} &=& \frac{K}{2 } [(1-f)z_\mathrm{nr}^{*} + f] - i \omega \alpha \\
	&&- \frac{K}{2} [(1-f) z_\mathrm{nr} + f] \alpha^{2},
	\label{eq:reduced_form}
\end{eqnarray}
with
\begin{equation}
    z_\mathrm{nr}= \int_{-\infty}^{\infty} \alpha^{*}(\omega , t) g(\omega) d \omega,
    \label{eq:order_para}
\end{equation}
and
\begin{equation}
    \label{eq:gaussian_dist}
    g(\omega)=\frac{1}{\sqrt{2\pi}}\exp\left[\frac{-(\omega-\omega_0)^2}{2}\right].
\end{equation}
In this case, the synchronization threshold of the bare Kuramoto model is obtained from Eq.~\eqref{eq:39} as $K_c=2\sqrt{2/\pi}$.

To arrive at an equation for the time evolution of $z_\mathrm{nr}$, we will follow the method outlined in Ref.~\cite{Campa_2022}, adapted to our system that includes subsystem resetting. To this end, we begin by expanding Eq.~\eqref{eq:gaussian_dist} in a power series to get
\begin{eqnarray}
	\label{eq:power_series}
	g(\omega)=\frac{1}{\sqrt{2\pi}}\left\{\sum_{l=0}^\infty \frac{1}{l!}\left[\frac{(\omega-\omega_0)^2}{2}\right]^l\right\}^{-1}.
\end{eqnarray}
To proceed further, we truncate the infinite series on the right hand side by retaining only the first six terms, leading to the approximate expression
\begin{eqnarray}
	\label{eq:power_series}
	g(\omega)=\frac{1}{\sqrt{2\pi}}\left\{\sum_{l=0}^\infty \frac{1}{l!}\left[\frac{(\omega-\omega_0)^2}{2}\right]^l\right\}^{-1}.
\end{eqnarray}

It is shown in Ref.~\cite{Campa_2022} in the context of the bare Kuramoto model that an approximation scheme that truncates a Gaussian $g(\omega)$ by retaining the first six terms yields results for the order parameter that are sufficiently close to the actual values obtained in numerical simulations of the bare dynamics. Eventually, in this section, we will also verify that a similar statement holds for our system by matching our analytical results with numerics. To start off with implementation of the mentioned approximation scheme, let us replace $g(\omega)$ in Eq.~\eqref{eq:order_para} with $\tilde g(\omega)$ of Eq.~\eqref{eq:approx_series} to get
\begin{equation}
	\label{eq:expanded_form}
	z_\mathrm{nr}=\frac{1}{\sqrt{2\pi}} \int_{-\infty}^{\infty} \alpha^{*}(\omega , t) \left\{\sum_{l=0}^6 \frac{1}{l!}\left[\frac{(\omega-\omega_0)^2}{2}\right]^l\right\}^{-1} d \omega.
\end{equation}

\begin{figure}
	\begin{subfigure}[b]{0.2555\textwidth}
		\includegraphics[width=\linewidth]{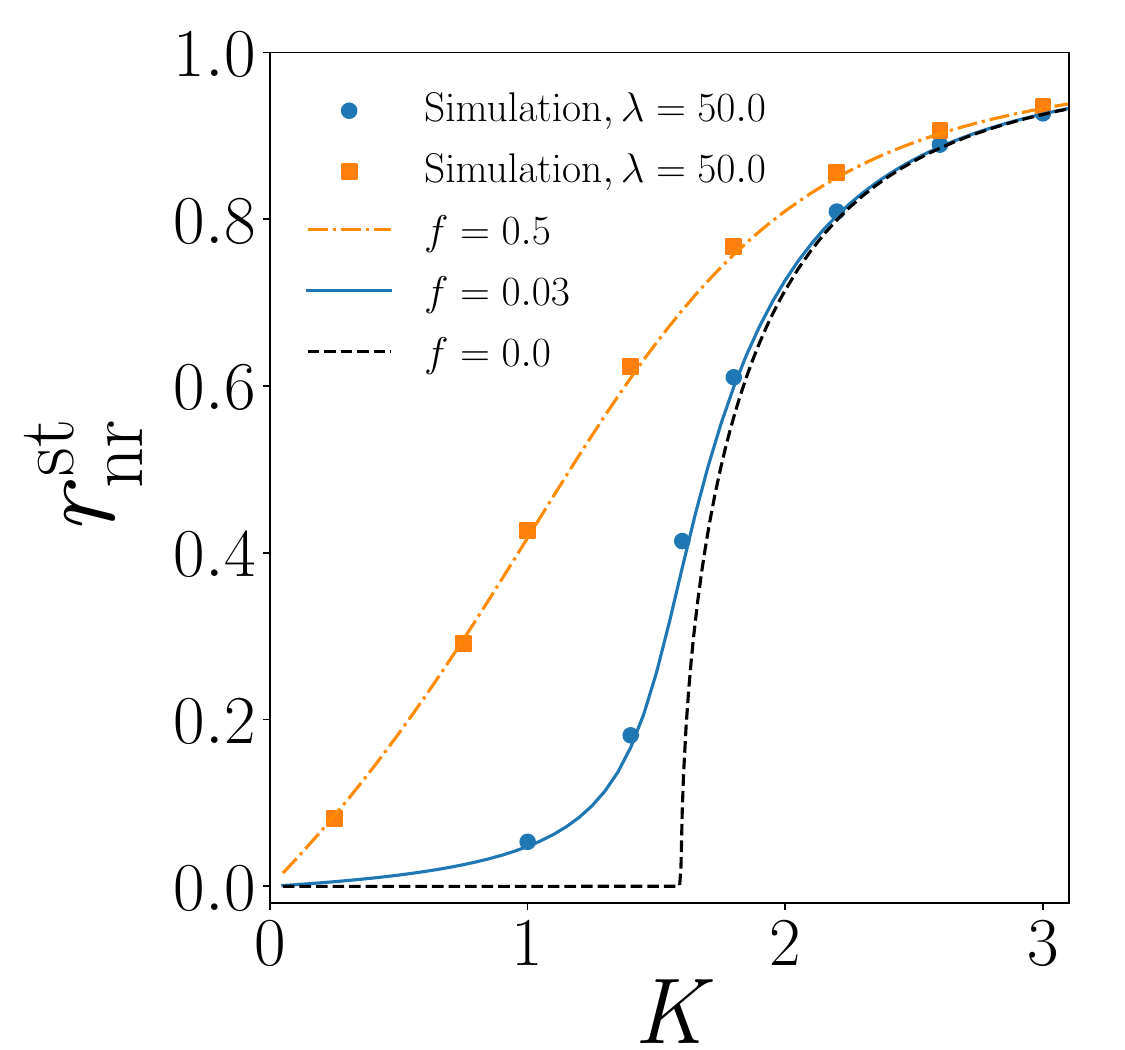}
		\caption{$\omega_0 = 0$}
		\label{fig:omegaeq0}
	\end{subfigure}%
	\begin{subfigure}[b]{0.241\textwidth}
		\includegraphics[width=\linewidth]{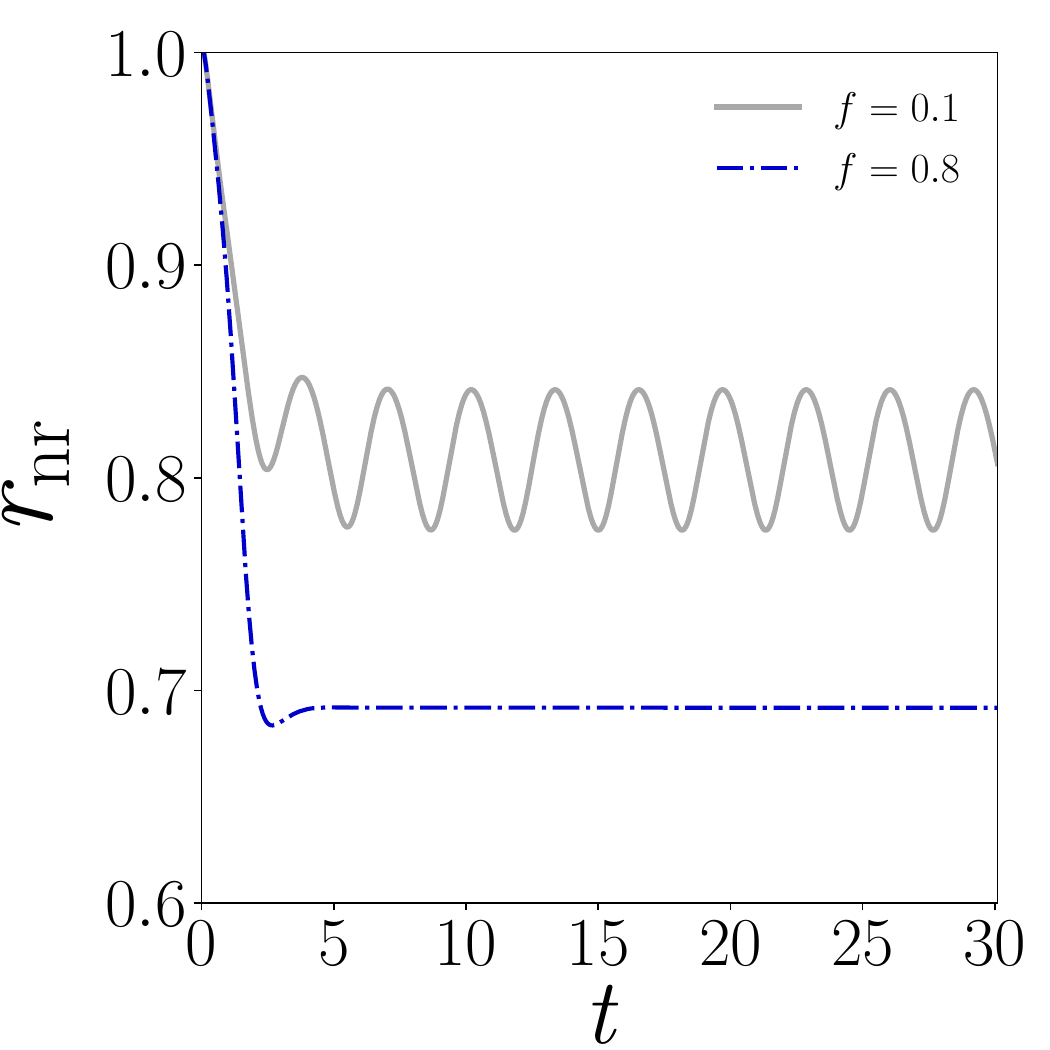}
		\caption{$\omega_0 = 2.0$}
		\label{fig:omeganeq0}
	\end{subfigure}
	\caption{\textbf{Subsystem resetting with $\lambda \to \infty$ and Gaussian $g(\omega)$ with both zero and non-zero mean $\omega_0$ and width $\sigma=1.0$}: Analytical results for the variation of $r^\mathrm{st}_\mathrm{nr}$ with $K$ for representative values of $f$ is presented in panel (a) for $\omega_0=0$. The analytical results are obtained by solving simultaneously Eqs.~\eqref{eq:order_para_sum} and~\eqref{eq:coupled_alpha}. The black dashed line corresponds to the bare Kuramoto dynamics. In this panel, we also show by points numerical simulation results for $f=0.03$ and $f=0.5$ and $\lambda=50.0$, demonstrating agreement with theory; the data correspond to the stationary state of a single realization of the dynamics for a system of $N=10^4$ oscillators with the integration time step equal to $0.01$. In panel (b), we present analytical results for $r_\mathrm{nr}$ versus time $t$ for $K = 2.5 > K_c=2\sqrt{2/\pi}$ in the case $\omega_0 = 2.0$, and for $f=0.1$ and $f=0.8$. }
	\label{fig:gaussian_dist}
\end{figure}

Similar to our analysis following Eq.~\eqref{eq:alpha-t0}, we will now extend the integral in Eq.~\eqref{eq:expanded_form} to the complex $\omega$-plane by assuming that $\alpha(\omega,t)$ can be analytically continued from real $\omega$ into the complex $\omega$-plane for all $t \geq 0$. Also, we assume that one has $|\alpha(\omega,t)| \to 0$ as Im$(\omega) \to - \infty$ and that $|\alpha(\omega,t)| \leq 1$ for real $\omega$. The integral in Eq.~\eqref{eq:expanded_form} will now have six poles, instead of a single pole as was the case with Lorentzian $g(\omega)$. Let the poles of $\tilde g(\omega)$ be denoted by $\Omega_m$ with $m=1,2,..,6$. Then, Eq.~\eqref{eq:expanded_form} can be evaluated to give
\begin{equation}
\label{eq:order_para_sum}
 z_\mathrm{nr}=-2\pi \rm{i}\sum_{m=1}^6 \alpha^{*}(\Omega_m , t) Res[\tilde g(\omega)]_{\omega=\Omega_m}, 
\end{equation}
where $\rm{Res}[\tilde g(\omega)]_{\omega=\Omega_m}$ denotes the residue of $\tilde g(\omega)$ at $\omega=\Omega_m$.
From Eq.~\eqref{eq:reduced_form}, we obtain a system of six coupled nonlinear differential equations, one for each pole $\Omega_m$ of $\tilde g(\omega)$ :
\begin{eqnarray}
	\label{eq:coupled_alpha}
	\frac{\partial \alpha(\Omega_m,t))}{\partial t} &=& \frac{K}{2 } [(1-f)z_\mathrm{nr}^{*} + f] - i \Omega_m \alpha(\Omega_m,t) \nonumber \\
	&&- \frac{K}{2} [(1-f) z_\mathrm{nr} + f] \alpha^{2}(\Omega_m,t),
\end{eqnarray}
with $m=1,2,..,6$ and $z_\mathrm{nr}$ given by Eq.~\eqref{eq:order_para_sum}. Equation~\eqref{eq:order_para_sum} together with Eq.~\eqref{eq:coupled_alpha} and the initial condition~\eqref{eq:alpha-t0} are solved numerically to obtain $r_{\rm{nr}}^{\rm{st}}(=|z_{\rm{nr}}^{\rm{st}}|)$ for our system for a given resetting fraction $f$ and a given value of the coupling constant $K$. 
The results so obtained are presented in Fig.~\ref{fig:gaussian_dist}, in which we also show a comparison with results from numerical simulations. For $\omega_0=0$, both our theory and numerical simulations verify stationary-state behavior of $r_\mathrm{nr}$ at long times, see Fig.~\ref{fig:gaussian_dist}(a). On the other hand, for $\omega_0\ne 0$, we find the existence of both oscillatory and stationary behavior of $r_\mathrm{nr}$ at long times, see Fig.~\ref{fig:gaussian_dist}(b). 

For a finite value of the resetting rate $\lambda$, the analysis is more intricate. One may proceed as in Sec.~\ref{sec:level6}, in which all equations up to Eq.~\eqref{eq:31} remain valid. However, unlike the Lorentzian case in which Eq.~\eqref{eq:31} leads to two coupled differential equations for $z_1, z_2$ given in Eq.~\eqref{eq:32}, we get instead twelve coupled differential equations arising from the six poles of $\tilde{g}(\omega)$. We have however checked in simulations that our results for a finite resetting rate and Gaussian $g(\omega)$ are qualitatively similar to those obtained for the case Lorentzian $g(\omega)$ in Sec.~\ref{sec:level5}.
 
\section{\label{sec:label10}Conclusions}
In this work, we highlight the effects of subsystem resetting in the Kuramoto model of coupled phase-only oscillators of distributed natural frequencies, which is a paradigmatic model employed in studying spontaneous synchronization. Subsystem resetting is a new protocol in the domain of resetting studies, whereby a fraction of the total number of system constituents in an interacting model system undergoes evolution according to the bare dynamics of the model and which is interspersed with simultaneous resetting at random times. By contrast, the remaining fraction evolves solely under the bare dynamics. The framework offers a contrasting dynamical scenario for the oft-studied case of global resetting, wherein the entire system undergoes simultaneous resets at random times. Nontrivial effects are expected on the basis of the fact that in the case of global resetting, each reset event initiates afresh the dynamics of the entire system, resulting in all memory of dynamical evolution getting completely washed away each time a reset happens. Consequently, what matters in determining the state of the system at any given time instant is when did the last reset event take place. This may be contrasted with what happens under the introduced protocol of subsystem resetting, where evidently the part of the system not undergoing resets retains memory of its entire dynamical evolution.  

Within the ambit of the Kuramoto model, we implement subsystem resetting by repetitively initializing a fraction of the total number of oscillators at random times to a synchronized state, which is followed by the bare dynamics of the Kuramoto model. These oscillators constitute the reset subsystem, while the non-reset subsystem is constituted by the remaining fraction of the total number of oscillators that undergoes solely the bare evolution of the Kuramoto model. Unlike the bare Kuramoto model, in our case of subsystem resetting, the mean $\omega_0$ of the natural frequency distribution of the oscillators plays a key role in determining the dynamics of the system at long times. When one has $\omega_0=0$, one has a synchronized stationary state at long times: the synchronization order parameter $r_\mathrm{nr}$ of the non-reset subsystem reaches a stationary state at long times and has then a non-zero value, irrespective of the fraction of the total number of oscillators that are being reset. On the other hand, when we have $\omega_0 \neq0$, one has at long times either a synchronized stationary state or an oscillatory synchronized state, with the latter characterized by an oscillatory behavior of $r_\mathrm{nr}$ at long times, with a non-zero time independent time average. One thus concludes that the non-reset subsystem is always synchronized at long times through the act of resetting of the reset subsystem. Remarkably, one is able to induce synchronization in the non-reset subsystem even in parameter regimes in which the bare dynamics does not support synchronization.  

In the infinite resetting rate limit, i.e.,  $\lambda\to\infty$, we map our non-reset subsystem to a system of forced Kuramoto oscillators. This feature helps us to apply the Ott--Antonsen ansatz to the non-reset subsystem and analytically derive an equation for the time evolution of  $r_\mathrm{nr}$, for two representative frequency distributions, namely, a Lorentzian and a Gaussian. The predictions of our analysis match well with numerical simulation results.

When $\lambda$ is finite, we are again able to derive an equation for the evolution of $r_\mathrm{nr}$ by applying the Ott--Antonsen ansatz separately to the reset and the non-reset subsystem and invoking an approximation that suitably decouples the evolution of the order parameters of the two subsystems. These equations are found to describe the stationary-state behaviour of $r_\mathrm{nr}$ quite accurately, as verified by a direct comparison with numerical simulation results. On the other hand, as regards the oscillatory behaviour of $r_\mathrm{nr}$, one has only a qualitative agreement between numerical simulation results and those predicted by our approximate theoretical analysis.  

In fine, we mention some interesting directions of research that may be pursued as immediate extensions of the current work. The first extension could be to systematically improve the approximate scheme employed by us in studying subsystem resetting for a finite value of the resetting rate, with the view to having a quantitative agreement with numerical simulation results for the case in which $r_\mathrm{nr}$ at long times oscillates as a function of time.
The subsystem resetting is applied here to a  Kuramoto model with mean-field (all-to-all) coupling. It would be interesting to investigate subsystem resetting in a version of the Kuramoto model defined on a network, which is a more realistic (but a more involved) dynamical set-up to study subsystem resetting. In such a scenario, additional complexity would be brought in by the network topology, but which would obviously generate interesting long-time properties carrying explicit signatures of the topology. 

\appendix
\section{\label{app:1} Details of numerical simulation of the Kuramoto model subject to subsystem resetting
}
Here, we detail the various steps involved in implementing numerical simulation of the Kuramoto model subject to subsystem resetting. The steps are as follows:
\begin{enumerate}
    \item Choose representative values of the total number of oscillators $N$, the fraction $f$ of reset oscillators, the resetting rate $\lambda$, the coupling constant $K$, and the total simulation time $\mathcal{T}$. Next, one chooses the frequency distribution $g(\omega)$ for which one wishes to study the dynamics. In our simulations, we choose $\mathcal{T}$ between $2 \times 10^1$ and $4 \times 10^2$ depending upon the value of $\lambda$. Also, we choose $N = 10^4$ and take the width of the frequency distribution to be $\sigma = 1.0$, which is tantamount to rescaling of time, as explained in the text following Eq.~\eqref{eq:13}.
    \item Choose a disorder realization $\{\omega_j \}$ of the individual oscillator frequencies, by appropriately sampling them independently from the given distribution $g(\omega)$.
    \item Choose the initial state $\{\theta_j(0)\}$ (which is also the reset state for the resetting oscillators) and the subset $n=fN$ of reset oscillators. According to our scheme of things detailed in the main text, we reset the oscillators labeled $j=1,2,\ldots,n$, while the ones with label $j=n+1,n+2,\ldots, N$ constitute the set of non-reset oscillators. In our implementation of the dynamics, we make the choice $\theta_j(0)=0~\forall~j$.
    \item Starting from the initial state, we let the system evolve under the bare Kuramoto dynamics~\eqref{eq:1} for a random time $\tau$ sampled from the exponential distribution~\eqref{eq:ptau}. This step is implemented in numerics by integrating the equations of motion of the bare model with a fourth-order Runge-Kutta algorithm. At the end of the evolution, the phases of the $n$ resetting oscillators are all reset back to the value zero instantaneously in time, while leaving unchanged the phases of the remaining oscillators. This sequence of bare evolution for a random time followed by an instantaneous reset is repeated the required number of times to ensure that the total duration of evolution equals the chosen value $\mathcal{T}$.
    \item Step 4 is repeated several times (typically of order $10^3$) to implement different realizations of the dynamics for the same disorder realization $\{\omega_j\}$. Note that the reset times vary across the different dynamical realizations. In this way, one obtains the values of $r_\mathrm{r}$ and $r_\mathrm{nr}$ as a function of time $t$ and averaged over different dynamical realizations, for a fixed disorder realization. With our choice of large $N$, the results do not change appreciably across different disorder realizations. 
\item    One may use a uniform random-number generator to sample a Lorentzian $g(\omega)$ by using standard procedure~\cite{whitlock1986monte}, while a Gaussian $g(\omega)$ may be sampled by using the standard Box-Muller algorithm~\cite{whitlock1986monte}. 
\end{enumerate}

\section{\label{app:2} Exact solution of Eq.~\eqref{eq:19}}
In this appendix, we solve Eq.~\eqref{eq:19} to obtain an exact expression for $\rm{r_{nr}^{st}}$, the stationary-state synchronization order parameter of the non-reset subsystem, in the limit $\lambda \to \infty$ and for the case of the mean frequency $\omega_0 = 0$. We achieve this by finding the roots of Eq.~\eqref{eq:19}. To unclutter the 
notations, we re-write the equation as 
\begin{eqnarray}
    (r^\mathrm{st}_\mathrm{nr})^3 + a (r^\mathrm{st}_\mathrm{nr})^2 + c~ r^\mathrm{st}_\mathrm{nr} -a =0, \label{eq:51}
\end{eqnarray}
where we have
\begin{eqnarray}
    a \equiv \left( \frac{f}{1-f} \right),~c \equiv  \left[ \frac{2}{K (1 - f)} 
    -1 \right] \label{eq:52}.
\end{eqnarray}
Equation~\eqref{eq:51} is a cubic equation, whose roots can be expressed as a function of the coefficients $a$ and $c$. From Cardano's formula~\cite{abramowitz1968handbook}, the general solution of  Eq.~\eqref{eq:51} can be written as
\begin{eqnarray}
   \nonumber &&r^\mathrm{st}_\mathrm{nr} = -\frac{a}{3}\\ 
    \nonumber &&+ \sqrt[3]{\left(\frac{a}{2}+\frac{ac}{6}-\frac{a^3}{27}  \right) + \sqrt{\left(\frac{a}{2}+\frac{ac}{6}-\frac{a^3}{27}  \right)^2+\left(\frac{c}{3}-\frac{a^2}{9}\right)^3}}\\
    \nonumber&& + \sqrt[3]{\left(\frac{a}{2}+\frac{ac}{6}-\frac{a^3}{27}  \right)- \sqrt{\left(\frac{a}{2}+\frac{ac}{6}-\frac{a^3}{27}  \right)^2+\left(\frac{c}{3}-\frac{a^2}{9}\right)^3}},\\
    \label{eq:53}
\end{eqnarray}
where $a$ and $c$ are given by Eq.~\eqref{eq:52}.

\section{\label{app:3} Implications of Eqs.~\eqref{eq:54} and~\eqref{eq:55}}
We show here by considering Eqs.~\eqref{eq:54} and~\eqref{eq:55} that in the case of $\omega_0 = 0 $, if we start from a state with $\psi_1(0) = \psi_2(0) = 0$, then we will have $\psi_1(t) = \psi_2(t) = 0~\forall~t$. 
Using $\omega_0=0$ in Eqs.~\eqref{eq:54} and~\eqref{eq:55}, we get
\begin{eqnarray}
    \frac{d\psi_1}{dt} &=& - K (1-f) \sin{\psi} \left(\frac{1 + r_1^2}{2 r_1}\right)  r_2, \label{eq:64}\\
    \frac{d\psi_2}{dt} &=& K f \sin{\psi} \left(\frac{1 + r_2^2}{2 r_2}\right)  r_1,\label{eq:65}
\end{eqnarray}
where we have $\psi = \psi_1 -\psi_2$. We can combine the above two equations to write
\begin{eqnarray}
    \nonumber
    \frac{ d \psi}{d t} &=& - K \left[ \left(\frac{1 + r_1^2}{2 r_1}\right) (1-f) r_2 + \left(\frac{1 + r_2^2}{2 r_2}\right) f r_1\right]\sin{\psi}.\\
    \label{eq:66}
\end{eqnarray}

Using the fact that $r_\mathrm{1}$~and~$r_2$ are intrinsically positive quantities, we find that Eq.~\eqref{eq:66} has the form $d\theta/dt=-\alpha \sin \theta$, with $\alpha>0$ and $\theta\in[0,2\pi)$, which evidently has one stable fixed point at $\theta=0$. We thus conclude from Eq.~\eqref{eq:66} that the long-time solution of $\psi$ is given by
\begin{eqnarray}
     \psi(t \to \infty) =  0=\psi(0).
\end{eqnarray}
Thus, $\psi(t)$ is zero both initially and at long times. If $\psi(t)$ takes nonzero values at intermediate times, it would imply then that $\psi(t)$ is a non-monotonic function of time. This in turn would imply that $d\psi/dt$ has multiple values at any given value of $\psi$, which is not allowed by the form of Eq.~\eqref{eq:66}. Hence, we conclude that one must have $\psi(t) = 0~\forall~t$. Then, from Eqs.~\eqref{eq:64}~and~\eqref{eq:65}, we get
\begin{eqnarray}
    \frac{d\psi_1}{dt} = \frac{d\psi_2}{dt} = 0 ~\forall~t,
\end{eqnarray}
which yields the desired result:
\begin{eqnarray}
    \psi_1(t) &=& \psi_1(0) = 0~\forall~t,\\
    \psi_2(t) &=& \psi_2(0) = 0~\forall~t.
\end{eqnarray}

\bibliography{paper}
\bibliographystyle{unsrt}

\end{document}